\documentclass[12pt, draftclsnofoot, onecolumn]{IEEEtran}
\IEEEoverridecommandlockouts
\usepackage{cite}
\usepackage{amsmath,amssymb,amsfonts}
\usepackage{algorithmic}
\usepackage{algorithm}
\usepackage{graphicx}
\usepackage{textcomp}
\usepackage{xcolor}
\usepackage{subfigure}
\usepackage{setspace}
\usepackage{hyperref}
\usepackage{soul}

\allowdisplaybreaks

\begin{document}


\title{\mbox{Play to Earn in Augmented Reality with} \mbox{Mobile\hspace{-1pt} Edge\hspace{-1pt} Computing\hspace{-1pt} over\hspace{-1pt} Wireless\hspace{-1pt} Networks:} \mbox{A Deep Reinforcement Learning Approach}}



\author{\IEEEauthorblockN{Terence Jie Chua, Wenhan Yu, and Jun Zhao} \vspace{-40pt}
\thanks{The authors are all with Nanyang Technological University, Singapore. Email: JunZHAO@ntu.edu.sg 
}
 \thanks{A 7-page short version containing partial results is accepted for the 2023 EAI GameNets ($12^{th}$ EAI International Conference on Game Theory and Advanced Communication Networks~\cite{ICC2023MALS}). IEEE Communications Society (ComSoc) writes ``It is possible to submit the journal and conference version at the same time'' at \url{https://www.comsoc.org/publications/journals/ieee-transactions-wireless-communications/conference-vs-journal}}
}

\maketitle

\begin{abstract}
 Play-to-earn (P2E) games have been gaining popularity as they enable players to earn in-game tokens which can be translated to real-world profits. With the advancements in augmented reality (AR) technologies, AR play-to-earn games become compute-intensive. In-game graphical scenes need to be offloaded from mobile devices to an edge server for computation. In this work, we consider an optimization problem where the Mobile edge computing Service Provider (MSP)'s objective is to reduce downlink transmission latency of in-game graphics, the latency of uplink data transmission, and the worst-case (greatest) battery charge expenditure of user equipments (UEs), while maximizing the worst-case (lowest) UE resolution-influenced in-game earning potential through optimizing the downlink  UE-Mobile edge computing Base Station (UE-MBS) assignment and the uplink transmission power selection. The downlink and uplink transmissions are executed asynchronously. We propose a multi-agent, loss-sharing (MALS) reinforcement learning model to tackle the asynchronous and asymmetric problem. We then compare the MALS model with other baseline models and show its superiority over other methods. Finally, we conduct multi-variable optimization weighting analyses and show the viability of using our proposed MALS algorithm to tackle joint optimization problems.

\end{abstract}


\section{Introduction}
\textbf{Background. }
Play-to-earn (P2E) games are a new breed of video games that allow players to earn virtual-world rewards or cryptocurrency by participating in gameplay. By combining entertainment with economic incentives, these games offer players the opportunity to turn their leisure time into a profitable endeavor. While P2E games are not restricted to Metaverse virtual worlds, the concept of play-to-earn has gained popularity within the Metaverse and blockchain gaming communities. Traditionally, players have little ownership of their in-game earnings, tokens, and characters as they belong to the game developers. However, blockchain contracts now allow players to own their in-game tokens and possessions, granting them real-world value~\cite{browne_2021}. Players can profit by playing these games, earning in-game tokens, acquiring possessions, and trading them for real-world currencies~\cite{browne_2021}. In this paper, we introduce the concept of ``earning potential", which represents the maximum earnings a player can achieve based on factors such as in-game graphics resolution. The earning potential is a key part of our problem formulation.


\vspace{-0.4cm}


\subsection{Problem \vspace{-0.2cm}}


\textbf{Compute-intensive Mobile Augmented Reality (MAR). }
Compute-intensive play-to-earn games, like \textit{Polkacity}~\cite{polka_city} and \textit{Reality Clash}~\cite{english}, utilize high-resolution graphics and complex interactions, driven by the increasing adoption of Mobile Augmented Reality (MAR)~\cite{roos_2022}. However, current mobile devices struggle to meet the demands of augmented reality (AR) applications, affecting gameplay smoothness in AR P2E games. Offloading game computation to edge computing servers is a potential solution. In the context of play-to-earn games, inefficient edge computing allocation and selection can lead to poor graphics, high latency, and signal interference, negatively impacting earning potential, gameplay smoothness, and play duration.

\textbf{Formulated Problem. }Developing an optimal strategy for maximizing P2E profitability, minimizing device energy consumption, and reducing transmission latency in the P2E games is complex. Key considerations include player profitability, in-game smoothness, and device battery levels. We formulate the problem in a way where the Mobile edge computing Service Provider (MSP) aims to minimize the (i) downlink transmission latency of graphical data, (ii) uplink transmission latency, (iii) and worst-case (greatest) UE battery charge expenditure, while maximizing (iv) the players' worst-case (lowest) in-game earning potential influenced by graphic resolution, by optimizing the UE-MBS allocation, and UE uplink transmit power. Our goal is to enhance player retention by improving in-game fluidity, experience, profitability, and battery life for extended play. We prioritize maximizing the worst-case earning potential and minimizing the worst-case battery charge expenditure across players, ensuring that no player's interests are neglected.
\vspace{-0.5cm}


\subsection{Our Approach and Rationale \vspace{-5pt}}\label{approach}To address these challenges, we propose a novel Multi-Agent Loss-Sharing (MALS) reinforcement learning (RL)-based orchestrator. It optimizes UE-MBS allocation and uplink transmission power to maximize the utility function of the MSP. Our approach involves a discrete-action space RL agent for the downlink (DL) UE-MBS allocation and a continuous-action space RL agent for UE uplink (UL) power selection. In the process of RL reward assignment, we break down the overarching objective function into parts so that the agents are assigned objectives within their control, avoiding confusion and providing clear goals. The DL and UL agents are asymmetric, with separate objectives and action space types. They execute asynchronously, with the DL agent performing UE-MBS allocation, followed by the UL agent's power selection. Although these processes seem distinct and asynchronous, optimizing both stages concurrently is crucial.


\textbf{Reinforcement learning approach over convex optimization. }
We proposed a complex play-to-earn mobile edge computing problem, which is non-convex due to the mixed-integer programming of UE-MBS allocation and uplink power transmission. Additionally, considering cumulative in-game profitability and UE battery charge expenditure adds sequential complexity, deterring solutions to be found at each transmission iteration.

\textbf{Play-to-earn under mobility. }The existence of P2E games on mobile devices would mean that players are expected to be on the move. As players move, their distance from the MBSs changes. This changing distance results in varying channel gain and hence effective rate of data transfer. In our proposed play-to-earn framework, we have to take into account players' (UEs') mobility.
\vspace{0.4cm}

\vspace{-1cm}\subsection{Existing Literature and Our Contributions \vspace{-5pt}}

In this subsection, we first present relevant literature and then highlight the contributions of this paper. We divide related studies into the following categories: 1) AR over wireless networks, 2) deep reinforcement learning for task offloading, 3) deep reinforcement learning tackling MINLP problems, and 4) multi-agent deep reinforcement learning.


\textbf{AR over wireless networks. }Numerous studies have investigated resource management and optimization challenges related to virtual, augmented, or extended reality over wireless networks, as evidenced by various works~\cite{chen2017resource,wang2021meta,liu2018edge,wang2020user}. Chen~\textit{et~al.}~\cite{chen2017resource} focus on addressing a resource allocation problem and propose a distributed machine learning algorithm to tackle it. In a different vein, Wang~\textit{et~al.}~\cite{wang2021meta} examine the optimization of VLC access points (VAP) and user-base station association to enhance virtual reality experiences in indoor wireless network environments.Liu~\textit{et~al.}~\cite{liu2018edge} propose an MAR over wireless network scenario and utilize convex optimization techniques to solve an object detection for MAR and latency of computation offloading problem. On the other hand,  Wang~\textit{et al.}~\cite{wang2020user} focus on reducing the energy consumption of users utilizing mobile augmented reality systems. Their objective is achieved through the optimization of both the configuration of the MAR (Mobile Augmented Reality) setup and the allocation of radio resources.



\textbf{Deep reinforcement learning for task offloading. }There are several works utilizing deep reinforcement learning to optimize task offloading~\cite{lu2020optimization,alfakih2020task}. \cite{lu2020optimization} introduces a novel approach using deep reinforcement learning (DRL) to address the offloading problem in large-scale heterogeneous mobile edge computing (MEC), incorporating LSTM network layer and candidate network set to enhance the DQN algorithm based on the MEC environment. \cite{alfakih2020task} propose a RL-SARSA algorithm which utilizes reinforcement learning to address the resource management problem in the edge server, enabling optimal offloading decisions that minimize system cost, including energy consumption and computing time delay.

\textbf{Deep RL for MINLP.} In addition to the works above, there are several existing works~\cite{qiao2019deep,bi2021lyapunov,truong2021partial,qiu2019online} that adopt deep RL approaches to tackle mixed-integer non-linear programming problems, as these problems are non-convex and may not yield good solutions with convex optimization approaches. \cite{qiao2019deep} present a cooperative edge caching scheme designed to enhance content placement and delivery within vehicular edge computing and networks. Their approach involves trilateral collaboration among a macro-cell station, roadside units, and smart vehicles. To address the joint optimization problem, the researchers formulate it as a double time-scale Markov decision process (DTS-MDP) and adopt a nature-inspired method based on the deep deterministic policy gradient (DDPG) framework. Through this framework, they obtain a computationally efficient suboptimal solution. \cite{bi2021lyapunov} address the joint binary offloading and system resource allocation problem by formulating it as a multi-stage stochastic MINLP problem. To overcome the coupling between decisions in different time frames, they propose LyDROO, a novel framework that integrates Lyapunov optimization and deep reinforcement learning (DRL) for effective solution. \cite{truong2021partial} introduce the ACDQN algorithm, a deep reinforcement learning approach, to jointly optimize computation offloading policy and channel resource allocation in a NOMA-assisted MEC network under dynamic network conditions. The algorithm effectively reduces computational overhead by considering partial computation offloading and a hybrid multiple access scheme that combines NOMA and orthogonal multiple access to cater to different user needs. \cite{qiu2019online} introduce a novel approach using model-free deep reinforcement learning for online computation offloading in blockchain-enabled mobile edge computing. Their method takes into account both mining tasks and data processing tasks. By formulating the problem as a Markov decision process, the researchers utilize deep reinforcement learning in conjunction with an adaptive genetic algorithm. Their objective is to maximize long-term offloading performance while effectively handling dynamic environments and reducing computational complexity. Nevertheless, these works do not consider a multi-agent transmission scenario.

\textbf{Multi-Agent Reinforcement Learning.} In contrast to these abovementioned works, there are several works which consider the concurrent optimization of several variables within their objective function and propose multi-agent reinforcement learning approaches to handle them. Guo \textit{et al.}~\cite{JO1} utilize a Centralized Training and Decentralized Execution (CTDE) framework to tackle handover and power selection. Similarly, He \textit{et al.}~\cite{JO2} adopt the CTDE~\cite{lowe2017multi} framework to address the user to channel allocation and transmission power selection problem. In our work, we utilize multiple agents, but traditional MARL algorithms (e.g., \cite{lowe2017multi, rashid2018qmix, JO1, JO2}) are not suitable for our proposed problem. These algorithms are based on centralized training and decentralized execution (CTDE) approach, where each agent has its own Critic considering all other agents' actions in a time step. However, these adaptations of MARL do not efficiently handle our problem. The simultaneous action selection and consideration of all agent actions lead to sparse observation spaces that do not capture sequential agent actions effectively. Additionally, our problem involves distinct agents with discrete and continuous actions. The existing MARL approaches are rendered impractical in tackling our proposed problem as the UL and DL transmissions in our work are considered asynchronously and the actions taken by our agents are asymmetric. Existing MARL approaches rely on agents sharing all states and actions through the Critic.



\textbf{Contributions.} We make the following contributions to the field:
\begin{itemize}
\item \emph{\textbf{Joint Optimization:}} We formulate a multi-variable optimization problem for    play-to-earn over mobile edge computing.

\item \emph{\textbf{Multi-Agent-Loss-Sharing:} }We propose a novel \textit{asymmetric} and \textit{asynchronous} multi-agent loss-sharing reinforcement learning-based orchestrator to tackle a novel asynchronous and asymmetric wireless communication problem.

\item \emph{\textbf{Comparison of our proposed methods against baseline models:}} We compare our (i) proposed method MALS with baseline methods such as (ii) independent dual-agent (IDA) and (iii) CTDE~\cite{lowe2017multi} reinforcement learning algorithm. The comparison shows the superiority of our proposed method in handling asynchronous executions.

\item \emph{\textbf{Analyses of weighted utility objective functions:}} We provide in-depth analyses of how the different weights on each factor in the joint optimization functions influence reward and other variables. These analyses provide insights into solving multi-agent joint-optimization problems.

\end{itemize}
The paper is organized as follows: Section~\ref{models} presents the play-to-earn in augmented reality (AR) with mobile edge computing optimization problem, while Section~\ref{RL} discusses its formulation using reinforcement learning. Our novel MALS deep reinforcement learning model is introduced in Section~\ref{methodss}, along with a detailed algorithm. Thorough experiments in Section~\ref{experiment} compare different methods, highlighting the superiority of the MALS algorithm for handling asymmetric and asynchronous problems. We also analyze the weighting of variables in utility functions to demonstrate the feasibility across various optimization scenarios. Finally, Section~\ref{section:conclusion} provides the paper's conclusion.

\textbf{Differences between this Journal and Our Conference Paper~\cite{ICC2023MALS}. }This journal builds upon the work in~\cite{ICC2023MALS}, focusing on the worst-case earning potential in the downlink stage and the worst-case battery charge expenditure in the uplink stage for optimization, unlike the conference version which considered the sum of potentials/expenditure. Additionally, a multi-variable optimization weighting analysis is included to demonstrate the algorithm's suitability for joint optimization problems, and more detailed rationale and explanations.


\vspace{-15pt}\section{System Model \vspace{-5pt}}
\label{models}
\textbf{Problem Scenario. }
Consider the real-time downlink (DL) and uplink (UL) transmission of $N$ players from a set $\mathcal{N}=\{1,2,...,N\}$ players (UE) moving about. In each complete transmission iteration, each UE $i \in \mathcal{N}$ begins downloading P2E in-game scenes from an MBS. We consider interference based on a Non-Orthogonal Multiple Access (NOMA) system in both the DL and UL transmission. After the DL of in-game graphical data, we consider the UL transmission of the same set of $N$ players' (UE) changes to in-game graphical scenes, to its allocated MBS $(\mathcal{M} = \{1,2,...,M\}$. Each UE $i \in \mathcal{N}$ will upload their in-game graphical scenes to their allocated MBS. We use iterations to refer to a complete round of data transmissions. Each iteration includes a downlink and an uplink data transmission. To simplify an already complex scenario, we consider the scenario in which users only make movement after each iteration of downlink and uplink data transmissions.


\begin{figure}[t]
\centering
\includegraphics[width=0.7\textwidth]
{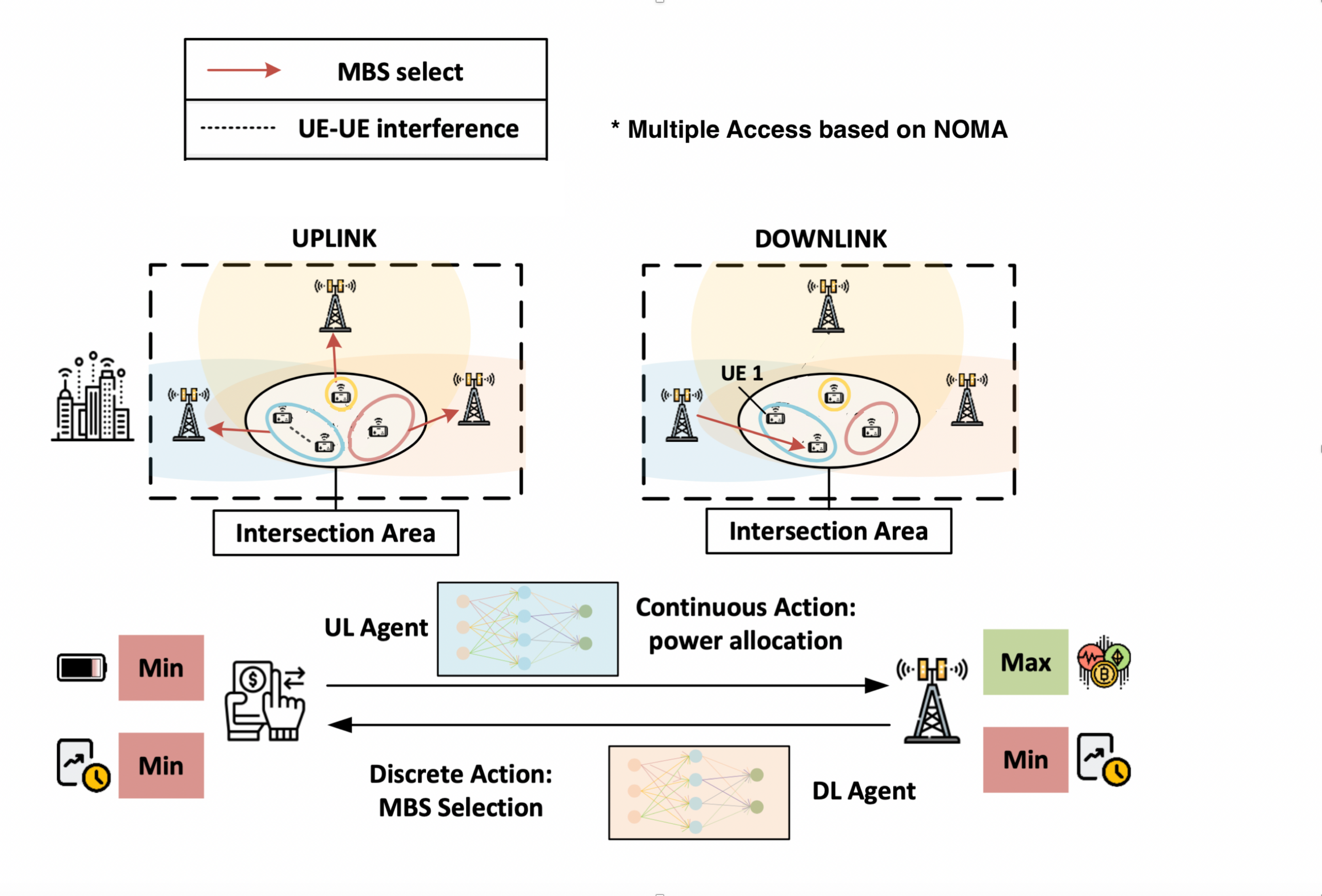}
\vspace{-0.5cm}
\caption{System model illustrating the interaction between the UEs and the MBSs, agents' objectives and variables-of-control.}
\label{fig:system_model}
\vspace{-0.8cm}
\end{figure}


\vspace{-15pt}\subsection{Downlink Communication \vspace{-5pt}}\label{sub:Communication-Model}


Each of the MBS from $\mathcal{M}=\{1,2,...,M\}$ will be allocated to some UEs from $\mathcal{N}=\{1,2,...,N\}$. We define the size of the in-game graphics to be downloaded from the MBSs to the UEs as $D^t = \{D^t_{1}, D^t_{2}, ..., D^t_{N}\}$, where $D^t_i$ represents the in-game data size to be downloaded from an MBS to UE $i \in \mathcal{N}$ at iteration $t$. In addition, the UE-MBS association is defined as: $\boldsymbol{c}^{t}=(c_1^t, ..., c_N^t)$. Each of the $c_i^t$ where $i \in \mathcal{N}$ is allocated to an MBS $v$ where $v \in \mathcal{M}$ at iteration $t$. We adopt the Non-Orthogonal Multiple
Access (NOMA) system as this work’s propagation model since it allows multiple users to share
the same frequency band~\cite{dai2018survey}. In a NOMA multiple access channel scheme, users (UEs) can download in-game graphics from MBSs efficiently and concurrently. When a UE initiates a request for downloading in-game graphics, the allocated MBS employs NOMA principles to allocate power levels and transmit the data to multiple users simultaneously. We define each MBS to be on a single distinct channel. Each MBS determines the power allocation based on the channel conditions of each of its assigned UE. Each MBS then superimposes the encoded graphics data for all its assigned UEs into a composite signal, incorporating the power allocation scheme. The composite signal is transmitted over the wireless channel, and each UE receives the signal along with interference from other UEs assigned to the same MBS. Through successive interference cancellation (SIC), each UE performs interference removal by canceling the contributions of other UEs' signals based on the power allocation and channel conditions. Each UE decodes their own intended graphics data while mitigating the interference from other UEs' signals. Once decoded, the UEs can perform necessary decompression and rendering operations to display the downloaded in-game graphics. 

The sequence of user devices (UEs) in the downlink (DL) is established by organizing them in a descending manner, with respect to their channel-to-noise ratios. We use $\Bar{d}_1,\Bar{d}_2,\ldots,\Bar{d}_{N}$ to denote the ordering, as referenced in \cite{NOMA}:
\begin{align}
    \frac{|g^t_{\Bar{d}_1,v}|^2}{(\sigma_{\Bar{d}_1,v}^t)^2}\geq 
    \frac{|g^t_{\Bar{d}_2,v}|^2}{(\sigma_{\Bar{d}_2,v}^t)^2}\geq \ldots \geq \frac{|g^t_{\Bar{d}_N,v}|^2}{(\sigma_{\Bar{d}_N,v}^t)^2}.
\end{align}
Consequently, the downlink rate $r_{i}^{\text{d},t}$ for UE $i$ is given by:
\begin{align}
    r_{i}^{\text{d},t}(\boldsymbol{c}^t) \hspace{-2pt}=\hspace{-2pt}
    B_v\hspace{-1pt}\log_2 \hspace{-2pt}\left(\hspace{-2.5pt} 1\hspace{-2pt}+\hspace{-2pt}\frac{p^{\text{d},t}_{c_i^t,i}|g_{i,v}^t|^2}
    {\sum\limits_{j=1}^{\iota'-1}\hspace{-1pt}p^{\text{d},t}_{c_i^t,\Bar{d_j}}|g_{i,v}^t|^2\hspace{-2pt}+\hspace{-2pt}B_v(\sigma_{i,v}^t)^2}\hspace{-2.5pt}\right)\hspace{-2.5pt},
    \label{eq:dlrate}
\end{align}
where $\iota'$ fulfils $c^{t}_{\iota'}=i$, indicating that UE $i$ holds the $\iota'$th position among the UEs assigned to MBS $v$ in terms of the magnitude of channel-to-noise ratio. $(\sigma_{i,v}^t)^2$ represents the power spectral density of the Additive White Gaussian Noise experienced by UE $i$ over the channel by MBS $v$ at time step $t$. $B_v$ corresponds to the bandwidth of MBS $v$. $p^{\text{d},t}_{c_i^t,i}$ represents the downlink transmission power provided by MBS $c_i^t$ to enable the transmission of in-game graphical scenes to UE $i$ during transmission step $t$ (the superscript "d" denotes downlink). $g_{i,v}^t$ denotes the channel gain between UE $i$ and MBS $v$ at transmission step $t$.

A higher downlink data transmission rate will reduce the downlink data transmission latency for a given in-game data size to be downloaded. The downlink transmission delay can be defined as:
\begin{align}
\ell^{\text{d},t}_i = \frac{D^{t}_i }{r^{\text{d},t}_i}
\label{eq:R5} 
\end{align}
where $D^{t}_i$ is the size of in-game data  to be downloaded from an MBS to UE $i$ at transmission step $t$ and the resolution-impacted earning potential function $\omega(r^{\text{d},t}_i)$ is influenced by the UE $i$'s signal-to-noise-ratio (SINR). Intuitively, a more efficient UE to MBS allocation results in a higher UE $i$ SINR, UE $i$ downlink transmission rate  $r^{\text{d},t}_{i}$ (data transmitted per unit of time), and lower  downlink transmission delay $\ell^{\text{d},t}_i$, at iteration $t$. A higher data transmitted per unit of time is translated to improved graphic resolution per unit of time, resulting in an enhancement in players' in-game visuals, which translates to better in-game performance and hence, better resolution-influenced earning potential.

\vspace{-0.3cm}



\vspace{-10pt}\subsection{Uplink Communication \vspace{-0.2cm}}\label{sub:Communication-up-Model}
In the uplink leg of the communication model, each player (UE) will upload its graphical changes to the previously downloaded P2E in-game scene data, to its assigned MBS. In the context of Play-to-Earn games, these in-game graphical changes could be induced by UEs' physical-world actions. These graphical changes to the scenes are denoted $F^t = \{F^t_{1}, F^t_{2}, ..., F^t_{N}\}$. 

In the uplink (UL) NOMA system, the players (UEs) assigned to a MBS are ordered in descending order based on their received signals. The resulting indices of the UEs are denoted as $\Bar{u}_1,\Bar{u}_2,\ldots, \Bar{u}_{N_v^t}$. According to this arrangement, the power levels and channel gains of the UEs satisfy the following inequality~\cite{dai2018survey}:
\begin{align}
    p^{\text{u},t}_{\Bar{u}_1}|g_{\Bar{u}_1,v}^t|^2 \geq p^{\text{u},t}_{\Bar{u}_2}|g_{\Bar{u}_2,v}^t|^2 \geq \ldots \geq p^{\text{u},t}_{\Bar{u}_{N_v^t}}|g_{\Bar{u}_{N_v^t},v}^t|^2,
\end{align}
where $g_{i,v}^t$ represents the channel gain between UE $i$ and MBS $v$ at iteration $t$. This study assumes equivalent UL and DL channel conditions. Based on the findings presented in \cite{NOMA}, the uplink rate $r_{i}^{\text{u},t}$ for UE $i$ is defined as follows:
\begin{align}
    r_{i}^{\text{u},t}(\boldsymbol{c}^t, \boldsymbol{p}^{\text{u},t}) \hspace{-2pt} = \hspace{-2pt} B_v\log_2 \hspace{-2pt} \left( \hspace{-2pt} 1 \hspace{-2pt} + \hspace{-2pt} \frac{p^{\text{u},t}_{c_i^t,i}|g_{i,v}^t|^2}
    {\sum_{j=\eta+1}^{N}p^{\text{u},t}_{c_i^t,\Bar{u}_j}|g_{\Bar{u}_j,v}^t|^2\hspace{-2pt} + \hspace{-2pt} B_v(\nu_{v}^t)^2} \hspace{-2pt} \right)\hspace{-2pt} ,
    \label{eq:ulrate}
\end{align}
where $\eta$ is determined such that $\Bar{u}_{\eta}=i$, indicating that UE $i$ is in the $\eta$-th position among the set of UEs, $\mathcal{N}_v^t$, ranked in terms of the magnitude of channel-to-noise ratio. In Equation \ref{eq:ulrate}, $(\nu_{v}^t)^2$ is the power spectral density of the Additive White Gaussian Noise experienced at MBS $v$. The term $p^{\text{u},t}_{c_i^t,i}$ (with the superscript "u" denoting uplink) corresponds to the transmission power of UE $i$ for facilitating the uplink transmission of graphical changes in-game data during iteration $t$.

The uplink transmission latency of UE $i$ at transmission step $t$ can be formulated as:
\begin{align}
\ell^{\text{u},t}_i = \frac{F^{t}_{i} }{r^{\text{u},t}_{i}}
\label{eq:R10} 
\end{align}
where $F^{t}_{i}$ is the uplink in-game graphics data size of UE $i$ at transmission step $t$, $r^{\text{u},t}_{i}$ is the uplink data transmission rate of UE $i$ at transmission step $t$. UE $i$'s energy expenditure for the uplink transmission of in-game graphics data at transmission step $t$ can be formulated as:
\begin{align}
E^t_{i} = p^{\text{u},t}_{c_i^t,i} \cdot \ell^{\text{u},t}_i.
\label{eq:R11} 
\end{align}



We define $Q^t_{i}$ to be the UE $i$ energy consumption as a percentage of the UE $i$'s initial battery amount, at transmission step $t$. $Q^t_{i}$ can be defined as: $Q^t_{i} = \frac{E^t_{i}}{Bat^0_{i}}\cdot 100$. In this context, $Bat^0_{i}$ represents the UE $i$'s initial battery amount while $E^t_{i}$ represents UE $i$'s energy consumption, at transmission step $t$. Utilizing a UE energy consumption as a percentage of the UE's initial battery charge is a natural choice. This is because normalized values are more conducive for training reinforcement learning agents. The control of uplink transmission power influences the uplink transmission latency, which influences fluidity. On the other hand, the selection of uplink transmission power also influences UE battery expenditure.

\vspace{-0.4cm}

\vspace{-5pt}\subsection{UE Mobility Model} \vspace{-5pt}

We consider the network area to be a two-dimensional coordinate system, $x$-dimension and $y$-dimension, where the UEs and MBSs are positioned. 
At each given iteration step $t$, UE $i$ is located at $(\mathcal{X}^t_i,\mathcal{Y}^t_i)$. The locations of UEs are restricted to an area, $\mathcal{Y}^t_i \in [0, Y_{max}]$ and $\mathcal{X}^t_i \in [0, X_{max}]$. $\mathcal{X}^t_i$ refers to UE $i$'s position in the $x$-dimension at transmission step $t$, while $\mathcal{Y}^t_i$ refers to UE $i$'s position in the $y$-dimension at transmission step $t$. $X_{max}$ is the defined boundary in the $x$-dimension while $Y_{max}$ is the defined boundary in the $y$-dimension. After each round of data transmission (downlink followed by uplink), UEs will take steps $x$ and $y$ in both $x$-dimension and $y$-dimension, with the step size uniformly sampled from $(-x_{max}, x_{max})$ and $(-y_{max}, y_{max})$, respectively. $x_{max}$ refers to the maximum allowed distance a UE can move in a single transmission iteration in the $x$-dimension. $y_{max}$ refers to the maximum allowed distance a UE can move in a single transmission iteration in the $y$-dimension. The $x$-dimension position $\mathcal{X}^t_i$ can be defined as:
\begin{align}
    \mathcal{X}^t_i =
    \begin{cases}
    0,& \text{if } \mathcal{X}^{t-1}_i + x\leq 0,\\
    X_{max},              & \text{if } \mathcal{X}^{t-1}_i + x\geq X_{max},\\
    \mathcal{X}^{t-1}_i + x,
    & \text{otherwise,}
    \label{eq:x_movement}
    \end{cases}
\end{align}
Likewise, the $y$-dimension position $\mathcal{Y}^t_i$ can be defined as:
\begin{align}
    \mathcal{Y}^t_i =
    \begin{cases}
    0,& \text{if } \mathcal{Y}^{t-1}_i + y\leq 0,\\
    Y_{max},              & \text{if } \mathcal{Y}^{t-1}_i + y\geq Y_{max},\\
    \mathcal{Y}^{t-1}_i + y,
    & \text{otherwise,}
    \label{eq:y_movement}
    \end{cases}
\end{align}
\vspace{-0.5cm}



\vspace{-10pt}\subsection{Problem formulation \vspace{-5pt}} \label{problemform}

To sum up, the MSP's DL objective is to find the optimal UE-MBS allocation $\boldsymbol{c}^{t}$ which minimizes the total downlink latency $\ell^{\text{d},t}_i$ while maximizing the worst-case (lowest) resolution-influenced UE earning potential $\min_{i \in \mathcal{N}}\sum_{t=1}^{T}\omega(r^{\text{d},t}_i)$ across all UEs. The number of transmission steps in an episode is denoted by $T$. We formulate our downlink utility function as:
\begin{align}
\setlength{\belowdisplayskip}{4pt plus 1pt minus 1.0pt}
\setlength{\belowdisplayshortskip}{4pt plus 1pt minus 1.0pt}
\setlength{\abovedisplayskip}{4pt plus 1pt minus 1.0pt} \setlength{\abovedisplayshortskip}{0.0pt plus 2.0pt}
\min_{\boldsymbol{c}^{t}} & \left [ q \cdot\left(\frac{1}{N}\sum_{i\in\mathcal{N}}  \sum_{t=1}^{T}  \ell^{\text{d},t}_i \right )- (1-q) \cdot b\cdot \min_{i \in \mathcal{N}}\sum_{t=1}^{T}\omega(r^{\text{d},t}_i)\right ]   , \label{eq:M1}\\
s.t.~~& c_{i}^{t} \in \mathcal{M}, \forall t \in \mathcal{T},~\forall i \in \mathcal{N},\label{eq:c2}\\
~~& p^{\text{u}}_{min} \leq p^{\text{u},t}_{i} \leq p^{\text{u}}_{max}, \forall t \in \mathcal{T},~\forall i \in \mathcal{N}.
\label{eq:c2.1}
\end{align}

 where $\mathcal{T}$ represents the set of all transmission steps in an episode. Our resolution-impacted in-game earning potential is motivated by a recent work \cite{feng2022resource} and is modeled as: $\omega(r^{\text{d},t}_i) = P\cdot \ln\left(1+{r^{\text{d},t}_i}\right )$, where $P$ is the player-specific profitability factor (using such logarithmic function as the utility also appears in the classical work on crowdsourcing~\cite{yang2012crowdsourcing}). The logarithmic function has a increasing first-order derivative and decreasing second-order derivative property. When applied to the P2E profit model, the increasing first-order derivative implies that UE in-game profits increases as the downlink transmission rate increases, while the decreasing second-order derivative implies that the rate of profit increment decreases as the downlink transmission rate increases. It is natural to model our profit model as such as a larger downlink data (resolution) per unit time improves UEs' visibility within the games, allowing them to perform better and reap higher profits. However, as the resolution improves, its impact on UE's visibility and hence earnings, decreases, resulting in a slowing rate of profit increment with increasing downlink transmission rate. The constraint~(\ref{eq:c2}) ensures that each UE is assigned to one MBS in each transmission step. $q$ is a factor which determines the weighting on $\sum_{i\in\mathcal{N}}  \sum_{t=1}^{T}\ell^{\text{d},t}_i$ and $b\cdot\min_{i \in \mathcal{N}}\sum_{t=1}^{T}\omega(r^{\text{d},t}_i)$. $b$ is a factor that places both terms $\sum_{i\in\mathcal{N}}  \sum_{t=1}^{T}\ell^{\text{d},t}_i$ and $b\cdot \min_{i \in \mathcal{N}}\sum_{t=1}^{T}\omega(r^{\text{d},t}_i)$ on the same measurement unit. Constraint~(\ref{eq:c2.1}) ensures that the uplink transmission power by UE falls between the feasible range of output powers.

On the other hand, MSP's UL objective is to obtain the optimal UE uplink power $p^{\text{u},t}$ that minimizes the total latency of the data uplink transmissions $\ell^{\text{u},t}_i$ and worst-case (greatest) UE device energy expenditure $Q^t_i$. We formulate our uplink utility function as:
\begin{align}
\setlength{\belowdisplayskip}{4pt plus 1pt minus 1.0pt}
\setlength{\belowdisplayshortskip}{4pt plus 1pt minus 1.0pt}
\setlength{\abovedisplayskip}{4pt plus 1pt minus 1.0pt} \setlength{\abovedisplayshortskip}{0.0pt plus 2.0pt}
\min_{\boldsymbol{p^{t}}} &\left [ h \cdot\left (\frac{1}{N} \sum_{i\in\mathcal{N}}\sum_{t=1}^{T}  \ell^{\text{u},t}_i \right ) + (1-h) \cdot f\cdot \max_{i \in \mathcal{N}}\sum_{t=1}^{T}Q^t_i  \right ], \label{eq:M3}\\
s.t.~~& \sum_{t=1}^{T}Q^t_i \leq 100, \forall i \in \mathcal{N},\label{eq:c3}\\
~~& p^{\text{d}}_{min} \leq p^{\text{d},t}_{i} \leq p^{\text{d}}_{max}, \forall t \in \mathcal{T},~\forall i \in \mathcal{N}.\label{eq:c3.1}
\end{align}

where $\ell^{\text{u},t}_i$ is the player (UE) $i$'s uplink data transmission delay at transmission iteration $t$. $h$ is a factor that determines the weighting on $\sum_{i\in\mathcal{N}}\sum_{t=1}^{T} \ell^{\text{u},t}_i$ and $f\cdot \max_{i \in \mathcal{N}}\sum_{t=1}^{T}Q^t_i$. Scaling factor $f$ in the objective function is to place both terms $\sum_{i\in\mathcal{N}}\sum_{t=1}^{T}\ell^{\text{u},t}_i$ and $f\cdot \max_{i \in \mathcal{N}}\sum_{t=1}^{T}Q^t_i$ on the same measurement unit. Constraint~(\ref{eq:c3}) ensures that the UE battery consumption has to be less than 100\% (or battery charge (\%) has to be greater than 0) for the uplink transmission. Finally, constraint~(\ref{eq:c3.1}) ensures that the MBS downlink power output falls within a feasible range.

To sum up, we aim to minimize the UL and DL transmission latency, worst-case P2E earning potential, worst-case UE battery charge expenditure. We write our overarching (overall) objective as:
\begin{subequations}
\begin{align}
    &\min\limits_{\boldsymbol{c}^t,\boldsymbol{p}^t} 
    \Bigg\{w_1 \left [ q \cdot\left(\frac{1}{N}\sum_{i\in\mathcal{N}}  \sum_{t=1}^{T}  \ell^{\text{d},t}_i \right )- (1-q) \cdot b\cdot \min_{i \in \mathcal{N}}\sum_{t=1}^{T}\omega(r^{\text{d},t}_i)\right ]+\notag\\
    &~~~~~~~~w_2\left [h \cdot\left (\frac{1}{N} \sum_{i\in\mathcal{N}}\sum_{t=1}^{T}  \ell^{\text{u},t}_i \right ) + (1-h) \cdot f\cdot \max_{i \in \mathcal{N}}\sum_{t=1}^{T}Q^t_i \right ]\Bigg\} \label{obj:eq1}\\
    &s.t.~\text{C1}:\sum_{t=1}^{T}Q^t_i \leq 100, \forall i \in \mathcal{N}, \\
    &~~~~~\text{C2}:c_{i}^{t} \in \mathcal{M}, \forall t \in \mathcal{T},~\forall i \in \mathcal{N}, \\
    &~~~~~\text{C3}:p^{\text{d}}_{min} \leq p^{\text{d},t}_{i} \leq p^{\text{d}}_{max}, \forall t \in \mathcal{T},~\forall i \in \mathcal{N}, \\
    &~~~~~\text{C4}:p^{\text{u}}_{min} \leq p^{\text{u},t}_{i} \leq p^{\text{u}}_{max}, \forall t \in \mathcal{T},~\forall i \in \mathcal{N}.
\end{align}
\end{subequations}
\vspace{-1.2cm}

\section{Reinforcement learning settings \vspace{-5pt}}
\label{RL}
We assign two reinforcement learning agents, one (DL agent) controlling the downlink transmission UE-MBS allocation, and the other (UL agent) controlling the uplink transmission power selection. The DL agent and UL agent serve the overarching objective (\ref{obj:eq1}). These reinforcement learning agents will be deployed as an overarching cloud orchestrator to manage resource requirements for the mobile edge computing (MEC) between MBSs and UEs. The DL and UL RL agents are first trained offline using simulations or in a controlled environment. The trained RL agents are then integrated into the MSP. This involves implementing the RL agents as software that can run on the server infrastructure. In the downlink transmission, the MBS sends channel gain information between UEs and MBS, as well as in-game scene data size, to the MSP cloud server. The DL agent at the cloud server makes decisions based on this information and communicates the UE-MBS allocation to the MBSs. Similarly, in the uplink transmission, the UEs transmit information about their battery life to the MSP cloud. The UL agent at the cloud server makes decisions based on this information and communicates the UE uplink power selection to the UEs. These information transmissions occur through dedicated channels. 
\vspace{-0.6cm}

\subsection{State \vspace{-5pt}}

For the DL agent's observation state $s^{\text{d}}$, we choose to include 1) channel gain between UEs and MBSs: $g^t_{v,i}$, 2) in-game scene DL data size at each transmission iteration $t$: $D^{t}_{i}$, as $D^{t}_{i}$ influences latency $\ell^{\text{d},t}_{i}$, and $g^t_{v,i}$ influences both data DL transmission latency $\ell^{\text{d},t}_{i}$ and worst-case (lowest) resolution-influenced in-game earning potential $\min_{i \in \mathcal{N}}\sum_{t=1}^{T}\omega(r^{\text{d},t}_i)$.

For the UL agent's observation state $s^{\text{u}}$, we choose to include 1) channel gain between UEs and MBSs: $g^t_{v,i}$, as it influences data UL transmission rate, latency $\ell^{\text{u},t}_{i}$ and uplink transmission battery consumption $Q^{t}_{i}$. As we consider the cumulative battery amount consumed by UEs and are enforcing that each UE has positive battery-charge to continue the uplink transmission at each transmission iteration step $t$, we have to include 2) UE battery life (\%): $\left [ \frac{Bat^{t}_{i}}{Bat^{0}_{i}}\cdot 100 \right ]$ , as UE battery life changes in each iteration.
\vspace{-0.6cm}


\vspace{-2pt}\subsection{Action \vspace{-5pt}}
In the DL communication model, the DL agent's action involves deciding the UE to MBS assignment. The UE to MBS allocation action at transmission step $t$ is defined as: $a^{\text{d},t}=\textbf{c}^{t}=\{c_1^t,...c_N^t\}.$ In the UL communication model, the UE power for uplink data transmission at transmission step $t$ is defined as: $a^{\text{u},t}=\textbf{p}^{\text{u},t}=\{p^{\text{u},t}_1, ... ,p^{\text{u},t}_N\}$. $p^{\text{u},t}_i$ represents UE $i$'s selected power for the uplink data transmission at transmission step $t$.
\vspace{-0.4cm}


\vspace{-5pt}\subsection{Reward \vspace{-5pt}}
\label{subsection:reward}

For the uplink communication model, the reward issued to the UL agent at transmission step $t$ is defined as:
\begin{align}
    R^{\text{u},t} = 
\begin{cases}
    -h\cdot\frac{ \sum_{i \in \mathcal{N}}\ell^{\text{u},t}_i}{N} + \left(1-h\right) \cdot f \cdot \min_{i \in \mathcal{N}}\sum_{\tau=1}^{t}Q^t_i,& \text{if } \sum_{\tau=1}^{t}Q^t_i \leq 100,~\forall i \in \mathcal{N}\\
    \varrho,              & \text{otherwise.}
    \label{eq:reward_up}
\end{cases}
\end{align}
The intuition for the downlink and uplink reward assignment follows the utility functions in Equations (\ref{eq:M1}) and (\ref{eq:M3}), respectively, and the overarching objective (\ref{obj:eq1}). For the \textbf{downlink} communication model, the reward given to the DL agent at transmission step $t$ is given as such:
\begin{align}
    R^{\text{d},t} = 
\begin{cases}
    - q\cdot\frac{ \sum_{i \in \mathcal{N}}\ell^{\text{d},t}_i }{N} +  \left(1-q\right) \cdot b\cdot\min_{i \in \mathcal{N}}\sum_{\tau=1}^{t}\omega(r^{\text{d},t}_i) + \varkappa R^{\text{u},t},& \text{if } \sum_{\tau=1}^{t}Q^t_i \leq 100,~\forall i \in \mathcal{N}\\
    \varrho,              & \text{otherwise.}
    \label{eq:reward_down}
\end{cases}
\end{align}
$\varrho$ is a very large penalty and $\tau$ denotes the summation of index. Note that the DL and UL agents are given only portions of the overarching objective that they control as this is to avoid ``confusing" agents with rewards that they do not control. The downlink reward comprises of a weighted component of the uplink reward as the DL agent's allocation decisions have consequences on the UL transmission latency and battery consumption as the UL transmission latency and energy consumption is controlled by the composition of UL transmission power selection and the UE-MBS allocation. By allocating the DL agent with a component of the UL reward, the DL agent can have a more comprehensive view of both the DL and UL process and optimize for both the DL and UL rewards. On the other hand, the selection of UL transmission power, and UE-MBS allocation influences UEs' battery life and whether further transmissions can be executed. Hence both the DL and UL agents are also issued a large penalty $\varrho$ should any UE have no remaining battery charge for transmission. In such circumstance, the episode ends and no further transmission continues. Following the logic above, in our work, we do not present the UL agent with rewards pertaining to downlink transmission latency as downlink transmission latency is not within the UL agent's control. The decomposition of rewards is inspired by~\cite{van2017hybrid}, and is known to be an effective method for training of reinforcement learning agents. Hence, from the above, we note that the UL and DL agents are deeply related, and the asynchronous execution of DL and UL transmissions must be optimized concurrently. 

\vspace{-0.1cm}

\vspace{-5pt}\section{\resizebox{.95\textwidth}{!}{Our~Multi-Agent~Loss-Sharing~(MALS)~Reinforcement~Learning\vspace{-7pt}~Algorithm}}
\label{methodss}

\textbf{Novelty of our MALS algorithm.}
Given that the MSP has multiple variables to optimize, rudimentary reinforcement learning agents face challenges in achieving their objectives and may encounter difficulties in achieving convergence. One reason is that independent agents act in their own interest, choosing the best actions to maximize their own rewards, even if it is at the expense of other agents. In a mixed-cooperative scenario as proposed in our study where each agent acts in its interest, agents' greedy actions may result in sub-optimal action solutions learned across all agents. To tackle this issue, we proposed a novel multiple-input, multiple-objective (head) critic which provides actors with actor-centric guidance, facilitating the actors in breaking down complex objectives into simpler ones. A (multi) dual-head critic refers to a critic model with (multiple) two distinct neural network input layers, each to accommodate to the input state of each agent, while sharing a common neural network backbone. The critic head serving the uplink actor beacons the UL actor in (i) minimizing the UL transmission delay and worst-case UE battery consumption in the uplink transmissions. The critic head serving the downlink actor beacons the DL actor in (ii) minimizing the DL transmission delay, maximizing the UE resolution-influenced in-game earning potential, minimizing the UL transmission delay and worst-case UE battery consumption in the uplink transmissions. To ensure that each agent involved learns a jointly optimal solution, the loss value of each critic's head is shared (weighted sum) for the joint update of the critic model. We refer to the above approach as Multi-Agent~Loss-Sharing~(MALS).
 
In the following section, we will introduce the PPO algorithm, its mechanism as it is the backbone algorithm in which our MALS algorithm is built upon.
\vspace{-0.5cm}

\begin{figure}[t] \label{fig:RL_structure}
\centering
\includegraphics[width=1\linewidth]{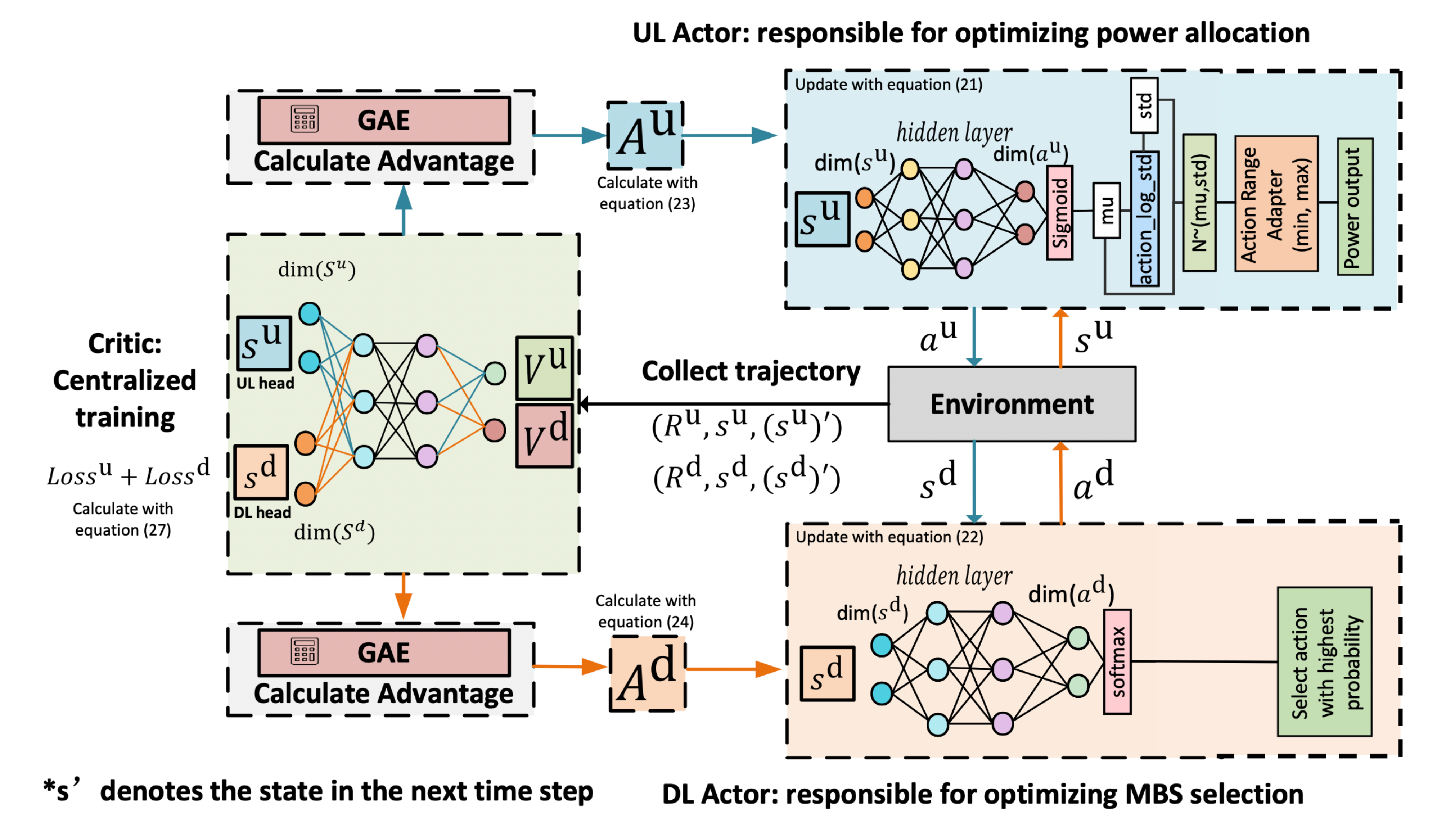}
\vspace{-30pt}\caption{Reinforcement Learning Multi-Agent-Loss-Sharing Proximal Policy Optimization (MALS-PPO) structure and model update.}
\label{fig:model}
\vspace{-25pt}
\end{figure}

\vspace{-0.17cm}\subsection{Proximal Policy Optimization (PPO)\vspace{-0.2cm}}
Our focus lies in the development of a multi-agent RL model that addresses tasks with asymmetrical characteristics in an asynchronous manner. In this context, ensuring policy stability becomes crucial. OpenAI introduced the Proximal Policy Optimization (PPO) algorithm~\cite{schulman2017proximal} as an improvement over the traditional policy gradient approach. PPO achieves better sample efficiency by utilizing separate policies for trajectory sampling and optimization, while also incorporating a policy constraint to enhance stability.


To summarize, PPO exhibits two key features within its policy network (Actor): (1) Enhanced sample efficiency: PPO employs two distinct policies, namely the evaluation policy denoted as $\pi_\theta$ and the data-sampling policy denoted as $\pi_{\theta^{'}}$, to improve the efficiency of sampling trajectories and optimization. The evaluation policy is controlled by the parameter $\theta$, while the data-sampling policy is controlled by $\theta_{'}$.

(2) \textit{Policy constraint.} After switching the data sampling policy from $\pi_{\theta}$ to $\pi_{\theta^{'}}$, an issue still remains. Although the expected value of the evaluation and data sampling policy is similar, their variances are starkly distinct. A policy constraint is implemented in the form of threshold clipping and the state-action pair expectation is defined as:
$\mathbb{E}_{(s^t,a^t)\sim\pi_{\theta^{'}}}[\min\{\frac{\pi_\theta(a^t|s^t)}{\pi_{{\theta'}}(a^t|s^t)}, \operatorname{clip}(\frac{\pi_\theta(a^t|s^t)}{\pi_{{\theta'}}(a^t|s^t)}, 1-\epsilon, 1+\epsilon)\}\cdot A^t]$, where $A^{t}$ is the advantage value. Gradient ascent is then used to improve the policy:
\begin{align}
    \Delta\theta = \mathbb{E}_{(s^t,a^t)\sim\pi_{\theta^{'}}}[\Delta\min\{\frac{\pi_\theta(a^t|s^t)}{\pi_{{\theta'}}(a^t|s^t)}, \operatorname{clip}(\frac{\pi_\theta(a^t|s^t)}{\pi_{{\theta'}}(a^t|s^t)}, 1-\epsilon, 1+\epsilon)\}\cdot A^t]. \label{eq:actorobj}
\end{align}
With regards to the critic network, all actors under the PPO algorithm share a similar critic network model. Hence, the authors of~\cite{schulman2017proximal} formulate the critic loss to be: 
\begin{align}
&L(\phi) = (V_\phi(s^t)-A^t+\gamma V_{\phi_{targ}}(s^{t}))^2. \label{eq:criticloss}
\end{align}
$\phi$ is the parameter of the critic model. The state-values $V_{\phi}(s^{t})$ and $V_{\phi_{targ}}(s^{t})$ are estimated with the critic model and the target critic model, respectively. Essentially, the critic model utilizes the loss function $L(\phi)$ to update the critic network.
\vspace{-0.3cm}

\vspace{-5pt}\subsection{Multi-Agent-Loss-Sharing (MALS) \vspace{-5pt}}

\textbf{What is Multi-Agent-Loss-Sharing? }We proposed a novel Multi-Agent-Loss-Sharing (MALS) algorithm based on actor-critic (AC) reinforcement learning. In this algorithm, each agent has an independent actor, similar to traditional interacting, independent dual agents (IDA). However, in MALS, agents share a critic model with multiple heads, allowing for personalized advantage values calculation and cooperative training. The sum of critic losses from both heads facilitates concurrent improvement of both agents' models and ensures cooperative training. MALS can accommodate asymmetric actors of different action spaces. We adopt PPO as the underlying algorithm of MALS and utilize two actors, one which has a discrete UE-MBS allocation action, and the other which has a continuous power selection action.


\textbf{Why not apply the CTDE framework? }In traditional Multi-Agent Reinforcement Learning (MARL) using the Centralized Training and Decentralized Execution (CTDE) framework~\cite{lowe2017multi}, each agent's actor operates independently during execution, while the shared critic model considers all agents' actions during training. However, CTDE is unsuitable for our asymmetric problem with distinct action spaces for DL and UL agents (discrete-valued UE-MBS allocation and continuous-valued UE uplink transmission power, respectively). Additionally, our asynchronous problem, where DL and UL agents alternate action executions, differs from the simultaneous action execution assumption in MARL, leading to computational inefficiency in CTDE~\cite{lowe2017multi}. Moreover, our asymmetric and asynchronous MEC resource allocation problem creates dependency between agents in a Partially Observable Markov Decision Process (POMDP). In a conventional single-agent environment, an agent's actions only affect the environment and rewards. However, in our two-agent asynchronous environment, an agent's transition probabilities are influenced by the other agent's decisions. This interdependence requires a new reinforcement learning structure to handle the asynchronous setting and the concurrent orchestration of UL and DL stages.

\textbf{Transmission mechanism}:
In a single iteration with one DL and one UL transmission, the downlink state $s^{\text{d},t}$ will be observed by the DL agent. The DL agent will then choose an action $a^{\text{d},t}$ based on its policy and observed state. This action causes an impact on the environment, and a DL reward $R^{\text{d},t}$ is issued to the DL agent in accordance to the merit of its actions. The altered environment consequently influences the UL state $s^{\text{u},t}$ observed by UL agent. Based on its policy, the UL agent will choose an action $a^{\text{u},t}$ which corresponds to the current observed state. This action by the UL agent influences the environment, and yields the UL agent a corresponding reward $R^{\text{u},t}$ and the next DL state $s^{\text{d},t+1}$. After several iterations, the rewards ($R^{\text{d},t},R^{\text{u},t}$) with states ($s^{\text{d},t},s^{\text{u},t}$) and their corresponding next states ($s^{\text{d},t+1},s^{\text{u},t+1}$) will be sent to the multi-head critic. The critic will then utilize these rewards, states, and next iteration states to calculate state-value $V$ and advantage value $A$ to update the actor network parameters. This transmission mechanism is illustrated in Fig \ref{fig:model}.

\textbf{Asymmetric Actor}:
In MALS, there are two Actors: one arranges UE-MBS allocation in the downlink transmission, and the other selects the UE power for the uplink transmission. The UE-MBS allocation is a discrete case selection and hence probabilities are computed for each action based on the DL agent policy~\cite{sutton1999policy}. The transmission power selection in the UL stage falls along a continuum and hence a probability distribution over the output power values are learnt by the UL agent.

\textbf{Discrete Action through Softmax. }In the discrete DL actor neural network, we adopt a softmax activation function in the final layer. Softmax is a commonly used activation function for the output layer of a neural network in multi-class classification tasks. The softmax function transforms the raw outputs of the network into a probability distribution over multiple actions. It maps the output values to a range between 0 and 1, ensuring that the sum of all probabilities is equal to 1. This allows the network to assign a probability to each action, indicating the likelihood of the input belonging to that particular action. The action with the highest probability is then chosen as the predicted class, where the UE-MBS allocation follows $c_{i}^{t}=\text{arg}\max_{v}\left [\text{prob}(a^t_i | s^t)\right ] ~\forall v \in \mathcal{M}, \forall t \in \mathcal{T},~\forall i \in \mathcal{N}$. By using softmax, the neural network can provide a meaningful and interpretable output for multi-action selection problems, enabling the identification of the most probable UE-MBS allocation among the available options. 

\textbf{Continuous Action through sampling. }The continuous PPO algorithm parameterizes a probability distribution over the continuous action space, allowing the agent to sample actions based on this distribution. The distribution is typically chosen to be a Gaussian distribution with a mean and standard deviation. The continuous output is then clamped to values between 0 and 1, and eventually scaled to a valid range with a continuous action space adapter, where $p^{\text{u}}_{min}$ is the minimal value, and $p^{\text{u}}_{max}$ being the maximum value, as per~\cite{PPO}.

\textbf{Actor network update. }In MALS, $\theta^{\text{d}}$ and $\theta^{\text{u}}$ are the network weights of DL agent and UL agent, and the gradients $\Delta\theta^{\text{d}}$, $\Delta\theta^{\text{u}}$ of the DL agent and UL agent are:
\begin{align}
    &\Delta\theta^{\text{u}} = \mathbb{E}_{(s^{\text{u},t},a^{\text{u},t})\sim\pi_{{\theta^{\text{u}}}'}}[\triangledown f^t(\theta^{\text{u}})A^{\text{u}}(s^{\text{u},t})], 
    \label{eq:gradient1}\\
    &\Delta\theta^{\text{d}} = \mathbb{E}_{(s^{\text{d},t},a^{\text{d},t})\sim\pi_{{\theta^{\text{d}}}'}}[\triangledown f^t(\theta^{\text{d}})A^{\text{d}}(s^{\text{d},t})],
    \label{eq:gradient2}
\end{align}
where $f^t(\theta)=\min\{\frac{\pi_\theta(a^t|s^t)}{\pi_{{\theta'}}(a^t|s^t)}, \operatorname{clip}(\frac{\pi_\theta(a^t|s^t)}{\pi_{{\theta'}}(a^t|s^t)}, 1-\epsilon, 1+\epsilon)\}$ in the case of PPO, in which 
 $\epsilon$ is the clipping parameter, and $A(\cdot)$ is the advantage function of both UL and DL agents.

\textbf{Advantage function. }We utilize the state-value function ($V$) and the truncated version of the Generalized Advantage Estimation (GAE)~\cite{GAE} method to calculate the advantage values. The policy is then executed for $\bar{T}$ time steps, where $\bar{T}$ is the number of steps in a trajectory. A trajectory of $\bar{T}$ number of steps represents the path of an agent through the state space up till horizon $\bar{T}$. The advantage value in our work can be calculated as:
\begin{align}
    &A^{\text{u},t} = \delta^{\text{u},t} + (\gamma\lambda)\delta^{\text{u},t+1}+...+(\gamma\lambda)^{\bar{T}-1}\delta^{\text{u},t+\bar{T}-1} ,\text{ for}~\delta^{\text{u},t}=R^{\text{u},t}+\gamma V_{\phi_{targ}}(s^{\text{u},t+1})-V_{\phi_{targ}}(s^{\text{u},t}),
\end{align}
\begin{align}
     &A^{\text{d},t} = \delta^{\text{d},t} + (\gamma\lambda)\delta^{\text{d},t+1}+...+(\gamma\lambda)^{\bar{T}-1}\delta^{\text{d},t+\bar{T}-1} ,\text{ for}~\delta^{\text{d},t}=R^{\text{d},t}+\gamma V_{\phi_{targ}}(s^{\text{d},t+1})-V_{\phi_{targ}}(s^{\text{d},t}),
\end{align}
where $\bar{T}$, $\gamma$, and $\lambda$ denote the trajectory length, discount factor, and GAE parameter, respectively. $\phi$ and $\phi_{targ}$ are the parameters of the critic model and the target model, respectively.

\textbf{Multi-head Critic}:
We proposed an asymmetric problem in our work, where each agent optimizes different variables and has different objectives. A multi-head critic caters to the asymmetric actors by providing personalized calculations of state-values $V_\phi^u, V_\phi^d$ using a neural network with two heads. Each critic head produces a loss value corresponding to the state-value that it produces, and the loss values of the critic heads are weighted-summed ($\kappa_1$ and $\kappa_2$) to give a conclusive loss value as follows:
\begin{align}
    &L^{\text{u}}(\phi) = (V_\phi(s^{\text{u},t})-\gamma V_{targ}^{\text{u},t}))^2, \\
    &L^{\text{d}}(\phi) = (V_\phi(s^{\text{d},t})-\gamma V_{targ}^{\text{d},t}))^2 ,\\
    &L(\phi) = \kappa_1 \times L^{\text{u}}(\phi) + \kappa_2 \times L^{\text{d}}(\phi) .\label{eq:HCloss}
\end{align}
The specific values of $\kappa_{1}$ and $\kappa_{2}$ will be specified in the experiments. The multi-head critic's model parameter is updated with the loss in Equation (\ref{eq:HCloss}) to improve the critics model in guiding the asymmetric actors. The MALS algorithm is detailed in Algorithm~\ref{alg:PPO}. The above explanations and equations, along with their connections, are clearly illustrated in Figure~\ref{fig:model}.
\vspace{-0.2cm}


\begin{figure}[!t] 
        \renewcommand{\algorithmicrequire}{\textbf{Initiate:}}
        \renewcommand{\algorithmicensure}{\textbf{Output:}}
        \begin{algorithm}[H]
            \caption{\label{alg:PPO}MALS-PPO (proposed RL structure)}
            \begin{algorithmic}[1]
                \REQUIRE critic model $\phi$ and target model $\phi_{targ}$, downlink actor model $\theta^{\text{d}}$ and sampling model $(\theta^{\text{d}})'$, uplink actor model $\theta^{\text{u}}$ and sampling model $(\theta^{\text{u}})'$, initial state $s^{\text{d},1}$. \STATE $s^{\text{d},t} \leftarrow s^{\text{d},1}$;
                \FOR{iteration  $t=1,2,...$}
                    \STATE DL agent executes action according to $\pi_{(\theta^{\text{d}})'}(a^{\text{d},t}|s^{\text{d},t})$;
                    \STATE Get reward $R^{\text{d},t}$ and uplink state $s^{\text{u},t}$;
                    \STATE UL agent executes action according to $\pi_{(\theta^{\text{u}})'}(a^{\text{u},t}|s^{\text{u},t})$;
                    \STATE Get reward $R^{\text{u},t}$ and next step downlink state $s^{\text{d},t+1}$;
                    \STATE DL agent executes action according to $\pi_{(\theta^{\text{d}})'}(a^{\text{d},t+1}|s^{\text{d},t+1})$;
                    \STATE Get reward $R^{\text{d},t+1}$ and uplink state $s^{\text{u},t+1}$;
                    \IF{iteration $t\geq 2$}
                    \STATE Sample ($s^{\text{d},t},a^{\text{d},t},s^{\text{u},t},a^{\text{u},t},R^{\text{d},t},R^{\text{u},t},s^{\text{d},t+1},s^{\text{u},t+1}$) iteratively till end of episode;
                    \ENDIF
                    \STATE $s^{\text{d},t}  \leftarrow s^{\text{d},t+1}$, $s^{\text{u},t} \leftarrow s^{\text{u},t+1}$,
                    $a^{\text{d},t} \leftarrow a^{\text{d},t+1}$,
                    $R^{\text{d},t} \leftarrow R^{\text{d},t+1}$;
                
                    \STATE Compute advantages $\{A^{\text{d},t},A^{\text{u},t}\}$
                    \STATE Compute target values \{$V^{\text{d},t}_{targ},V^{u,t}_{targ}$\}
    
                    \FOR{$k$ = $1,2,...,K$}
                        \STATE Assign samples into training groups;
                        \FOR{each training group}
                            \STATE Compute UL and DL actor model gradients:
                            $\triangledown \theta^{\text{d}}, \triangledown \theta^{\text{u}}$
                            \STATE Apply gradient ascent on $\theta^{\text{d}}$ using $\triangledown \theta^{\text{d}}$ by Eq.~(\ref{eq:gradient2})
                            \STATE Apply gradient ascent on $\theta^{\text{u}}$ using $\triangledown \theta^{\text{u}}$ by Eq.~(\ref{eq:gradient1})
                            \STATE Update the critic model with loss from Eq.~(\ref{eq:HCloss})
                        \ENDFOR
                    \ENDFOR
                    \STATE Update the critic target model parameters $\phi_{targ}$ with the critic model parameters $\phi$ every $C$ steps where $C$ is the critic model update interval;
                \ENDFOR
            \end{algorithmic}
        \end{algorithm}
\vspace{-1.5cm}
\end{figure}

\vspace{-0.3cm}\section{Experiments \vspace{-5pt}}
\label{experiment}
We conduct several experiments and showcase the superiority of our MALS RL structure in handling asymmetric and asynchronous problems. We first discuss the numerical settings for the experiments we conduct. We then show how the proposed MALS algorithm is effective in improving resource management when compared against commonly adopted metrics. We then compare the proposed MALS model with baseline algorithms (i) independent dual agent (IDA)~\cite{zhang2021multi} and (ii) Centralized Training and Decentralized Execution multi-agent model (CTDE)~\cite{foerster2018counterfactual}. Finally, we vary the weighting on each variable in the DL and UL agents' objective functions to study the performance of the MALS model on varying variable weights and their implications on the model performance.
\vspace{-1.5mm}

\begin{figure}[t]
    \vspace{-0.3cm}
    \centering
    \includegraphics[width=1\linewidth]{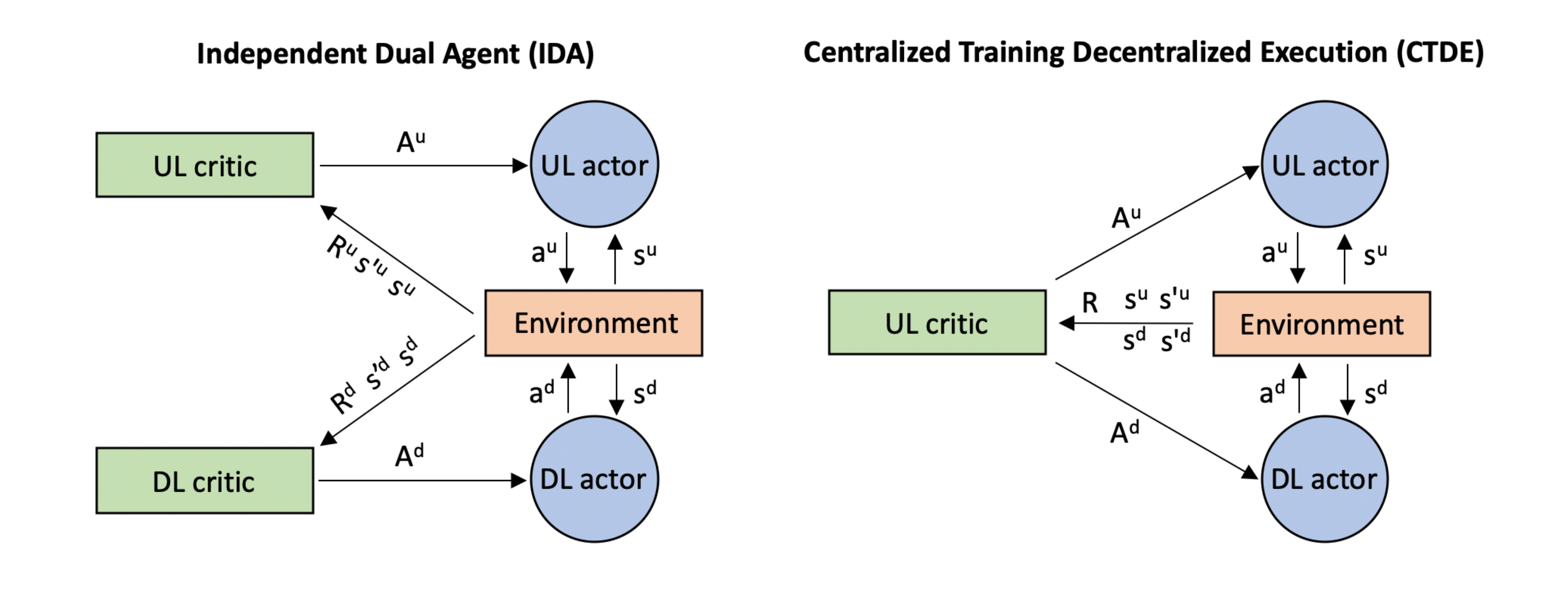}
    \vspace{-1cm}
    \caption{Independent Dual Agent (IDA) (left) and Centralized Training Decentralized Execution (CTDE) (right).}
    \label{fig:algorithmothers}
    \vspace{-0.8cm}
\end{figure}

\vspace{-13pt}\subsection{Baseline algorithms \vspace{-5pt}}
We study how (i) our proposed model (MALS model structure) compared to other reinforcement learning-based structures. We adopt two reinforcement learning-based model structures; the most intuitive (ii) independent dual agent (IDA) (which employs two individual agents in which information transfer is only via the states), and (iii) CTDE (which is the framework used for multi-agent reinforcement learning~\cite{lowe2017multi}). We pit the baseline algorithms (IDA and CTDE) against our proposed MALS model in tackling the asymmetric and asynchronous task. The three models - Independent Dual Agent (IDA), Centralized Training and Decentralized Execution (CTDE), and Multi-agent Loss Sharing (MALS) - all have two separate actors influencing the environment. The IDA model has two separate actors and two critics, with each actor of each agent outputting action $a^{\text{d}}$, then $a^{\text{u}}$,  in each iteration. The environment is influenced by the actions of the two agents, and returns state $s^{\text{d}}$ (after $a^{\text{d}}$) and $s^{\text{u}}$ (after $a^{\text{u}}$) to its corresponding actors. Each agent's critic updates with trajectories of the states and its agent's reward assignment $R^{\text{u}}$ and $R^{\text{d}}$. The CTDE model also has two separate actors, but they share one critic model. The environment returns $s^{\text{d}}$ and $s^{\text{u}}$ to the corresponding actors, and a single, common reward $R$ is returned to each agent. Note that the reward $R$ assigned for the CTDE model is not decomposed, and corresponds to the overarching objective in equation~\ref{obj:eq1}. The sole critic receives the states $s^{\text{d}}$ and $s^{\text{u}}$ and reward $R$ and updates its model. On the other hand, our proposed MALS model has two actors and each outputs actions $a^{\text{d}}$ and $a^{\text{u}}$, correspondingly that influence the environment, with each agent receiving separate rewards $R^{\text{d},t}$ and $R^{\text{u},t}$. States $s^{\text{d}}$ and $s^{\text{u}}$ and separate rewards are sent to the central, multi-headed critic for computation of separate advantage $A^{\text{d},t}$ and $A^{\text{u},t}$ values. Finally, the computed advantage values for each agent is used to update their actor. Illustrations of the model structures are shown in Figure~\ref{fig:model} and \ref{fig:algorithmothers}. Note that we do not compare our proposed model with a multi-input, multi-headed, single agent. The states presented to the agent at each transmission is different and a single agent is not able to accommodate differing states sizes.
\vspace{-5mm}



\subsection{Configuration} 
\vspace{-0.3cm}\label{Configuration}

\begin{figure*}[t]

\centering
\subfigtopskip=2pt
\subfigbottomskip=2pt

\subfigure[Average downlink delay.]{
\begin{minipage}[t]{0.24\linewidth}
\centering
\includegraphics[width=1\linewidth]{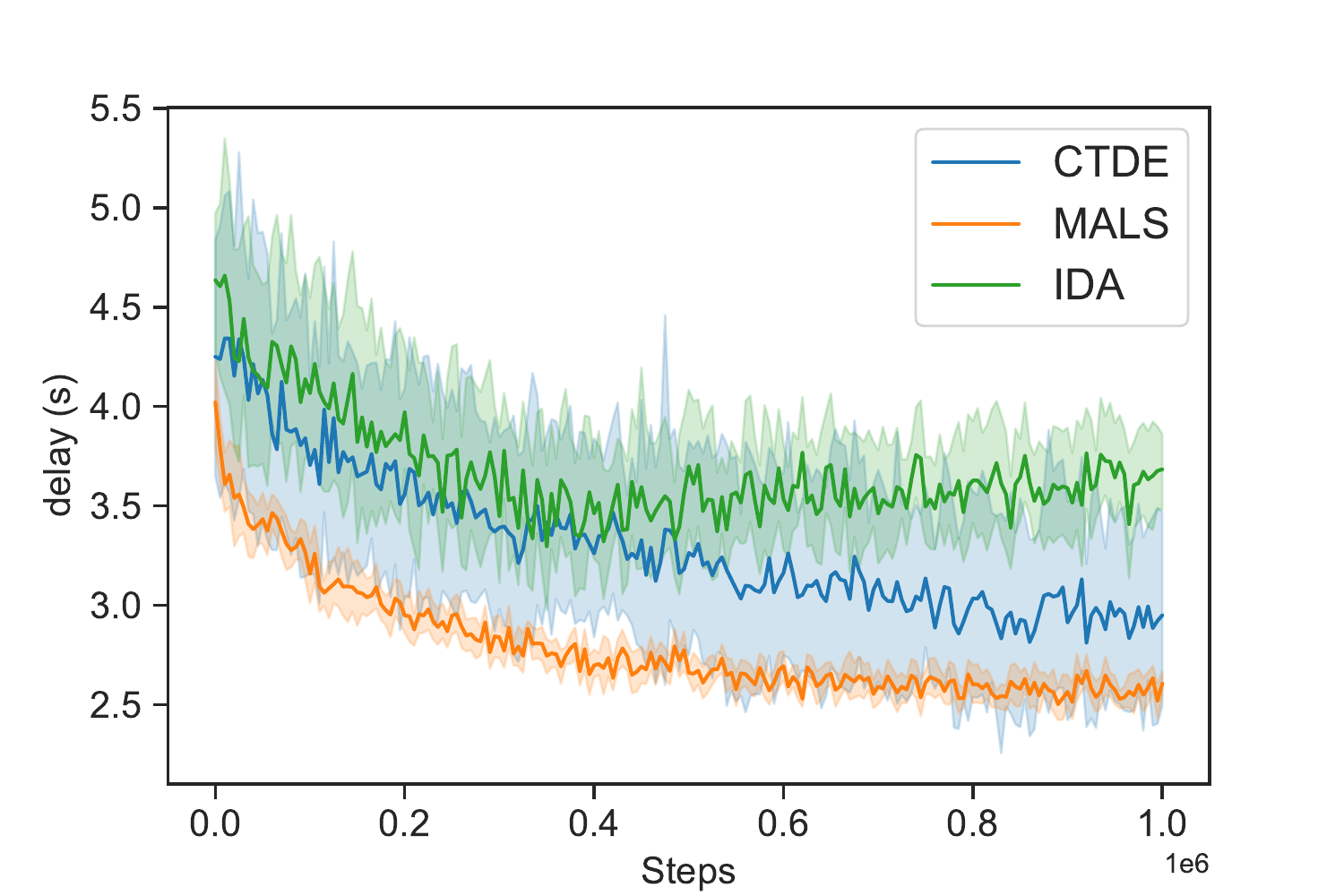}
\label{fig:delay_down}
\vspace{-10mm}
\end{minipage}%
}%
\subfigure[Average UE earning potential.]{
\begin{minipage}[t]{0.24\linewidth}
\centering
\includegraphics[width=1\linewidth]{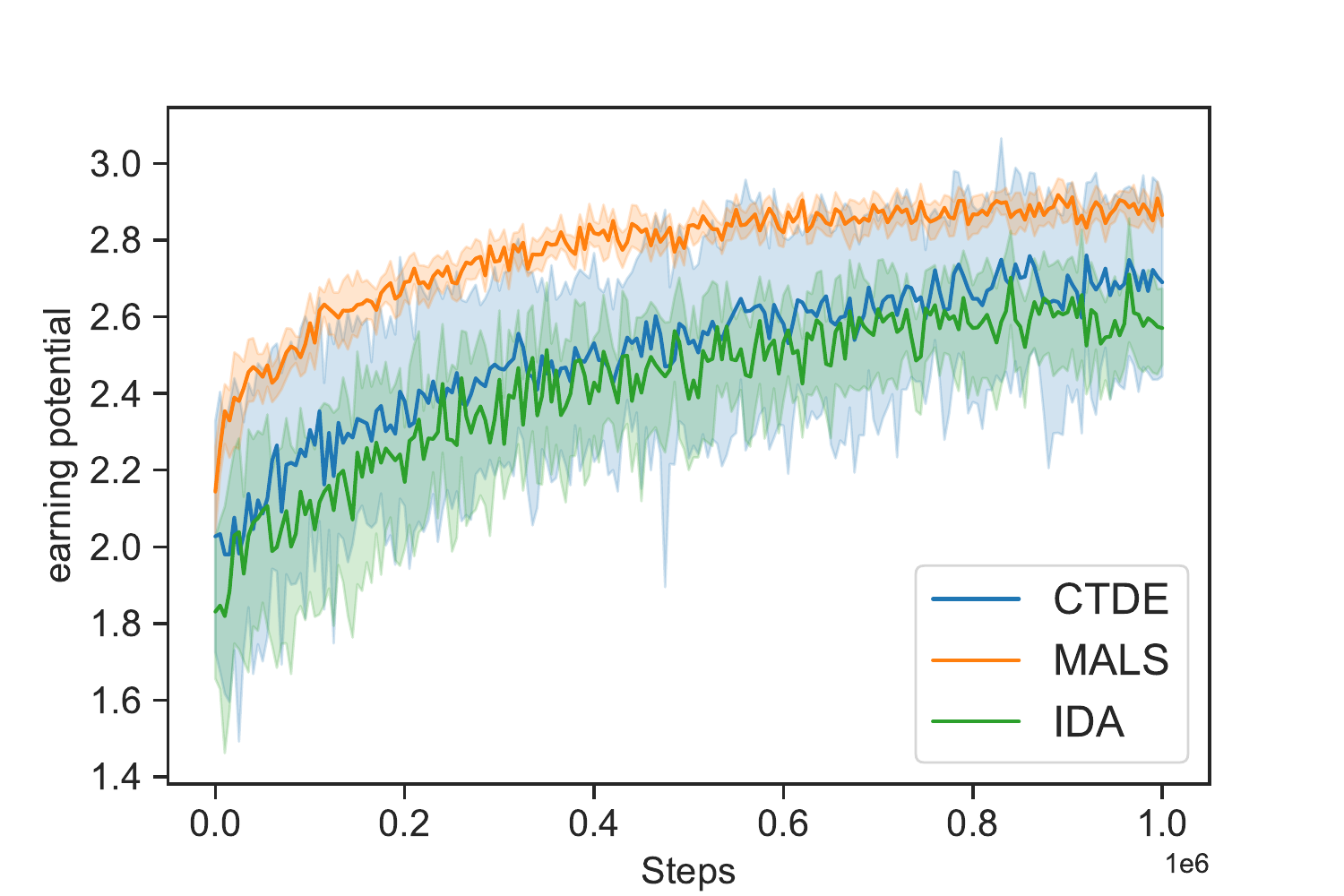}
\label{fig:earning_ability}
\vspace{-10mm}
\end{minipage}%
}%
\subfigure[Average uplink delay.]{
\begin{minipage}[t]{0.24\linewidth}
\centering
\includegraphics[width=1\linewidth]{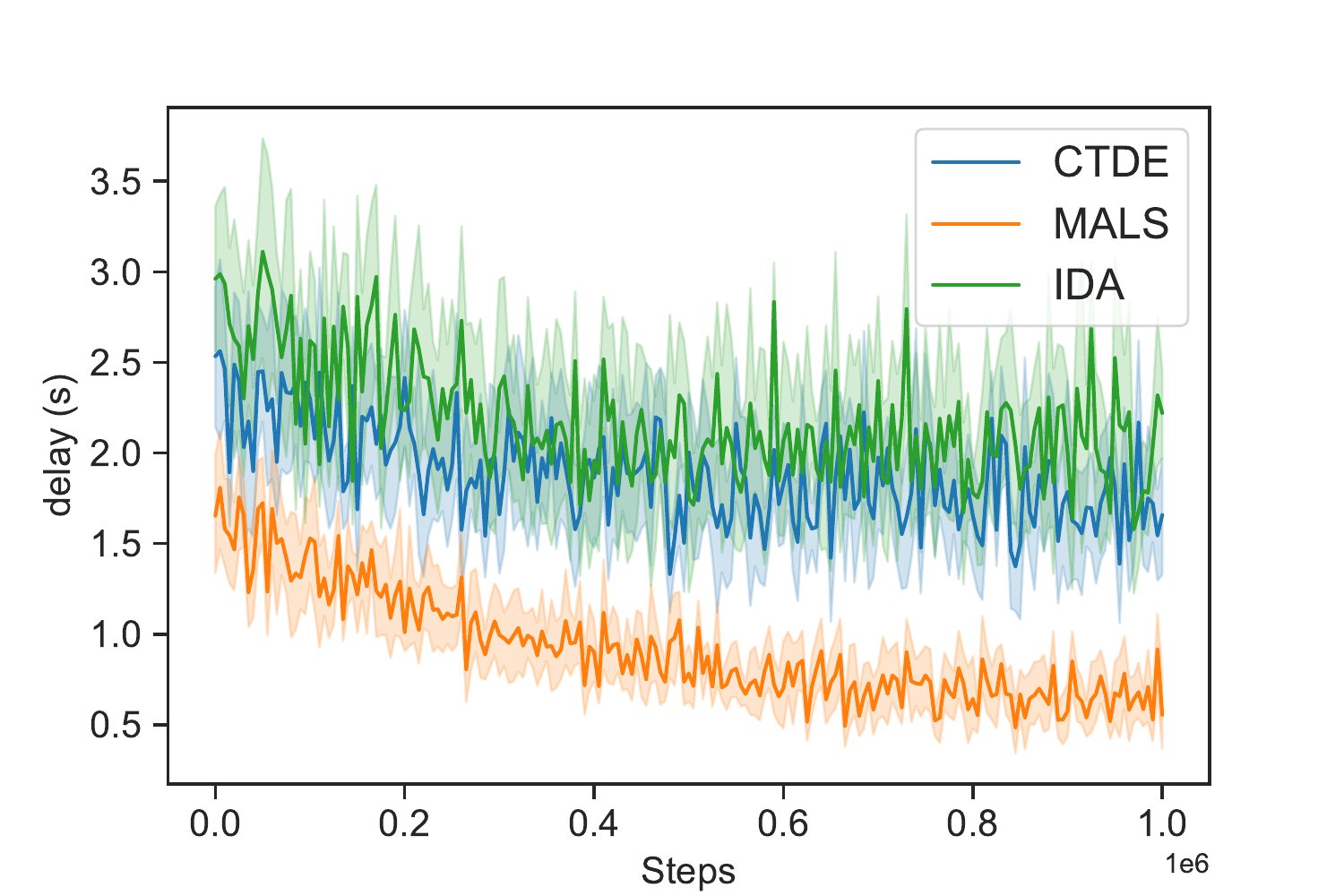}
\label{fig:log_delay_up}
\vspace{-10mm}
\end{minipage}
}%
\subfigure[Average battery consumption.]{
\begin{minipage}[t]{0.24\linewidth}
\centering
\includegraphics[width=1\linewidth]{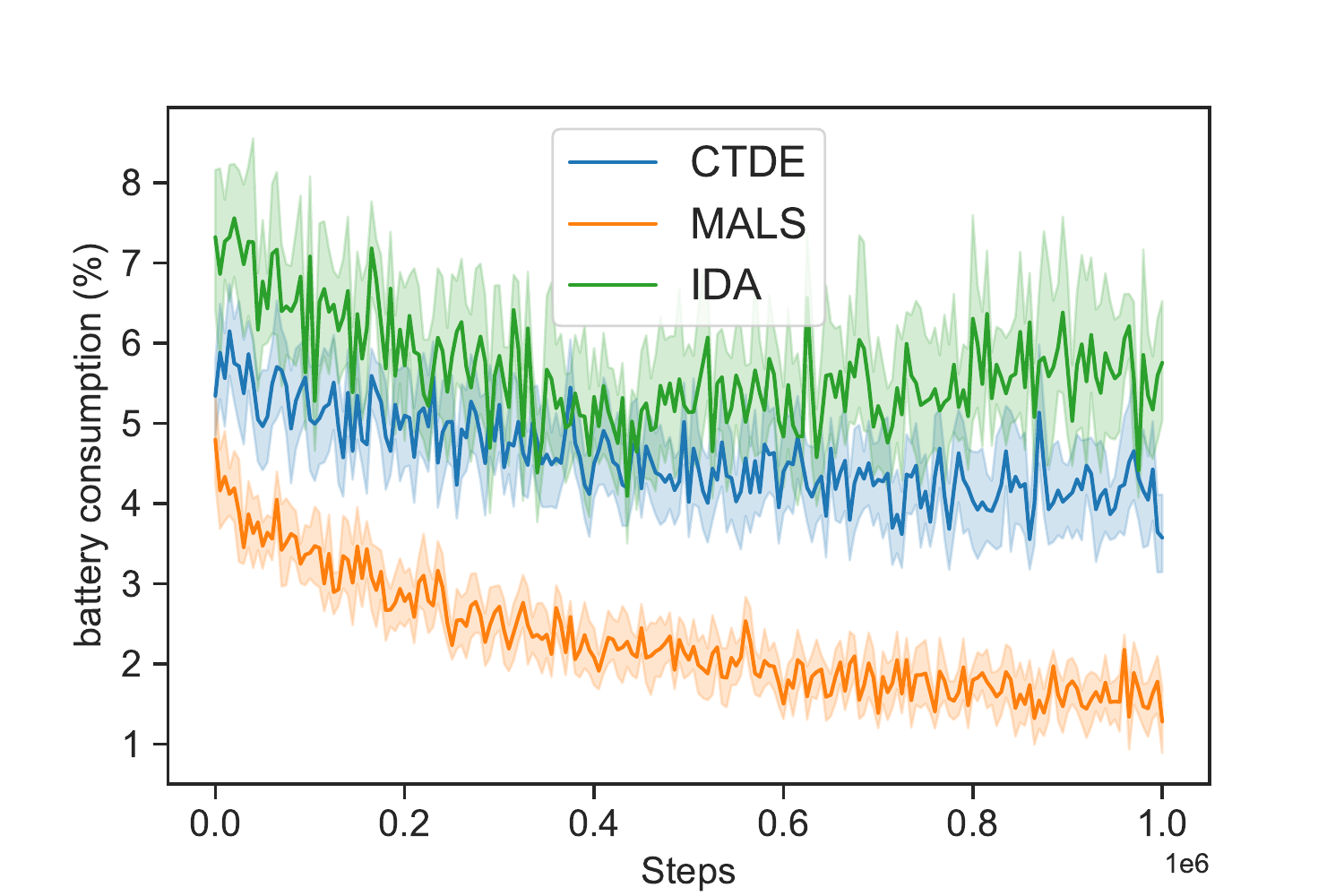}
\label{fig:log_battery_percent_consumed}
\vspace{-10mm}
\end{minipage}
}%
\caption{Key metrics performance obtained by MALS across training steps for the 4 MBS, 8 UE configuration.}
\label{fig:complete_weight_1}
\vspace{-0.5cm}
\end{figure*}





\begin{figure*}[htb]

\centering
\subfigtopskip=2pt
\subfigbottomskip=2pt

\subfigure[Downlink reward for 4 MBS, 6 UE.]{
\begin{minipage}[t]{0.33\linewidth}
\centering
\includegraphics[width=1\linewidth]{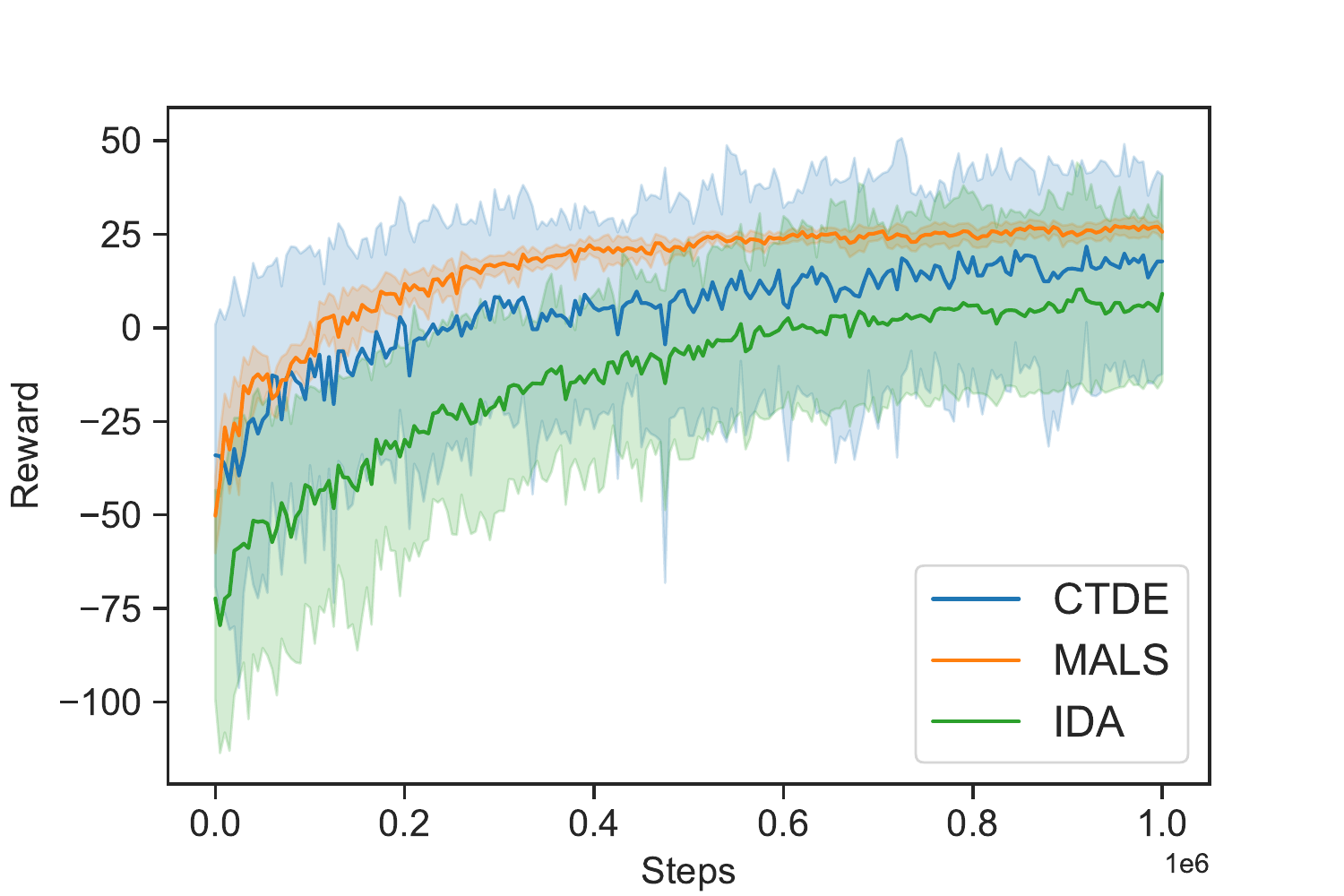}
\label{fig:4a}
\vspace{-10mm}
\end{minipage}%
}%
\subfigure[Downlink reward for 4 MBS, 7 UE.]{
\begin{minipage}[t]{0.33\linewidth}
\centering
\includegraphics[width=1\linewidth]{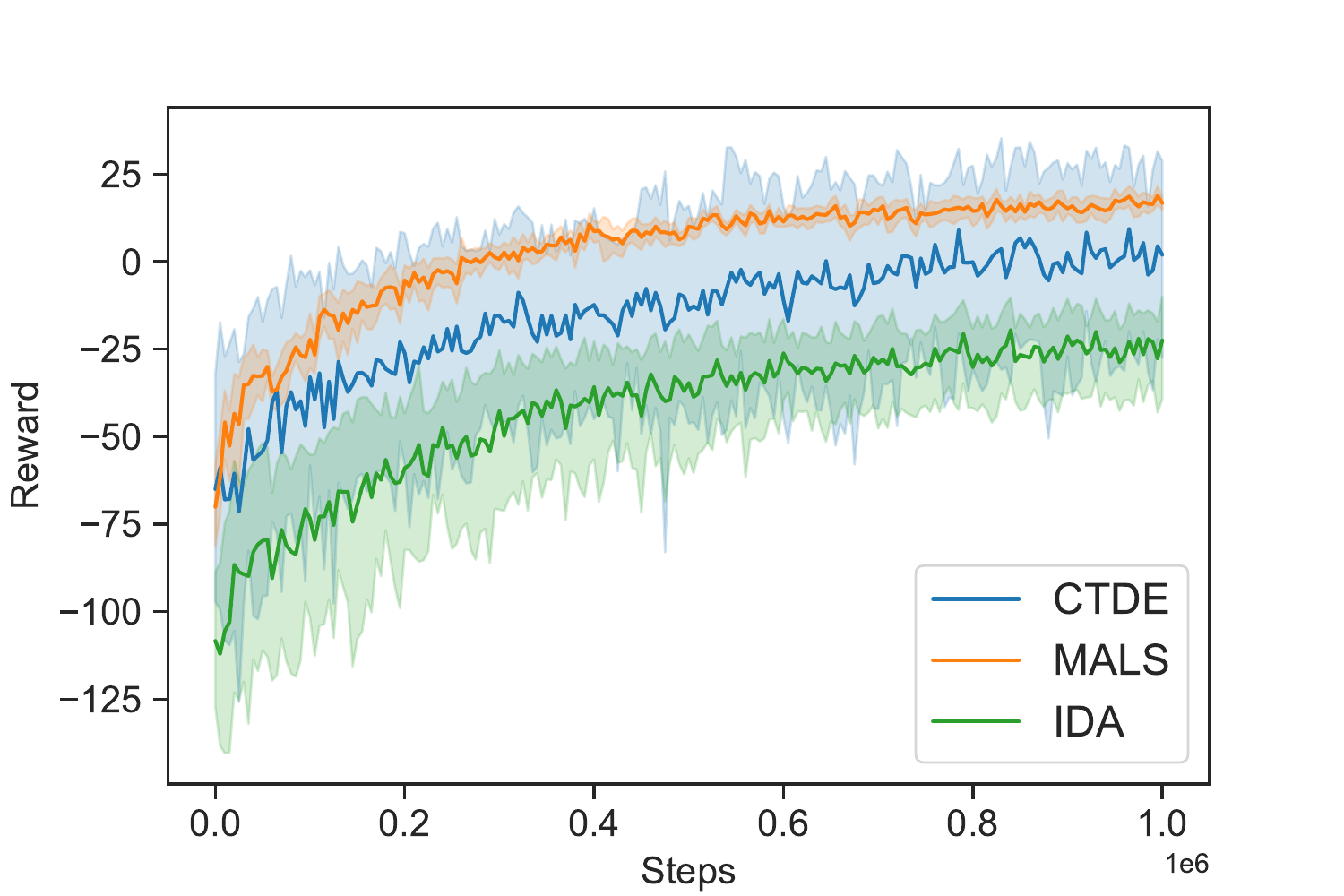}
\label{fig:4b}
\vspace{-10mm}
\end{minipage}
}%
\subfigure[Downlink reward for 4 MBS, 8 UE.]{
\begin{minipage}[t]{0.33\linewidth}
\centering
\includegraphics[width=1\linewidth]{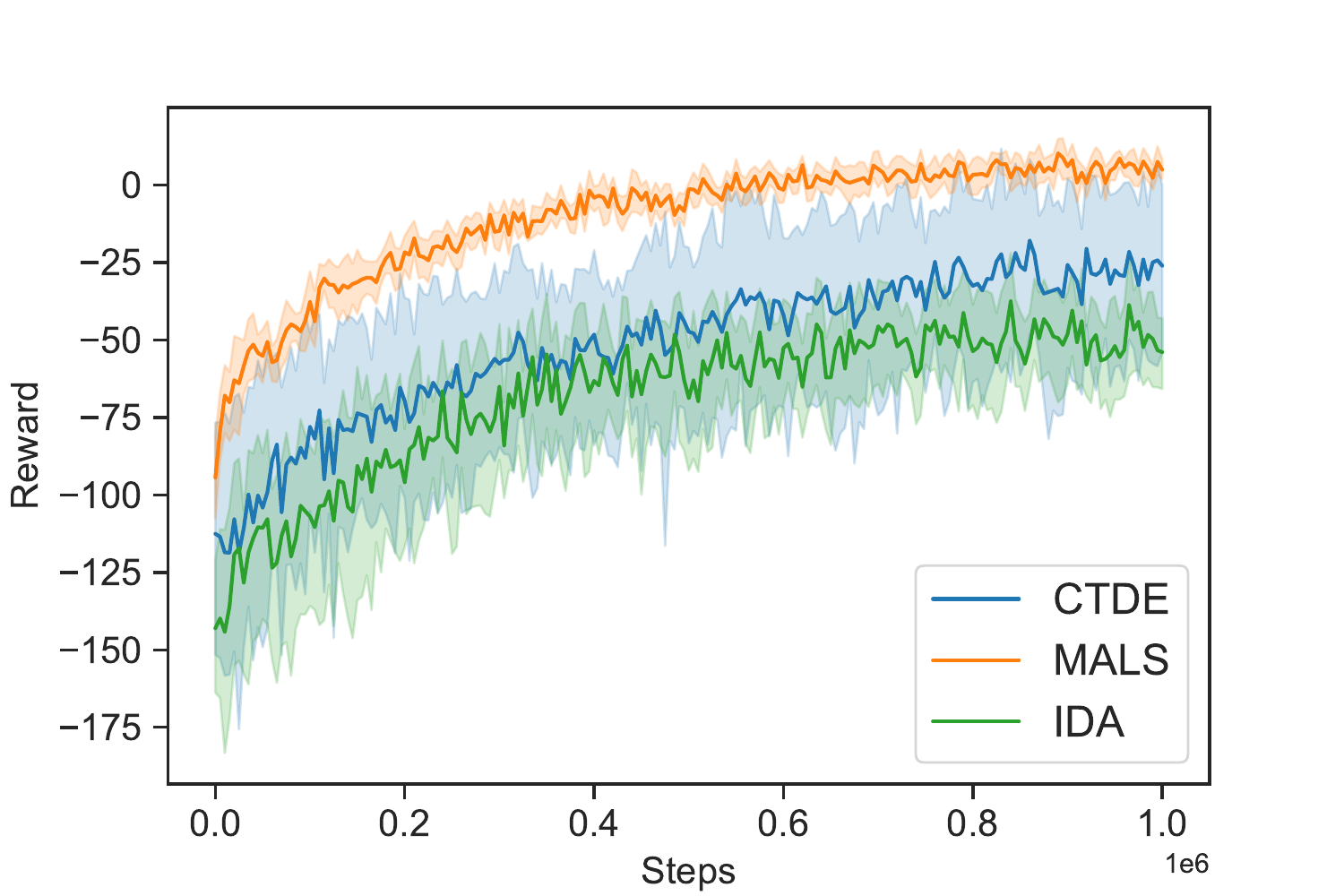}
\label{fig:4c}
\vspace{-10mm}
\end{minipage}%
}%

\subfigure[Uplink reward for 4 MBS, 6 UE.]{
\begin{minipage}[t]{0.33\linewidth}
\centering
\includegraphics[width=1\linewidth]{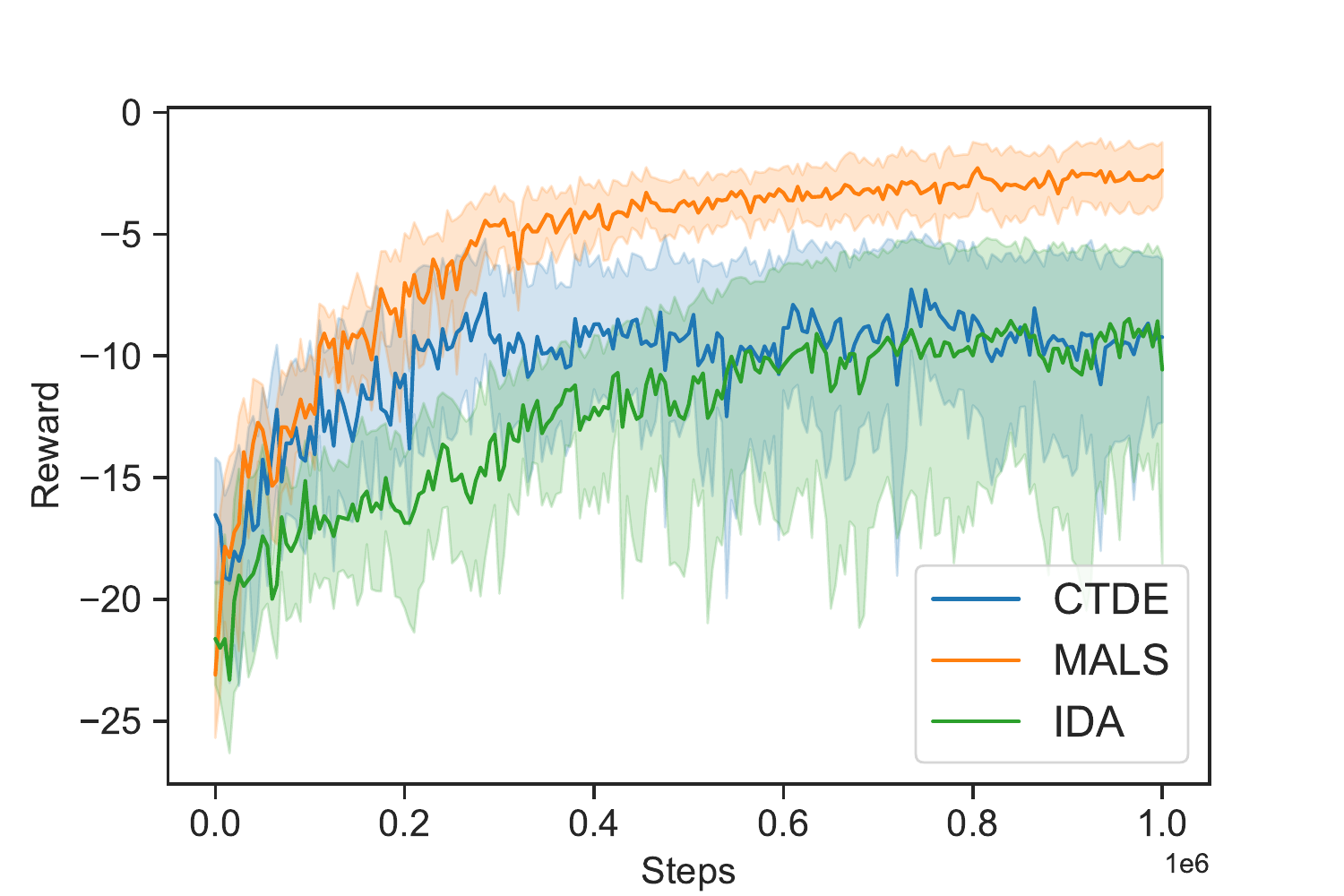}
\label{fig:4d}
\vspace{-10mm}
\end{minipage}
}%
\subfigure[Uplink reward for 4 MBS, 7 UE.]{
\begin{minipage}[t]{0.33\linewidth}
\centering
\includegraphics[width=1\linewidth]{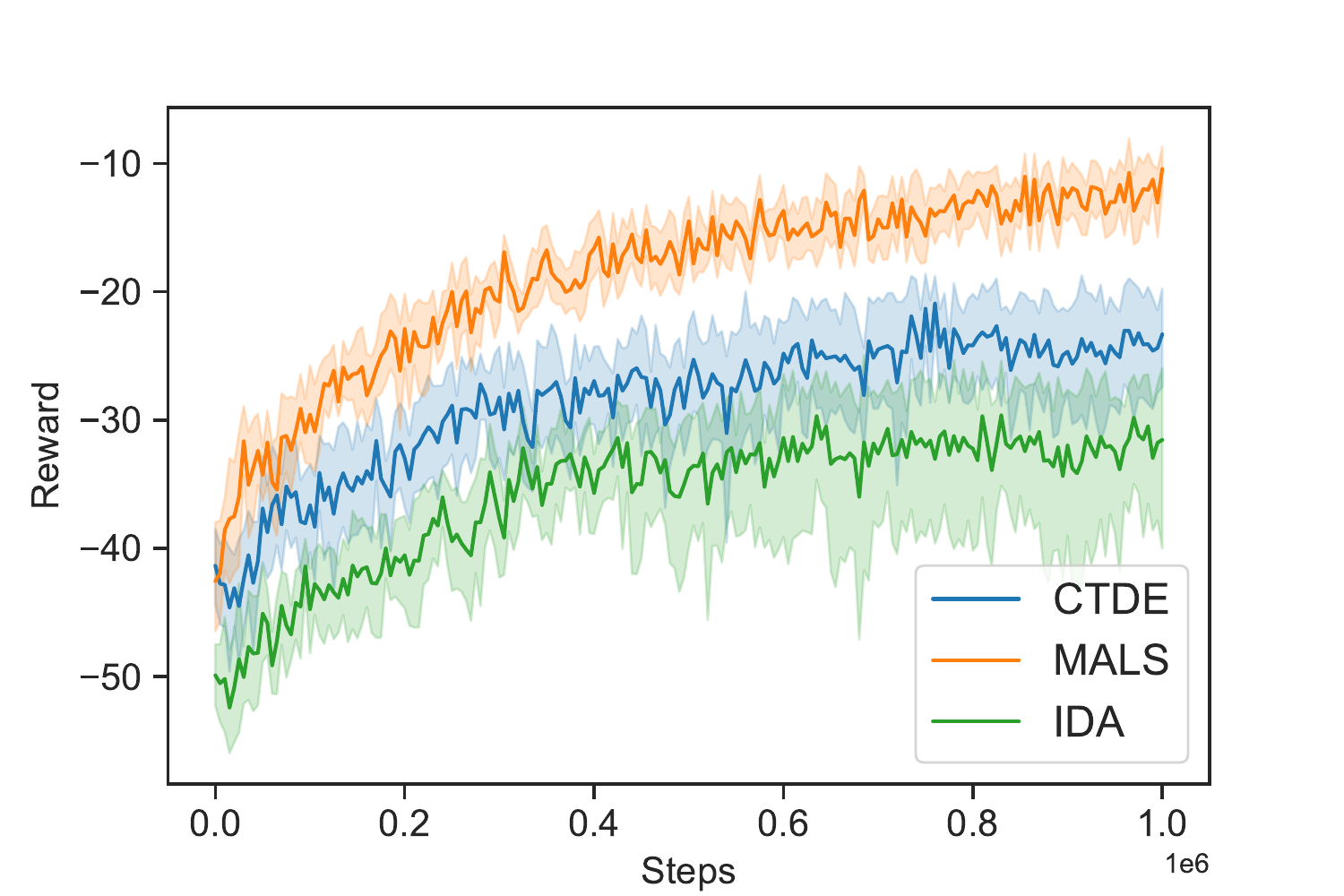}
\label{fig:4e}
\vspace{-10mm}
\end{minipage}
}%
\subfigure[Uplink reward for 4 MBS, 8 UE.]{
\begin{minipage}[t]{0.33\linewidth}
\centering
\includegraphics[width=1\linewidth]{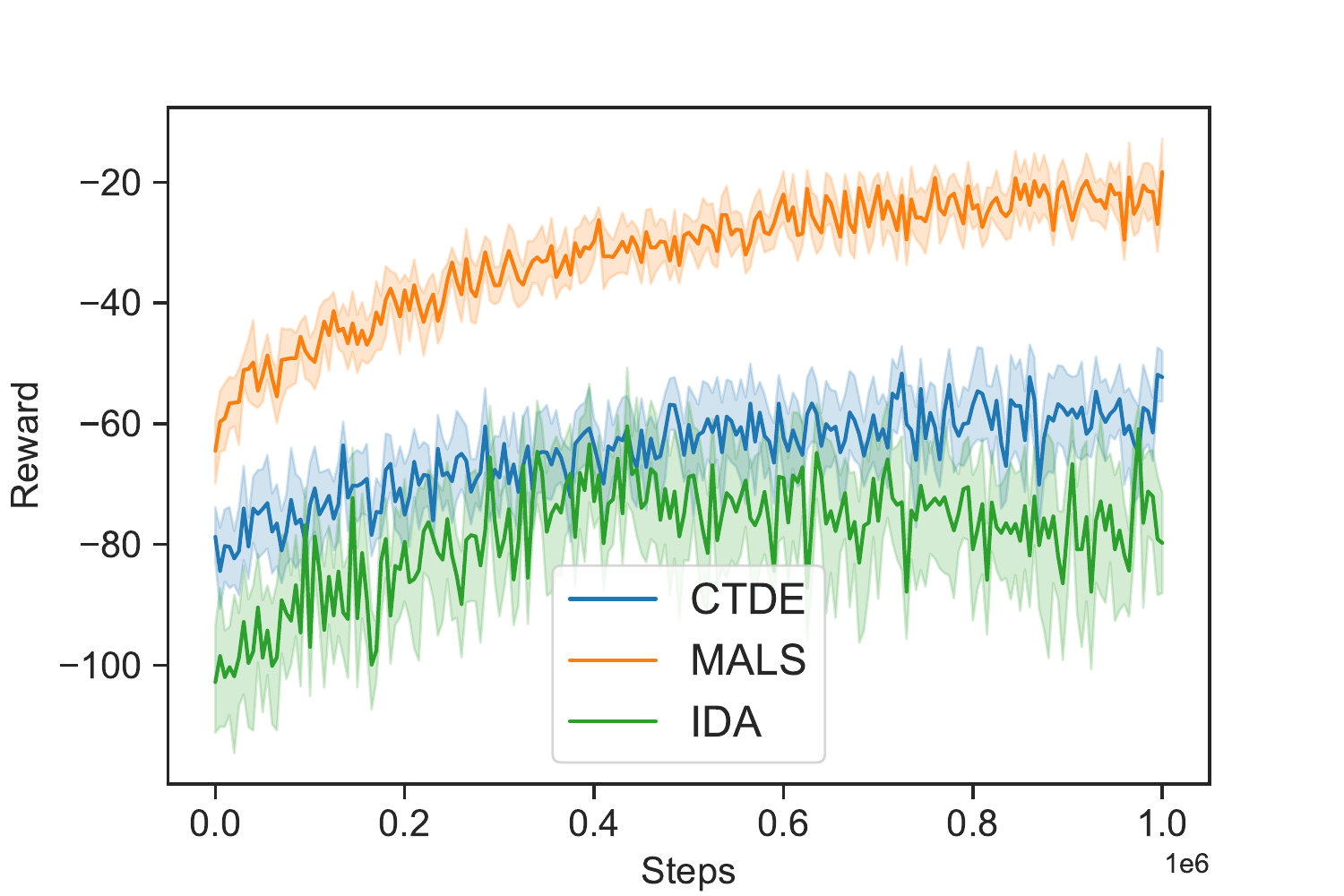}
\label{fig:4f}
\vspace{-10mm}
\end{minipage}
}%
\caption{Uplink and downlink agent rewards obtained for MALS model, Independent Dual Agent model, CTDE model, across configuration settings and seed 0 to 9. Bands within the graph indicate the range of reward values obtained across the seeds.}
\label{fig:complete_weight_2}
\vspace{-0.8cm}
\end{figure*}


We design an experiment with three scenarios: 4 MBSs and 6 to 8 UEs to test our proposed Multi-Agent-Loss-Sharing (MALS) reinforcement learning-based method. $T$ which is defined as the number of transmission steps within an episode is set as $100$. We limit our experiments to configurations with no more than 4 MBSs and 8 UEs. Even this configuration poses a significant computational challenge for our RL agents. The complexity of discrete actions increases exponentially with the number of MBSs and UEs, reaching $4^8$ in our most complex scenario per iteration. Since we run the transmission process for 100 iterations, where the actions selected in each iteration influences subsequent iterations, the overall complexity becomes $4^{8^{100}}$. Consequently, adding more MBSs or UEs would render the computation infeasible due to the exponentially expanding action space. Additionally, since the UL and DL agents influence each other, introducing more complex configurations beyond our proposed limits would make the RL training process difficult and prevent the agents from achieving convergence. Exhaustive search is an infeasible method to optimize the UE-MBS assignment as the sequential nature of our proposed problem renders the complexity to be $(M^{N})^{T}$, where $M$ is the number of MBS, $N$ is the number of UEs, and $T$ is the number of transmission iterations in an episode. The bandwidth $B_{v}$ is set as $10$ Ghz~\cite{yang20196g} as P2E applications are bandwidth-intensive and a wider band permits higher data transfer rates. We have designed all MBS to have similar frequency band as (1) we are only considering a single application (P2E), (2) and sharing a similar frequency band simplifies the network planning and deployment process and enables seamless handover between MBSs as UEs move. The noise $\sigma^2$ is set as $100$ dBm/Hz. The output power by the MBS for the downlink transmission is set to be within $(1.5,2.0)$ Watt. We initialize the locations of the MBSs and UEs to be random, within a 100m by 100m area. The P2E in-game graphic size $D_i^t$ of each UE at every step is uniformly distributed between $800$ and $1000$ Mb. We set $P$, $b$ and $f$ to be 10, 1 and 1, respectively, after empirical tuning, and $h$ and $q$ to be 0.5. $b$, $f$, $h$ and $q$ are defined in subsection~\ref{problemform}. $p^{\text{d}}_{min}, p^{\text{d}}_{max}, p^{\text{u}}_{min}, p^{\text{u}}_{max}$ are set to be 5, 20, 3, 10 watts, respectively. We set the values of both $\kappa_{1}$ and $\kappa_{2}$ from Equation (\ref{eq:HCloss}) to be $0.5$ and $\varkappa$ from Equation (\ref{eq:reward_down}) to be 0.3. $\varrho$ is set to $-50$. We set the maximum distance UEs can move in each iteration ($x_{max}$ and $y_{max}$) to be 10m. The ADAM optimizer~\cite{adam} is utilized for the algorithms we study. We trained the models for 1,000,000 iterations (steps) for all congestion settings. We conducted the training and simultaneous evaluation of the models for each configuration at different seed settings: seed 0 to seed 9.
\vspace{-0.7cm}

\subsection{Channel Attenuation\vspace{-0.2cm}}
We adopt a Rician fading as the small-scale fading, in terms of channel gain, following the works of~\cite{ricianRIS}. It is given by:
\begin{align}
    &g_{i,v}^t = \sqrt{\beta_n^t}\zeta_{i,v}^t,
\end{align}
where
\begin{align}
    &\beta_i^t = \beta_0 (\mathcal{L}_i^t)^{-\alpha}, \\
    &\zeta_{i,v}^t = \sqrt{\frac{K}{K+1}}\bar{g} + \sqrt{\frac{1}{K+1}}\tilde{g}, \\
    &\mathcal{L}_i =\sqrt{(\mathcal{X}_i-\mathcal{X}_{v})^2 + (\mathcal{Y}_i-\mathcal{Y}_{v})^2}.
\end{align}
UE $i$'s position is indicated by $(\mathcal{X}_i,\mathcal{Y}_i)$. The distance between UE $i$ and MBS $v$ is denoted as $\mathcal{L}_i^t$, while $\beta_i^t$ represents UE $i$'s large-scale channel gain during iteration $t$. The components $\bar{g}$ represents the Line-Of-Sight (LOS) component while $\tilde{g}$ represents the Non-LOS (NLOS) components. We assign variable $\tilde{g}$ to follow a standard complex normal distribution with a mean of 0 and a standard deviation of 1. $\beta$ is the channel gain at a $\mathcal{L}_0 = 1$m. In this context, the Rician factor $K$ and the path-loss exponent $\alpha$ is simulated to be 3 and 2, respectively.




\vspace{-5pt}\subsection{Result analyses\vspace{-0.2cm}}
\subsubsection{\textbf{MALS model convergence}}
In training our MALS model, all three configurations (i) 4 MBS, 6 UE, (ii) 4 MBS, 7 UE and (iii) 4 MBS, 8 UE showed a general increment of achieved test-time reward for both the downlink and the uplink agents (shown in Figure~\ref{fig:complete_weight_2}) and improvement in the underlying metrics (shown in Figure~\ref{fig:complete_weight_1},~\ref{fig:complete_weight_appendix} and ~\ref{fig:complete_weight_appendix_36}). The solid line in the charts represents the mean reward of the different seed settings, for each configuration. The color hue bands around the solid line represent the minimum (lower bound) and maximum (upper bound) of the rewards received. The training of RL agents converged quicker in the 4 MBS, 6 UE configuration as opposed to the 4 MBS, 7 UE and 4 MBS, 8 UE as there is less complexity in its scenario, allowing the model to converge quicker and achieve a higher eventual reward. Conversely, a more complex configuration like 4 MBS, 8 UE converges slower due to the higher complexity of the action space and having more UEs to manage.  Overall, the improvement in both the downlink and uplink reward signifies that the overall utility function for both agents are improving, indicating that the training of the downlink agent and uplink agent improves its UE-MBS allocation and UL power allocation, respectively.




\textbf{Downlink. }It is observed that as training progresses, there is a decrease in in-game downlink transmission delay (shown in Figure~\ref{fig:delay_down}, \ref{fig:delay_down_37}, \ref{fig:delay_down_36}) and increase in worst-case (lowest) earning potential of the UEs (shown in Figure~\ref{fig:earning_ability}, \ref{fig:earning_ability_37}, \ref{fig:earning_ability_36}) with each training step. These observations directly reflect that the downlink agent is allocating UE-MBS more efficiently, achieving a higher downlink data transfer rate which decreases the downlink latency and improves the worst-case (lowest) earning capability of the UEs. In addition, the improvement of worst-case battery charge expenditure and uplink transmission delay (shown in Figure~\ref{fig:log_delay_up_37},~\ref{fig:log_delay_up_36} and Figure~\ref{fig:log_battery_percent_consumed_37},~\ref{fig:log_battery_percent_consumed_36}, respectively), as training progress, is also in part attributed to better UE-MBS allocation, which results in less battery charge expended and reduced uplink transmission latency. The improvement of the metrics, as training progress, contribute to an improving reward obtained by the DL agent.



\textbf{Uplink. }Similarly, we note that as training progresses, a more optimal use of UE transmission output power for the uplink transmission results in a lower average uplink delay (shown in Figure.~\ref{fig:log_delay_up}) and lower worst-case (greatest) battery charge expenditure. This is reflected in Figure \ref{fig:log_battery_percent_consumed}. These improvements in uplink delay and lower worst-case battery charge expenditure contributes to higher reward obtained by the UL agent. Furthermore, a higher downlink reward is also in part, attributed to more efficient uplink battery charge expenditure by the UL agent, as an efficient UL power selection enables continual iterations of transmissions, and lesser likelihood for receiving battery charge depletion penalty. 


\subsubsection{\textbf{Comparison with other reinforcement learning model structures}} \label{rlstructcompare} 


Our proposed model (MALS) achieves a smooth convergence for each configuration, exhibiting a narrow range of downlink and uplink rewards, across different seeds (shown in Figure \ref{fig:complete_weight_2}). We observe that our proposed model obtains a significantly higher reward for both the uplink and downlink rewards. The prowess of the MALS model is more notable in more complex configurations (i.e., 4 MBS, 7 UE and 4 MBS, 8 UE) where it outperforms other baseline models and obtain significantly higher rewards, when compared to both the IDA and CTDE baseline models. The narrower bands indicate that the proposed MALS model is more stable across the different seed settings, for each of the configurations.




 \begin{figure*}[t]

\centering
\subfigtopskip=2pt
\subfigbottomskip=2pt

\subfigure[Earning potential wrt $q$.]{
\begin{minipage}[t]{0.25\linewidth}
\centering
\includegraphics[width=1\linewidth]{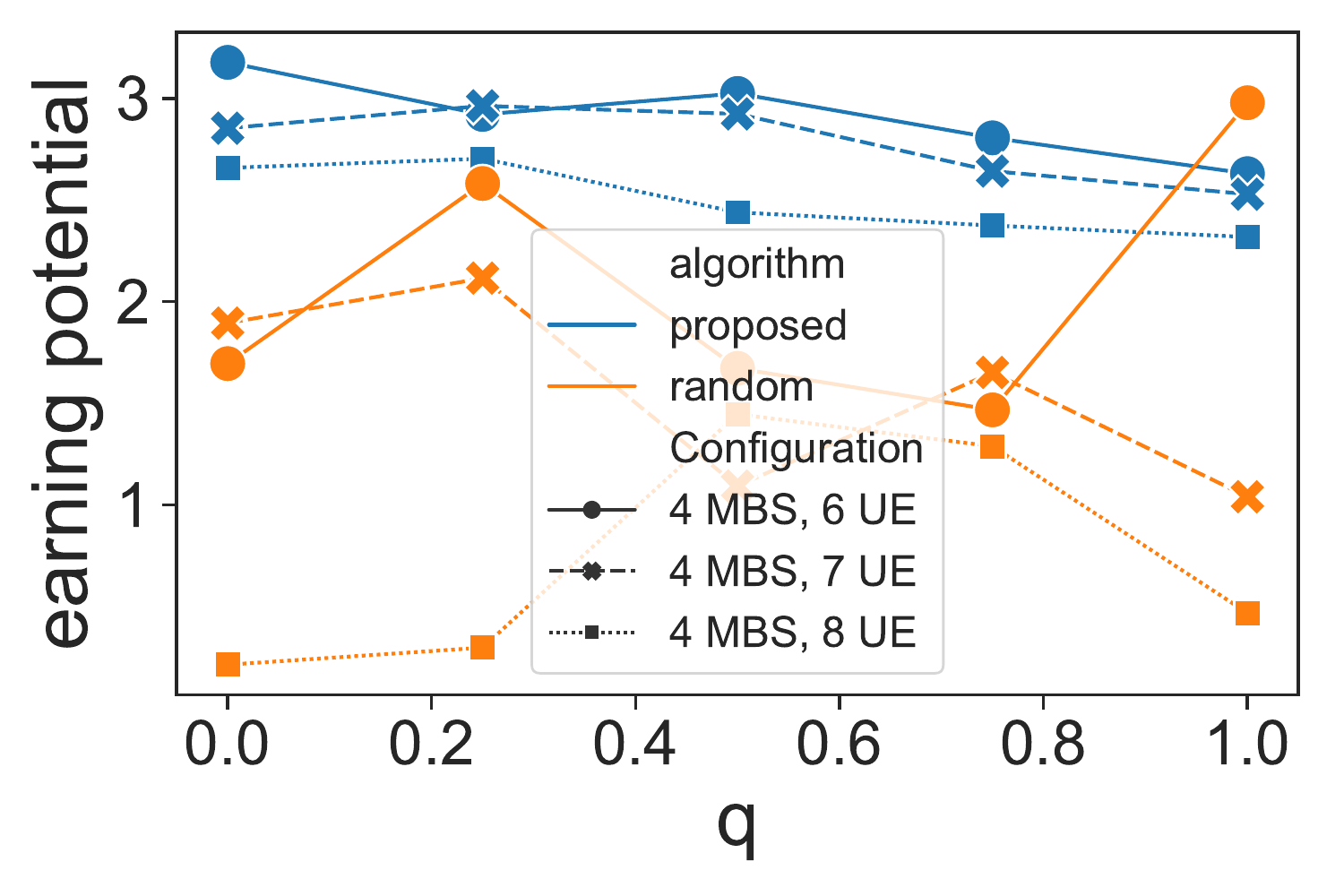}
\label{fig:earning1}
\vspace{-10mm}
\end{minipage}%
}%
\subfigure[Downlink delay wrt $q$.]{
\begin{minipage}[t]{0.25\linewidth}
\centering
\includegraphics[width=1\linewidth]{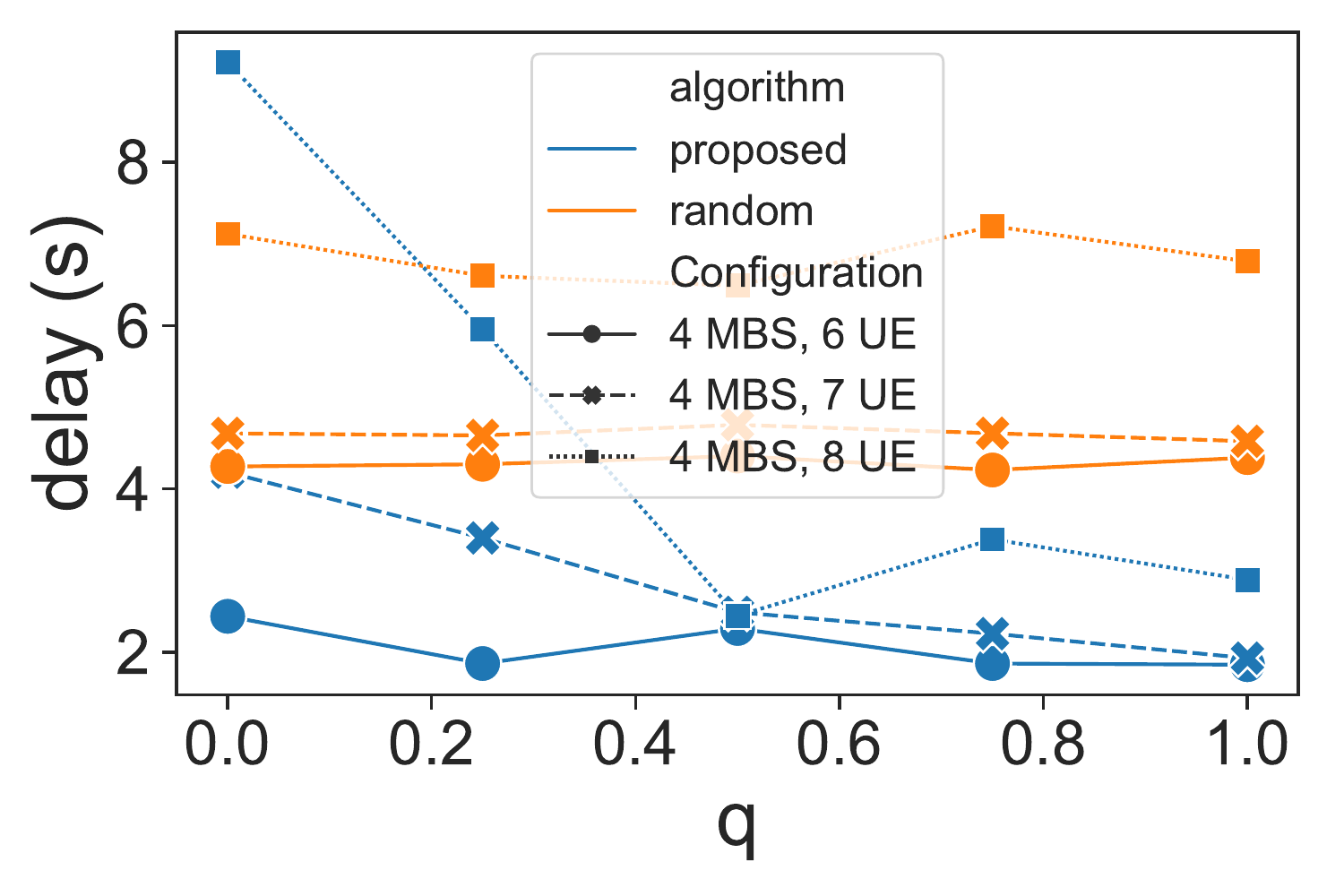}
\label{fig:delay_down1}
\vspace{-10mm}
\end{minipage}%
}%
\subfigure[Battery consumption   wrt $q$.]{
\begin{minipage}[t]{0.25\linewidth}
\centering
\includegraphics[width=1\linewidth]{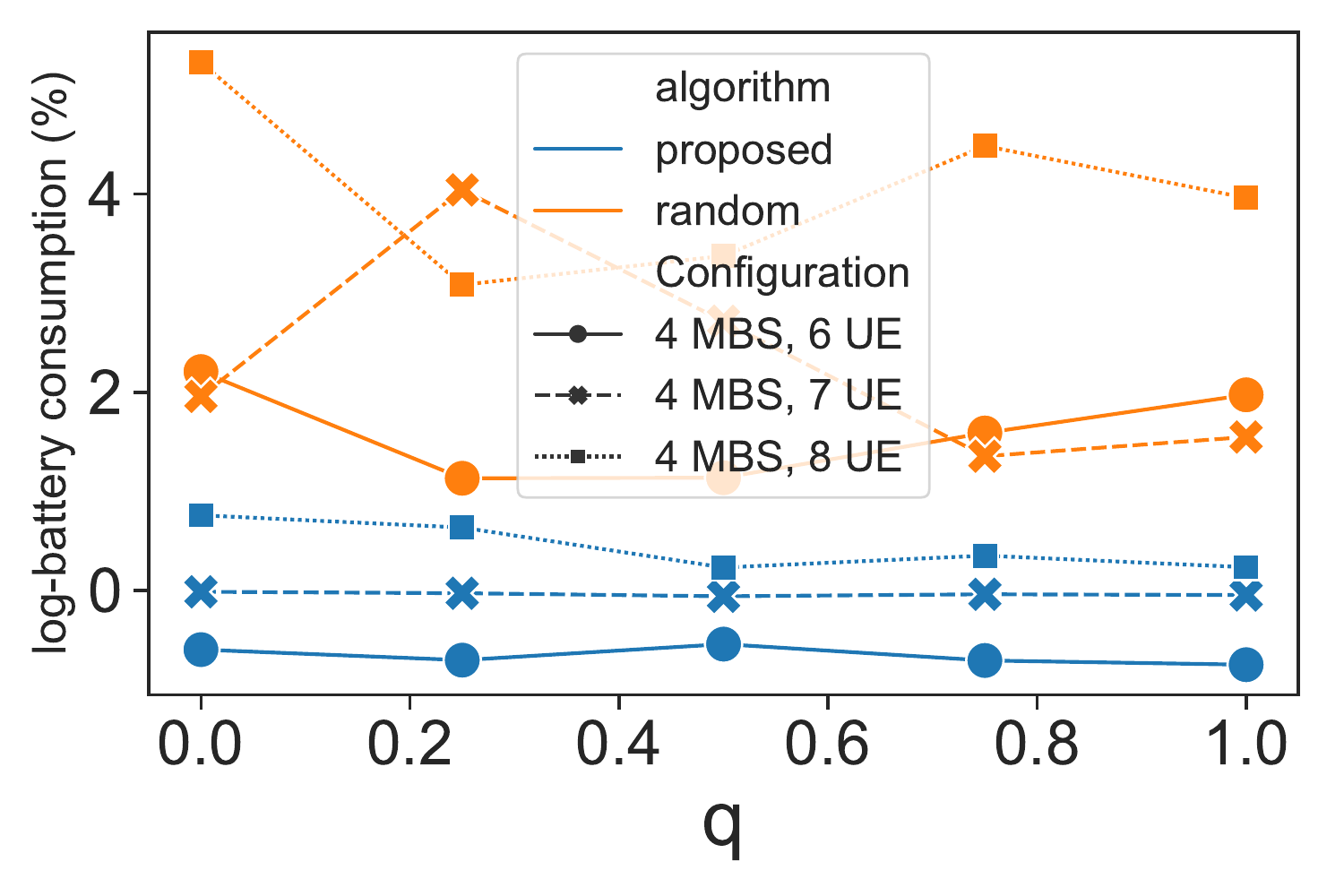}
\label{fig:log_bat1}\vspace{-10mm}
\end{minipage}%
}%
\subfigure[Uplink delay values wrt $q$.]{
\begin{minipage}[t]{0.25\linewidth}
\centering
\includegraphics[width=1\linewidth]{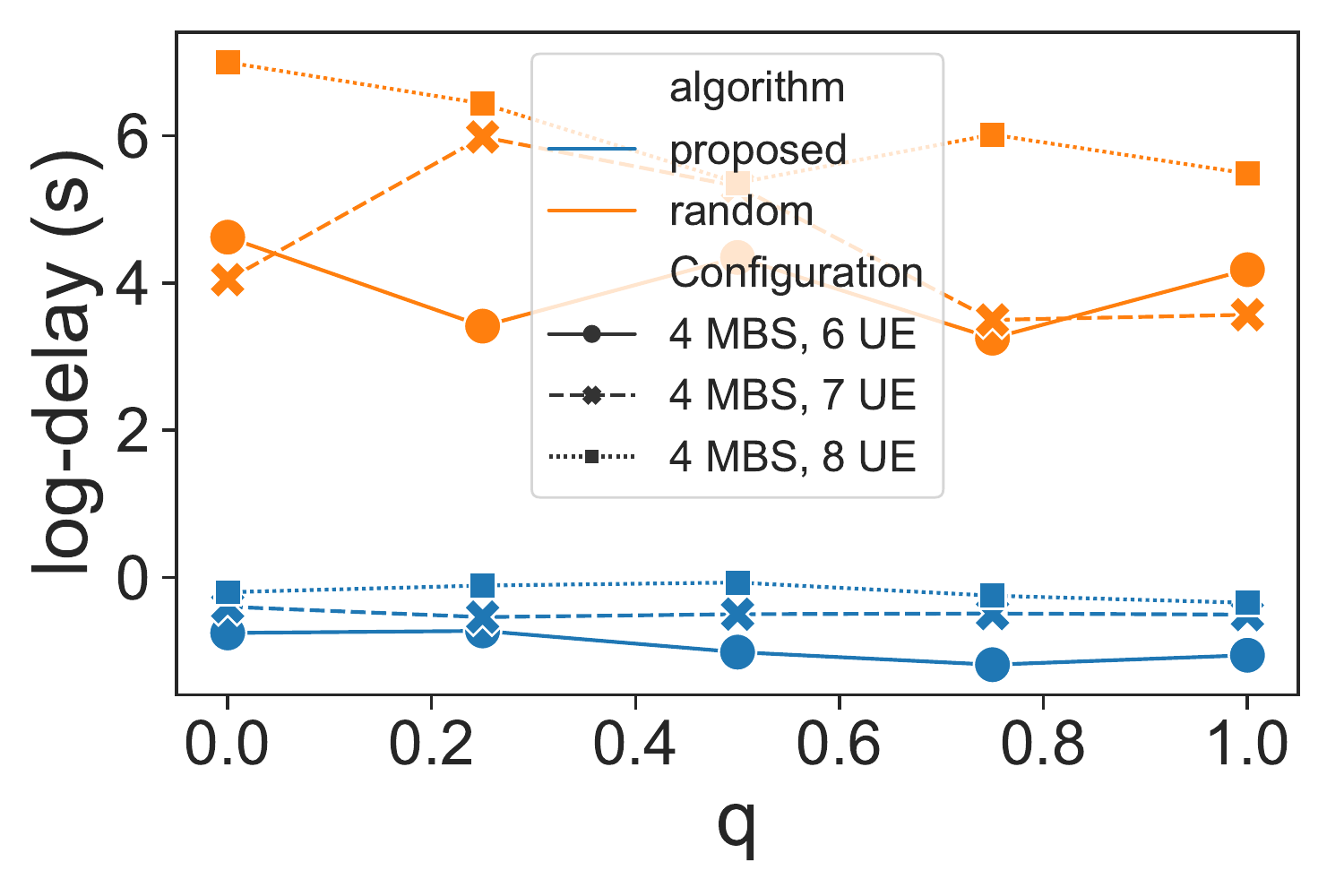}
\label{fig:log_del_up1}
\vspace{-10mm}
\end{minipage}%
}%

\subfigure[Earning potential wrt $h$.]{
\begin{minipage}[t]{0.25\linewidth}
\centering
\includegraphics[width=1\linewidth]{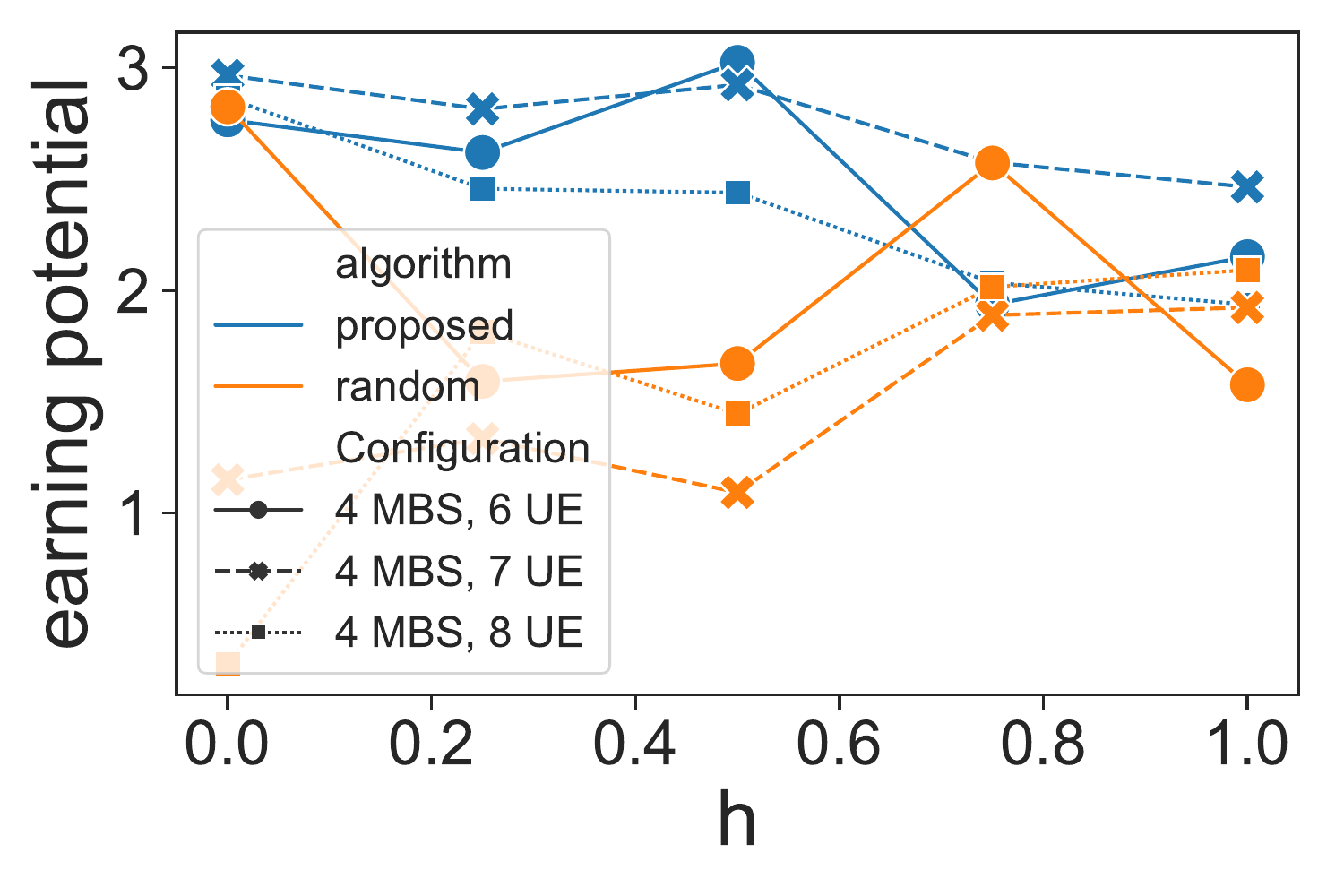}
\label{fig:earning2}
\vspace{-10mm}
\end{minipage}%
}%
\subfigure[Downlink delay wrt $h$.]{
\begin{minipage}[t]{0.25\linewidth}
\centering
\includegraphics[width=1\linewidth]{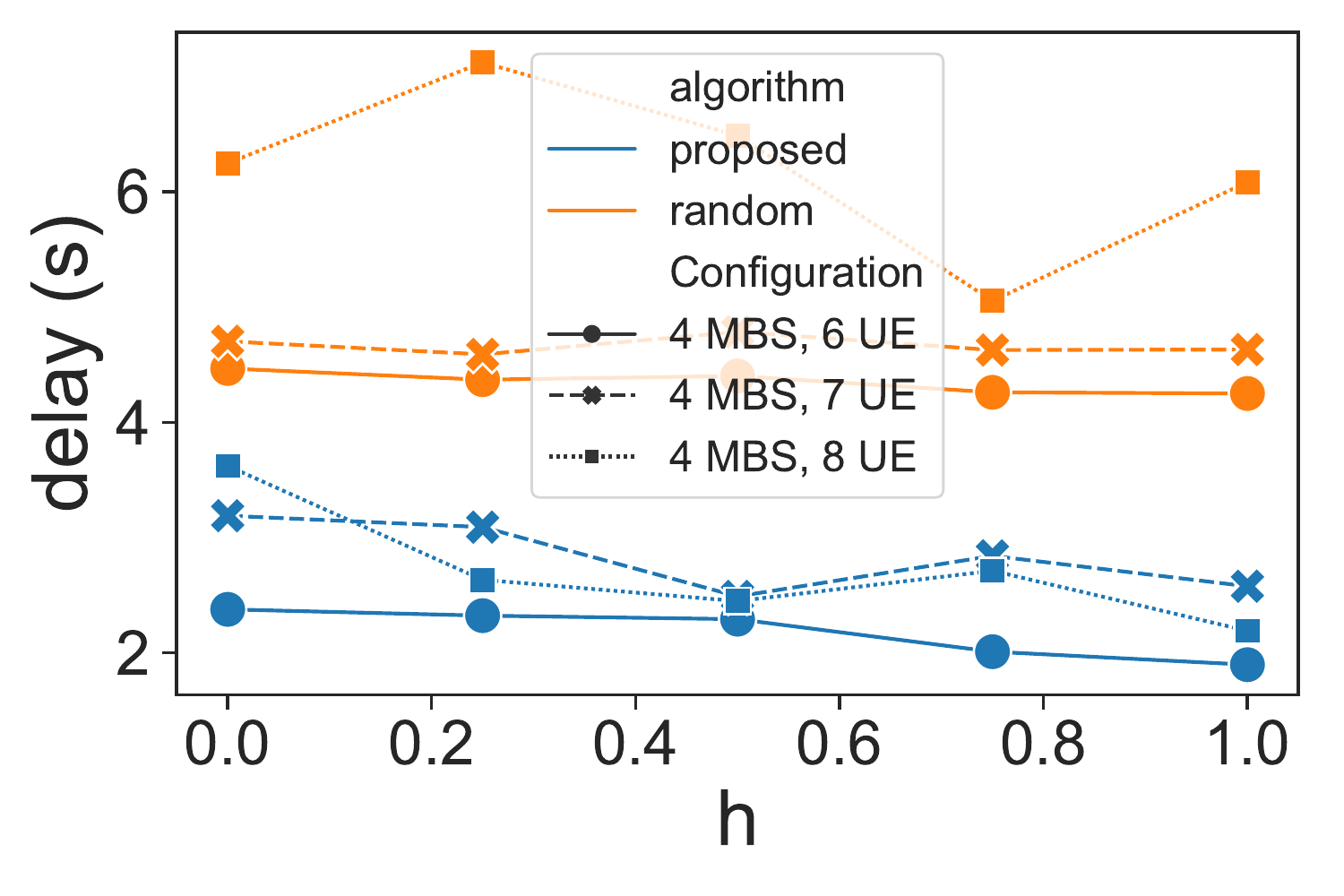}
\label{fig:delay_down2}
\vspace{-10mm}
\end{minipage}%
}%
\subfigure[Battery consumption  wrt $h$.]{
\begin{minipage}[t]{0.25\linewidth}
\centering
\includegraphics[width=1\linewidth]{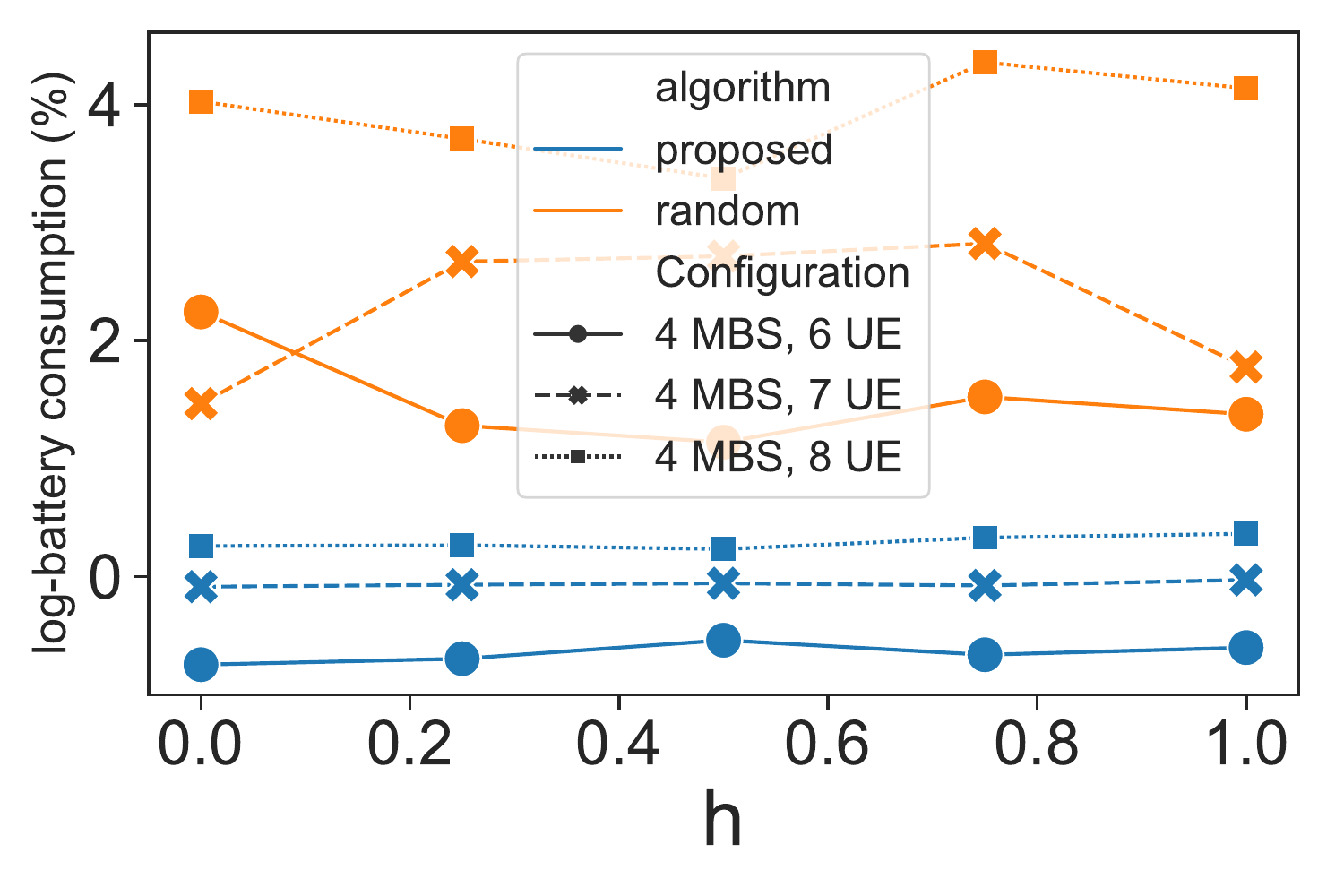}
\label{fig:log_bat2}\vspace{-10mm}
\end{minipage}%
}%
\subfigure[Uplink delay   wrt $h$.]{
\begin{minipage}[t]{0.25\linewidth}
\centering
\includegraphics[width=1\linewidth]{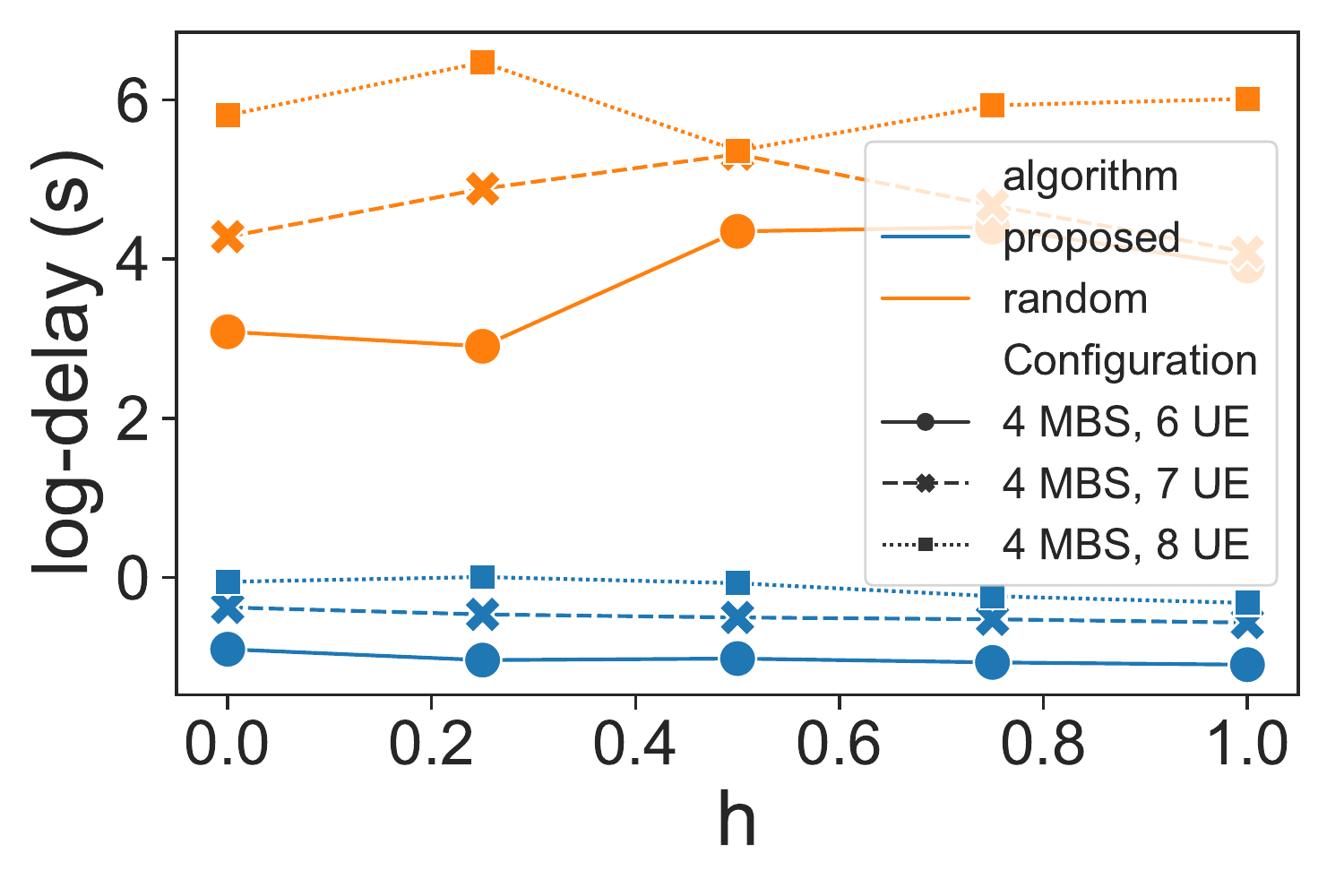}
\label{fig:log_del_up2}
\vspace{-10mm}
\end{minipage}%
}%
\caption{Metric values with respect to (wrt)  different $q$ and $h$.}
\label{fig:complete}
\vspace{-0.8cm}
\end{figure*}

On the other hand, the (ii) IDA model converged at a much lower reward for both the UL and DL agents (shown in Figure \ref{fig:complete_weight_2}), across all configurations, and took longer number of training steps to achieve similar rewards obtained by MALS earlier in the training process. We observe that the independent dual agents model obtained significantly lower downlink and uplink rewards across all configurations, when compared to both the proposed MALS model and the CTDE model. Furthermore, the variance of both downlink and uplink rewards obtained is much larger for the independent dual agents model. This is notable across all configurations. This indicates instability and hyper-sensitivity of the model when handling different seed settings and minute changes in environment. In addition, IDA shows signs of training instability. From Figure \ref{fig:4d}, \ref{fig:4e}, \ref{fig:4f}, we observe that the IDA model have a slight improvement in the uplink rewards in the early training steps, followed by a subsequent decline in rewards obtained, across all configurations (indicated by the lower bands). Some seeds show an evident decline in uplink reward obtained as training proceeds. This shows a convergence failure for some of these seeds in the independent dual-agent model. It is apparent that the downlink agent's rewards are prioritised over the uplink agent's rewards in this case, which renders the IDA model unsuitable for real-world deployment.


The (iii) CTDE model achieves slightly poorer performance to our proposed methods (MALS) in each configuration, but better than the IDA model for both the downlink and uplink rewards (shown in Figure \ref{fig:4a}, \ref{fig:4d}), but achieved lackluster performance for the 4 MBS, 7 UE and 4 MBS, 8 UE configurations for both the downlink and uplink rewards (shown in Figure \ref{fig:4b}, \ref{fig:4c}, \ref{fig:4e}, \ref{fig:4f}). The CTDE produces a large variance of downlink and uplink reward outputs across the different seeds, albeit not as severe as the IDA model, as the model still converges in all seeds. However, the large variance and sub-par rewards obtained renders the CTDE inferior to our proposed methods due to its poorer performance, instability and unreliability.

\subsubsection{\textbf{Weighting Emphasis On Joint Utility function}}
 When dealing with joint optimization problems, it is essential for us to observe how our proposed model handles the varying weights on variables and their influence on variables of interest, as this provides future research insights on multi-agent joint utility function design.

Firstly, we vary the weight $q$ in the downlink reward $R^{\text{d},t}$ and vary the weight $h$ in the uplink reward $R^{\text{u},t}$, that are defined in subsection~\ref{subsection:reward}, with values from the set of \{0, 0.25, 0.50, 0.75, 1\}. While we vary $q$, we keep $h$ at a value of 0.5, and vice versa. We then take the average results of the last 20,000 training steps of the different variables of interest (downlink delay, worst-case (lowest) earning potential, uplink transmission delay, worst-case (greatest) battery charge expenditure). These averaged results are then compared with a random policy that assigns the downlink UE-MBS allocation and uplink power output randomly. We compare our proposed model with a random policy to show how our proposed model adds value in optimizing UE-MBS allocation and power selection. Please refer to the Appendix (shown in Fig. \ref{fig:complete1} and \ref{fig:complete2}) for the complete comparison with the reinforcement learning model structures discussed (IDA and CTDE). We do not focus on the 
 weighted analysis of IDA and CTDE as they have been shown to produce unreliable performance with high variances.
 
 

\textbf{Downlink weight analysis. }
The earnings of our proposed MALS model consistently outperform random allocation across all $q$ weight values, except for $q = 1$ (refer to Figure \ref{fig:earning1}). This indicates that our model significantly improves the earning capability of the UEs. It is worth noting that the random policy agent allows higher earnings for UEs at $q = 1$ compared to our model. This is because higher $q$ values imply more emphasis on minimizing downlink delay and less emphasis on the worst-case (lowest) UE earning capability. At $q = 1$, the downlink agent in our proposed model focuses solely on minimizing downlink and uplink latency, as well as worst-case uplink battery expenditure, without considering the maximization of the worst-case (lowest) UE earning potential. Additionally, it is observed that the worst-case (lowest) UE earning potential decreases as $q$ increases across each scenario configuration in our proposed method. This decrease is expected since higher $q$ values prioritize minimizing downlink latency over maximizing the worst-case (lowest) UE earning potential.

We observe that our proposed downlink latency in various configurations and different $q$ values is significantly lower compared to the random policy, except for the 4 MBS, 8 UE configuration at $q = 0$ (refer to Figure \ref{fig:delay_down1}). It is not surprising to see poorer performance in this specific configuration, as at $q = 0$, there is no weight emphasis on minimizing downlink delay and the entire focus is on maximizing the worst-case (lowest) UE earning potential. Another noteworthy observation is that the overall downlink latency for each configuration decreases as $q$ increases. This trend can be attributed to the fact that a higher value of $q$ increases the weight emphasis on minimizing agent downlink delay. Extensive results are shown in Fig. \ref{fig:complete1}

\textbf{Uplink weight analyses. } Our findings reveal that, across different $h$ values and configurations, the worst-case (highest) battery charge expenditure and uplink delay obtained by our proposed model is significantly lower than that of the random policy (refer to Figure \ref{fig:log_bat2} and \ref{fig:log_del_up2}). This outcome highlights the substantial contribution of our proposed model in the power allocation process. Upon closer examination of the changes in the worst-case (highest) battery charge expenditure with respect to the weight $h$, we observe a general increase in the worst-case battery charge consumption for each configuration, and decrease in uplink delay, as $h$ increases. This behavior can be attributed to the higher weight emphasis placed on minimizing uplink latency by the uplink agent as $h$ increases, at the expense of worst-case (highest) battery charge expenditure. Extensive results are shown in Figure \ref{fig:complete2}.




\vspace{-0.3cm}

\vspace{-7pt}\section{Conclusion \vspace{-7pt}}
\label{section:conclusion}

In this work, we consider a play-to-earn mobile edge computing framework and formulate an asymmetric (discrete-continuous) and asynchronous (alternating DL and UL) multi-variable optimization where the Mobile edge computing Service Provider's objective is to minimize in-game graphics downlink, uplink data transmission latency, and worst-case (greatest) battery charge expenditure, while maximizing the UEs' in-game resolution-influenced worst-case (lowest) earning potential. We then propose a novel multi-agent loss-sharing (MALS) RL model to tackle the abovementioned asynchronous and asymmetric problem, and demonstrate its superiority in performance over other methods. Finally, we conduct joint optimization weighting analyses and show the viability of utilizing our proposed MALS algorithm to tackle joint-optimization problems across different variable weighting emphases.



\begin{spacing}{1.18}
\renewcommand{\refname}{~\\[-25pt]References\vspace{0pt} }

\bibliographystyle{IEEEtran}
\bibliography{ref}

\begin{thebibliography}{10}
\providecommand{\url}[1]{#1}
\csname url@samestyle\endcsname
\providecommand{\newblock}{\relax}
\providecommand{\bibinfo}[2]{#2}
\providecommand{\BIBentrySTDinterwordspacing}{\spaceskip=0pt\relax}
\providecommand{\BIBentryALTinterwordstretchfactor}{4}
\providecommand{\BIBentryALTinterwordspacing}{\spaceskip=\fontdimen2\font plus
\BIBentryALTinterwordstretchfactor\fontdimen3\font minus \fontdimen4\font\relax}
\providecommand{\BIBforeignlanguage}[2]{{%
\expandafter\ifx\csname l@#1\endcsname\relax
\typeout{** WARNING: IEEEtran.bst: No hyphenation pattern has been}%
\typeout{** loaded for the language `#1'. Using the pattern for}%
\typeout{** the default language instead.}%
\else
\language=\csname l@#1\endcsname
\fi
#2}}
\providecommand{\BIBdecl}{\relax}
\BIBdecl

\bibitem{ICC2023MALS}
\BIBentryALTinterwordspacing
T.~J. Chua, W.~Yu, and J.~Zhao, ``Play to earn in the metaverse over wireless networks with deep reinforcement learning,'' \emph{submitted to the 2023 IEEE International Conference on Communications (ICC)}. [Online]. Available: \url{https://personal.ntu.edu.sg/JunZhao/ICC2023MALS.pdf}
\BIBentrySTDinterwordspacing

\bibitem{browne_2021}
\BIBentryALTinterwordspacing
R.~Browne, ``Cash grab or innovation? the video game world is divided over {NFTs},'' Dec 2021. [Online]. Available: \url{https://www.cnbc.com/2021/12/20/cash-grab-or-innovation-the-video-game-world-is-divided-over-nfts.html}
\BIBentrySTDinterwordspacing

\bibitem{polka_city}
\BIBentryALTinterwordspacing
``Polka city.'' [Online]. Available: \url{https://www.polkacity.io/}
\BIBentrySTDinterwordspacing

\bibitem{english}
\BIBentryALTinterwordspacing
``Reality clash.'' [Online]. Available: \url{https://realityclash.com/}
\BIBentrySTDinterwordspacing

\bibitem{roos_2022}
\BIBentryALTinterwordspacing
N.~Roos, ``Best {P}lay-to-{E}arn games with {NFT} or crypto,'' November 2022. [Online]. Available: \url{https://www.playtoearn.online/games/}
\BIBentrySTDinterwordspacing

\bibitem{chen2017resource}
M.~Chen, W.~Saad, and C.~Yin, ``Resource management for wireless virtual reality: Machine learning meets multi-attribute utility,'' in \emph{GLOBECOM 2017-2017 IEEE Global Communications Conference}.\hskip 1em plus 0.5em minus 0.4em\relax IEEE, 2017, pp. 1--7.

\bibitem{wang2021meta}
Y.~Wang, M.~Chen, Z.~Yang, W.~Saad, T.~Luo, S.~Cui, and H.~V. Poor, ``Meta-reinforcement learning for immersive virtual reality over {THz/VLC} wireless networks,'' in \emph{ICC 2021-IEEE International Conference on Communications}.\hskip 1em plus 0.5em minus 0.4em\relax IEEE, 2021, pp. 1--6.

\bibitem{liu2018edge}
Q.~Liu, S.~Huang, J.~Opadere, and T.~Han, ``An edge network orchestrator for mobile augmented reality,'' in \emph{IEEE INFOCOM 2018-IEEE Conference on Computer Communications}.\hskip 1em plus 0.5em minus 0.4em\relax IEEE, 2018, pp. 756--764.

\bibitem{wang2020user}
H.~Wang and J.~Xie, ``User preference based energy-aware mobile ar system with edge computing,'' in \emph{IEEE INFOCOM 2020-IEEE Conference on Computer Communications}.\hskip 1em plus 0.5em minus 0.4em\relax IEEE, 2020, pp. 1379--1388.

\bibitem{lu2020optimization}
H.~Lu, C.~Gu, F.~Luo, W.~Ding, and X.~Liu, ``Optimization of lightweight task offloading strategy for mobile edge computing based on deep reinforcement learning,'' \emph{Future Generation Computer Systems}, vol. 102, pp. 847--861, 2020.

\bibitem{alfakih2020task}
T.~Alfakih, M.~M. Hassan, A.~Gumaei, C.~Savaglio, and G.~Fortino, ``Task offloading and resource allocation for mobile edge computing by deep reinforcement learning based on sarsa,'' \emph{IEEE Access}, vol.~8, pp. 54\,074--54\,084, 2020.

\bibitem{qiao2019deep}
G.~Qiao, S.~Leng, S.~Maharjan, Y.~Zhang, and N.~Ansari, ``Deep reinforcement learning for cooperative content caching in vehicular edge computing and networks,'' \emph{IEEE Internet of Things Journal}, vol.~7, no.~1, pp. 247--257, 2019.

\bibitem{bi2021lyapunov}
S.~Bi, L.~Huang, H.~Wang, and Y.-J.~A. Zhang, ``Lyapunov-guided deep reinforcement learning for stable online computation offloading in mobile-edge computing networks,'' \emph{IEEE Transactions on Wireless Communications}, vol.~20, no.~11, pp. 7519--7537, 2021.

\bibitem{truong2021partial}
T.~P. Truong, T.-V. Nguyen, W.~Noh, S.~Cho \emph{et~al.}, ``Partial computation offloading in noma-assisted mobile-edge computing systems using deep reinforcement learning,'' \emph{IEEE Internet of Things Journal}, vol.~8, no.~17, pp. 13\,196--13\,208, 2021.

\bibitem{qiu2019online}
X.~Qiu, L.~Liu, W.~Chen, Z.~Hong, and Z.~Zheng, ``Online deep reinforcement learning for computation offloading in blockchain-empowered mobile edge computing,'' \emph{IEEE Transactions on Vehicular Technology}, vol.~68, no.~8, pp. 8050--8062, 2019.

\bibitem{JO1}
D.~Guo, L.~Tang, X.~Zhang, and Y.-C. Liang, ``Joint optimization of handover control and power allocation based on multi-agent deep reinforcement learning,'' \emph{IEEE Transactions on Vehicular Technology}, vol.~69, no.~11, pp. 13\,124--13\,138, 2020.

\bibitem{JO2}
C.~He, Y.~Hu, Y.~Chen, and B.~Zeng, ``Joint power allocation and channel assignment for {NOMA} with deep reinforcement learning,'' \emph{IEEE Journal on Selected Areas in Communications}, vol.~37, no.~10, pp. 2200--2210, 2019.

\bibitem{lowe2017multi}
R.~Lowe, Y.~I. Wu, A.~Tamar, J.~Harb, O.~Pieter~Abbeel, and I.~Mordatch, ``Multi-agent actor-critic for mixed cooperative-competitive environments,'' \emph{Advances in {N}eural {I}nformation {P}rocessing {S}ystems}, vol.~30, 2017.

\bibitem{rashid2018qmix}
T.~Rashid, M.~Samvelyan, C.~Schroeder, G.~Farquhar, J.~Foerster, and S.~Whiteson, ``Qmix: Monotonic value function factorisation for deep multi-agent reinforcement learning,'' in \emph{International conference on machine learning}.\hskip 1em plus 0.5em minus 0.4em\relax PMLR, 2018, pp. 4295--4304.

\bibitem{dai2018survey}
L.~Dai, B.~Wang, Z.~Ding, Z.~Wang, S.~Chen, and L.~Hanzo, ``A survey of non-orthogonal multiple access for 5g,'' \emph{IEEE communications surveys \& tutorials}, vol.~20, no.~3, pp. 2294--2323, 2018.

\bibitem{NOMA}
------, ``A survey of non-orthogonal multiple access for 5g,'' \emph{IEEE communications surveys \& tutorials}, vol.~20, no.~3, pp. 2294--2323, 2018.

\bibitem{feng2022resource}
J.~Feng and J.~Zhao, ``Resource allocation for augmented reality empowered vehicular edge metaverse,'' \emph{arXiv preprint arXiv:2212.01325}, 2022.

\bibitem{yang2012crowdsourcing}
D.~Yang, G.~Xue, X.~Fang, and J.~Tang, ``Crowdsourcing to smartphones: Incentive mechanism design for mobile phone sensing,'' in \emph{Annual International Conference on Mobile Computing and Networking (MobiCom)}, 2012, pp. 173--184.

\bibitem{van2017hybrid}
H.~Van~Seijen, M.~Fatemi, J.~Romoff, R.~Laroche, T.~Barnes, and J.~Tsang, ``Hybrid reward architecture for reinforcement learning,'' \emph{Advances in {N}eural {I}nformation {P}rocessing {S}ystems}, vol.~30, 2017.

\bibitem{schulman2017proximal}
J.~Schulman, F.~Wolski, P.~Dhariwal, A.~Radford, and O.~Klimov, ``Proximal policy optimization algorithms,'' \emph{arXiv preprint arXiv:1707.06347}, 2017.

\bibitem{sutton1999policy}
R.~S. Sutton, D.~McAllester, S.~Singh, and Y.~Mansour, ``Policy gradient methods for reinforcement learning with function approximation,'' \emph{Advances in {N}eural {I}nformation {P}rocessing {S}ystems}, vol.~12, 1999.

\bibitem{PPO}
J.~Schulman, F.~Wolski, P.~Dhariwal, A.~Radford, and O.~Klimov, ``Proximal policy optimization algorithms,'' \emph{arXiv preprint arXiv:1707.06347}, 2017.

\bibitem{GAE}
J.~Schulman, P.~Moritz, S.~Levine, M.~Jordan, and P.~Abbeel, ``High-dimensional continuous control using generalized advantage estimation,'' \emph{arXiv preprint arXiv:1506.02438}, 2015.

\bibitem{zhang2021multi}
K.~Zhang, Z.~Yang, and T.~Ba{\c{s}}ar, ``Multi-agent reinforcement learning: A selective overview of theories and algorithms,'' \emph{Handbook of Reinforcement Learning and Control}, pp. 321--384, 2021.

\bibitem{foerster2018counterfactual}
J.~Foerster, G.~Farquhar, T.~Afouras, N.~Nardelli, and S.~Whiteson, ``Counterfactual multi-agent policy gradients,'' in \emph{Proceedings of the AAAI conference on artificial intelligence}, vol.~32, no.~1, 2018.

\bibitem{yang20196g}
P.~Yang, Y.~Xiao, M.~Xiao, and S.~Li, ``{6G} wireless communications: Vision and potential techniques,'' \emph{IEEE Network}, vol.~33, no.~4, pp. 70--75, 2019.

\bibitem{adam}
D.~P. Kingma and J.~Ba, ``Adam: A method for stochastic optimization,'' \emph{arXiv preprint arXiv:1412.6980}, 2014.

\bibitem{ricianRIS}
I.~Yildirim, A.~Uyrus, and E.~Basar, ``Modeling and analysis of reconfigurable intelligent surfaces for indoor and outdoor applications in future wireless networks,'' \emph{IEEE Transactions on Communications}, vol.~69, no.~2, pp. 1290--1301, 2021.

\end{thebibliography}

\end{spacing}



\vspace{-7pt}\begin{center}
{Appendix: Model performance and weighting comparisons between models \vspace{-0.4cm}} 
\end{center}


\begin{figure*}[h]

\centering
\subfigtopskip=2pt
\subfigbottomskip=2pt



\subfigure[Downlink delay with 6 UEs.]{
\begin{minipage}[t]{0.25\linewidth}
\centering
\includegraphics[width=1\linewidth]{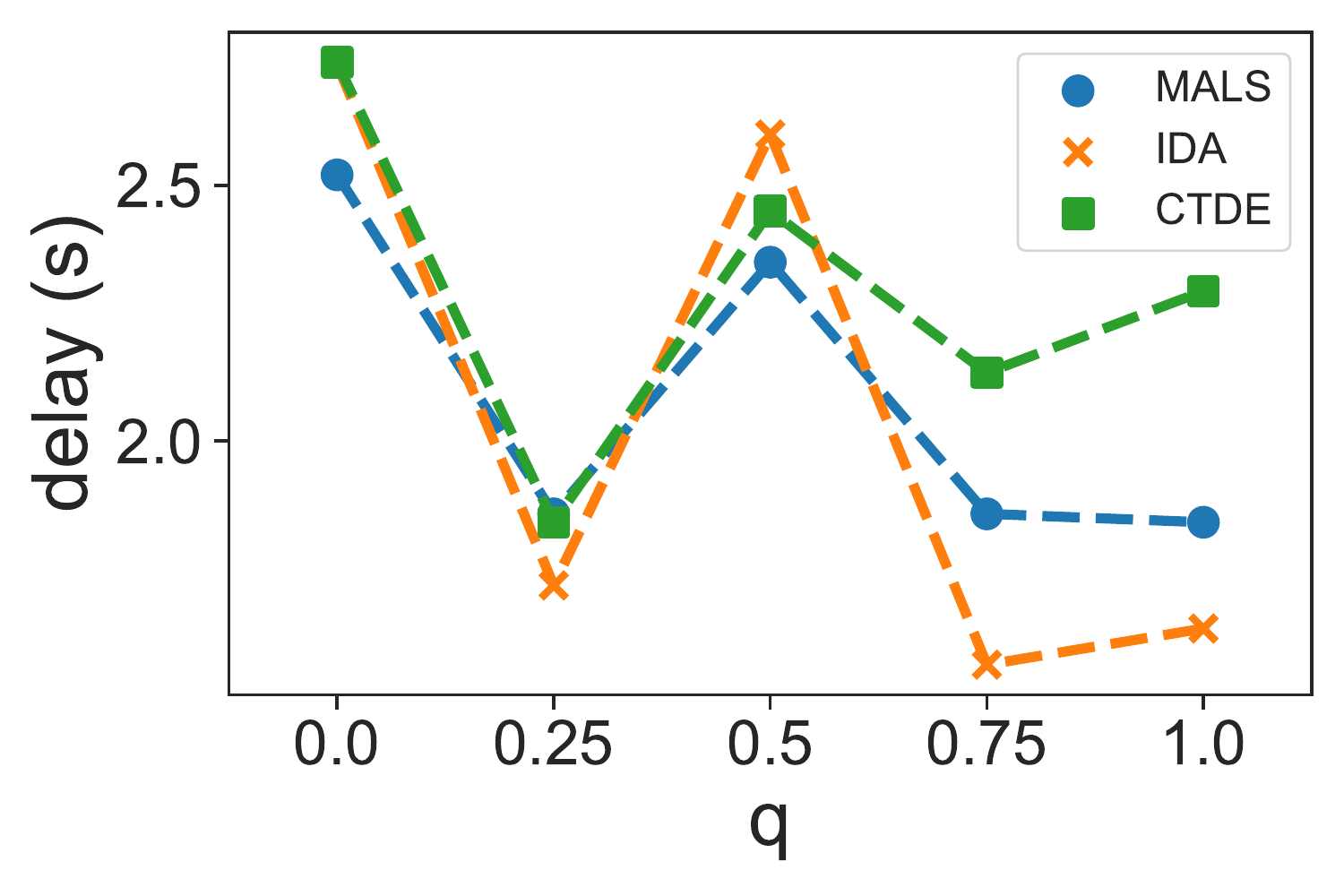}
\vspace{-10mm}
\end{minipage}
}%
\subfigure[Earning potential with  6 UEs.]{
\begin{minipage}[t]{0.25\linewidth}
\centering
\includegraphics[width=1\linewidth]{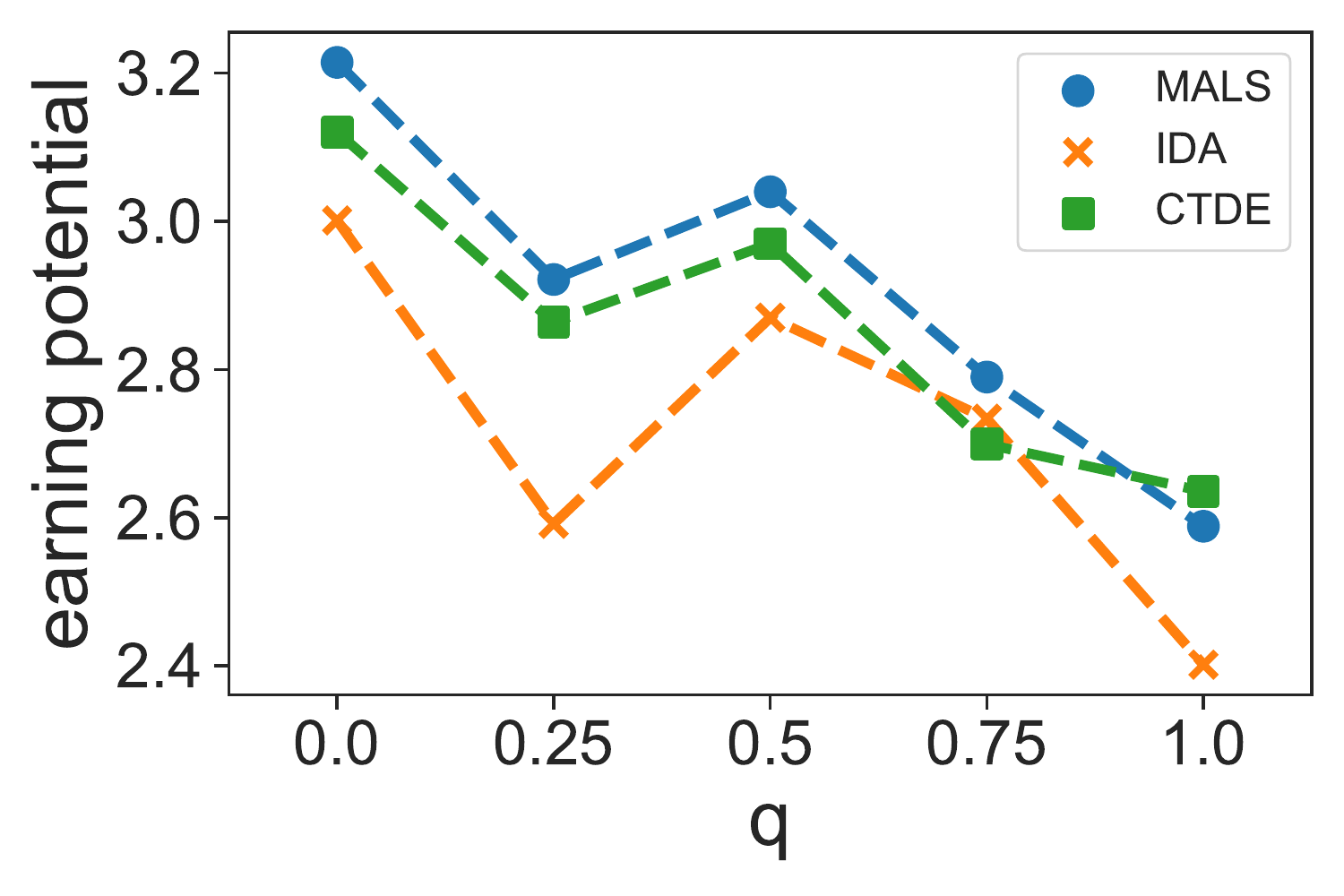}
\vspace{-10mm}
\end{minipage}
}%
\subfigure[Up-link delay with 6 UEs.]{
\begin{minipage}[t]{0.25\linewidth}
\centering
\includegraphics[width=1\linewidth]{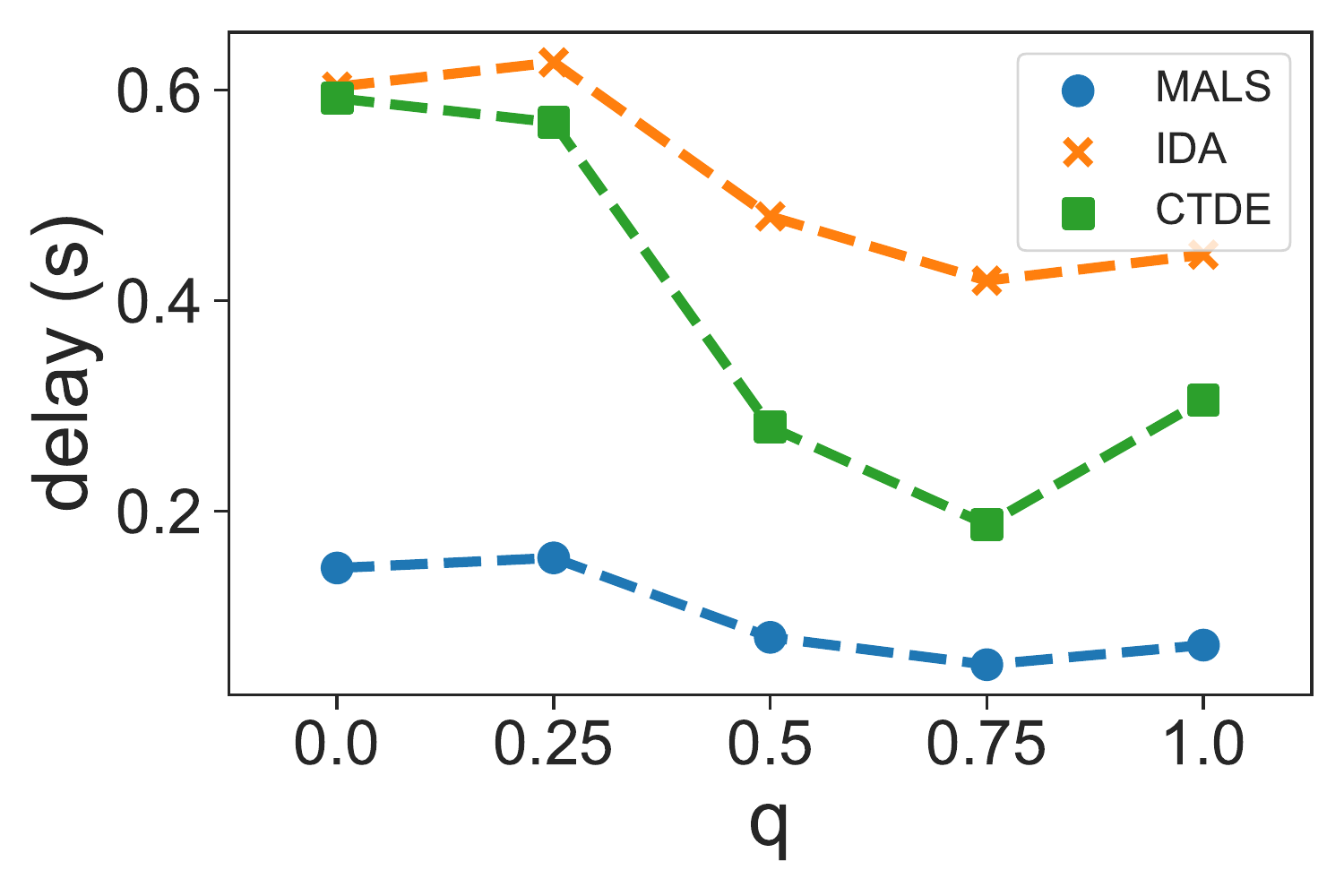}
\vspace{-10mm}
\end{minipage}
}%
\subfigure[Battery usage with 6 UEs.]{
\begin{minipage}[t]{0.25\linewidth}
\centering
\includegraphics[width=1\linewidth]{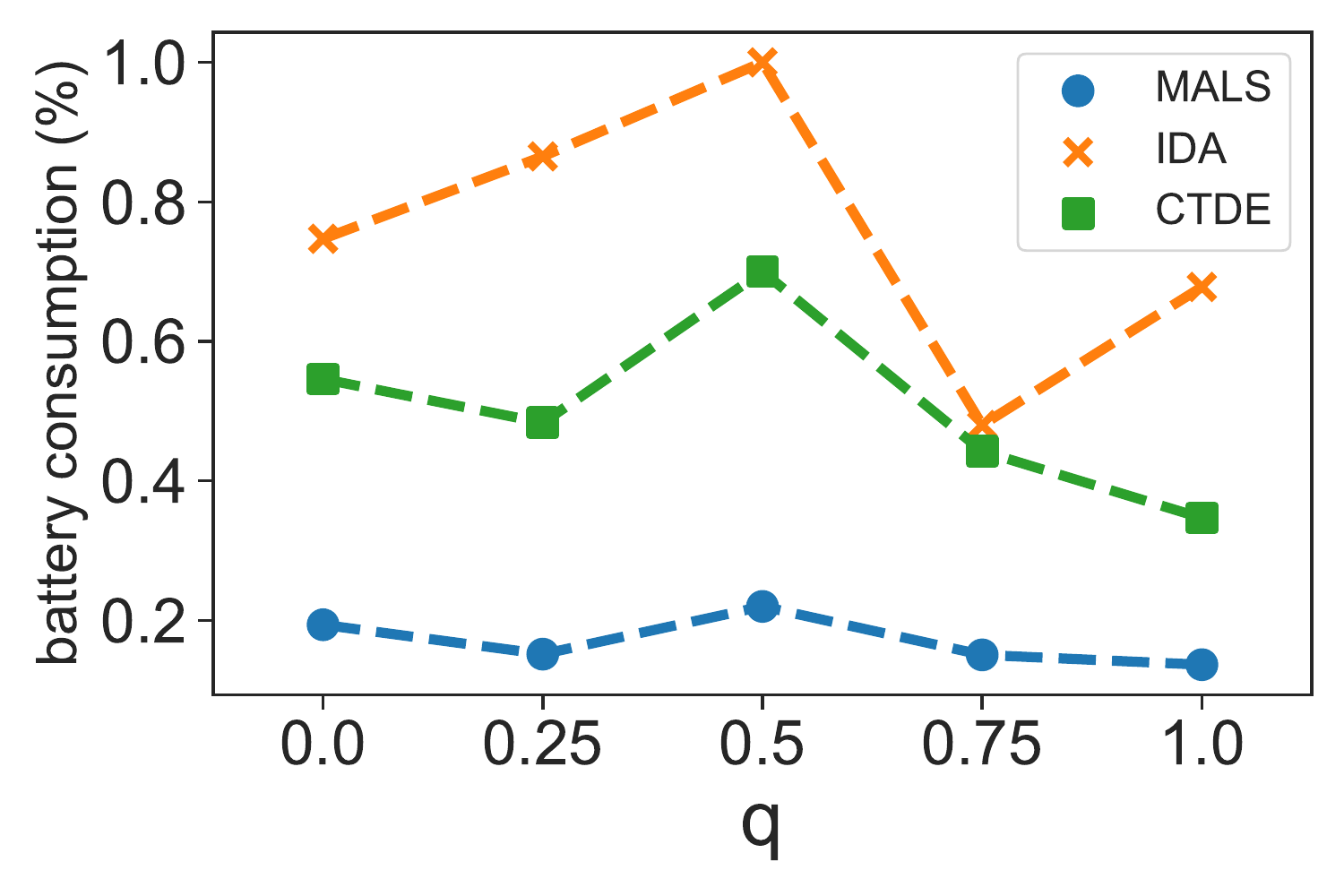}
\vspace{-10mm}
\end{minipage}%
}%

\subfigure[Downlink delay with 7 UEs.]{
\begin{minipage}[t]{0.25\linewidth}
\centering
\includegraphics[width=1\linewidth]{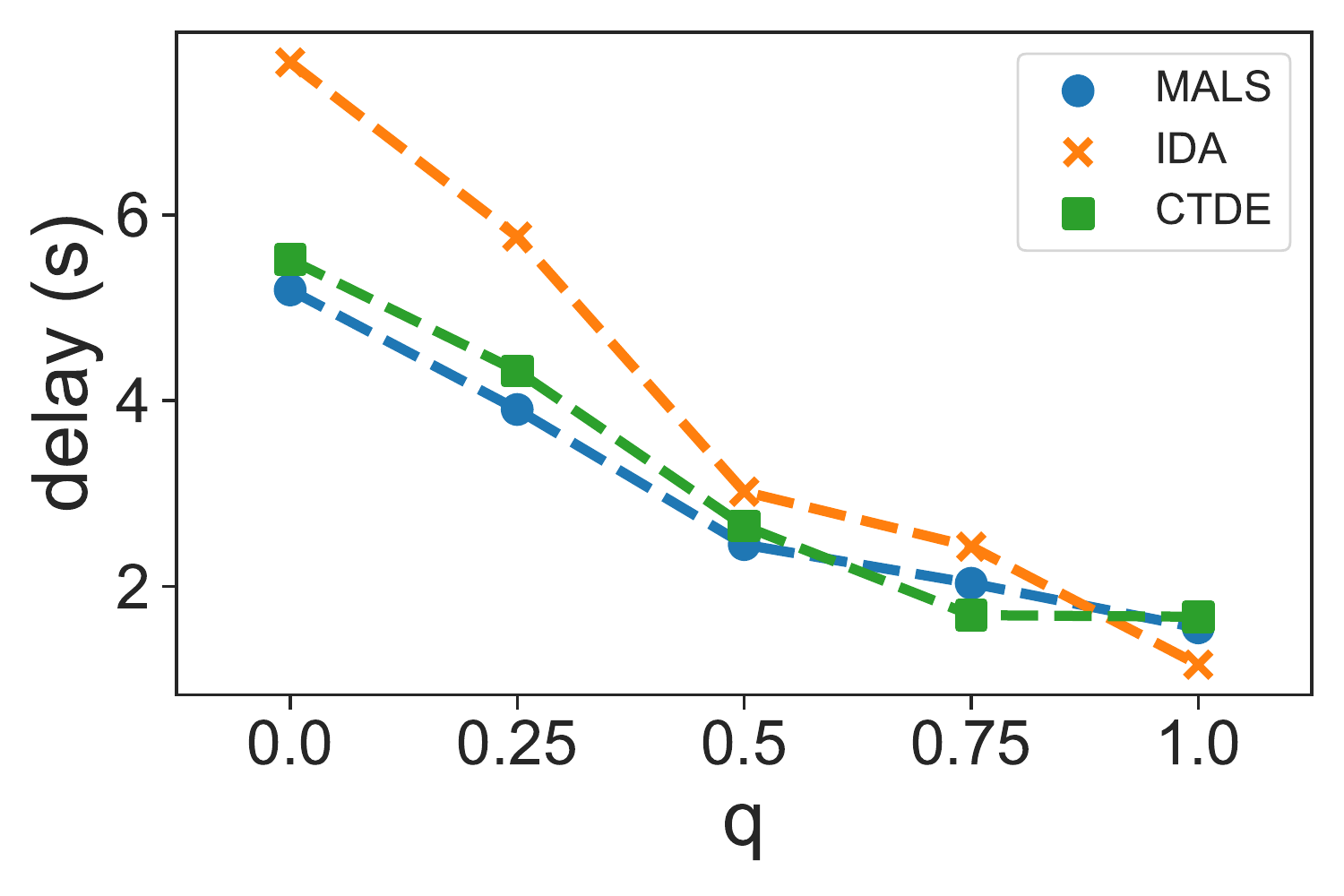}
\vspace{-10mm}
\end{minipage}
}%
\subfigure[Earning potential with  7 UEs.]{
\begin{minipage}[t]{0.25\linewidth}
\centering
\includegraphics[width=1\linewidth]{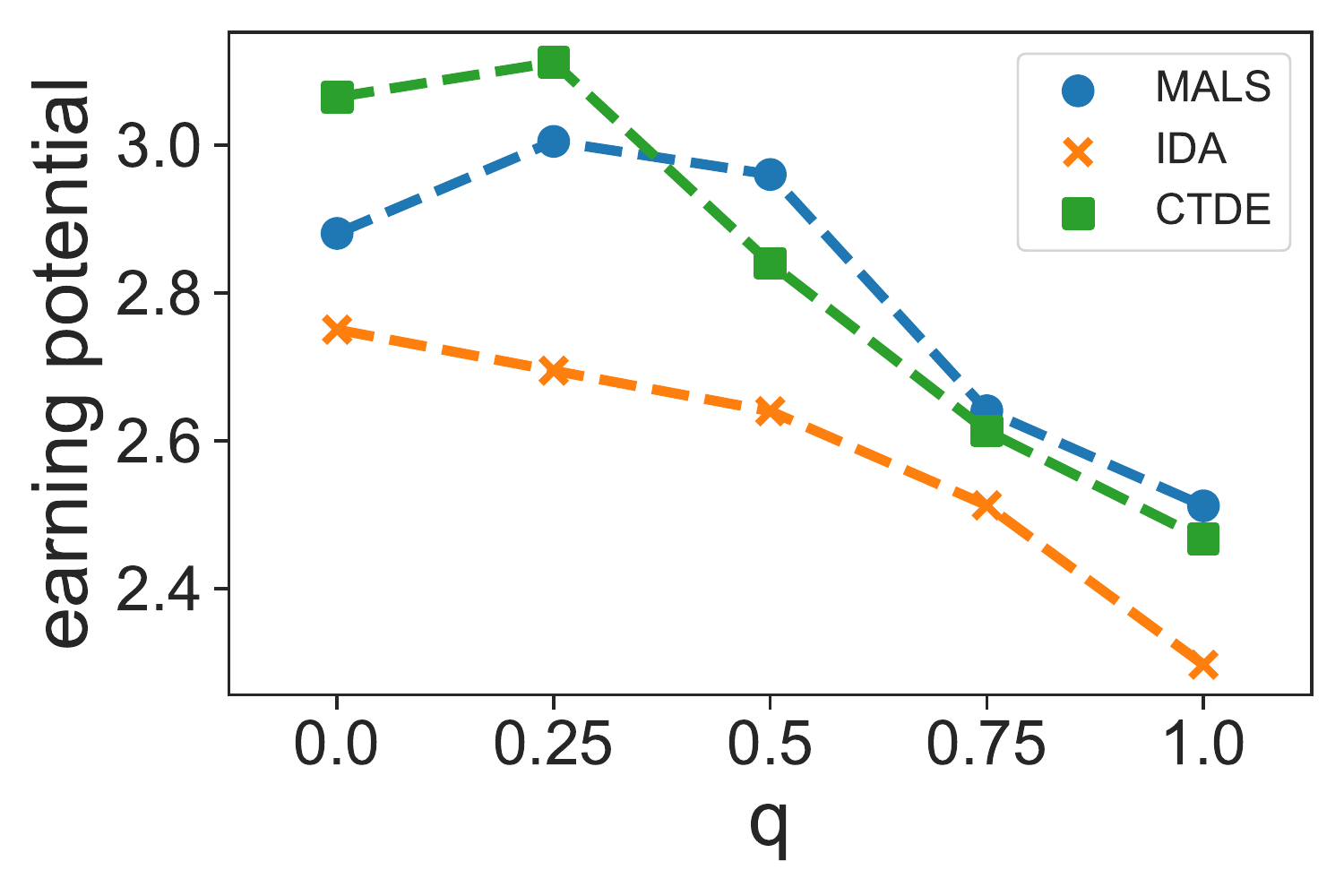}
\vspace{-10mm}
\end{minipage}
}%
\subfigure[Up-link delay with 7 UEs.]{
\begin{minipage}[t]{0.25\linewidth}
\centering
\includegraphics[width=1\linewidth]{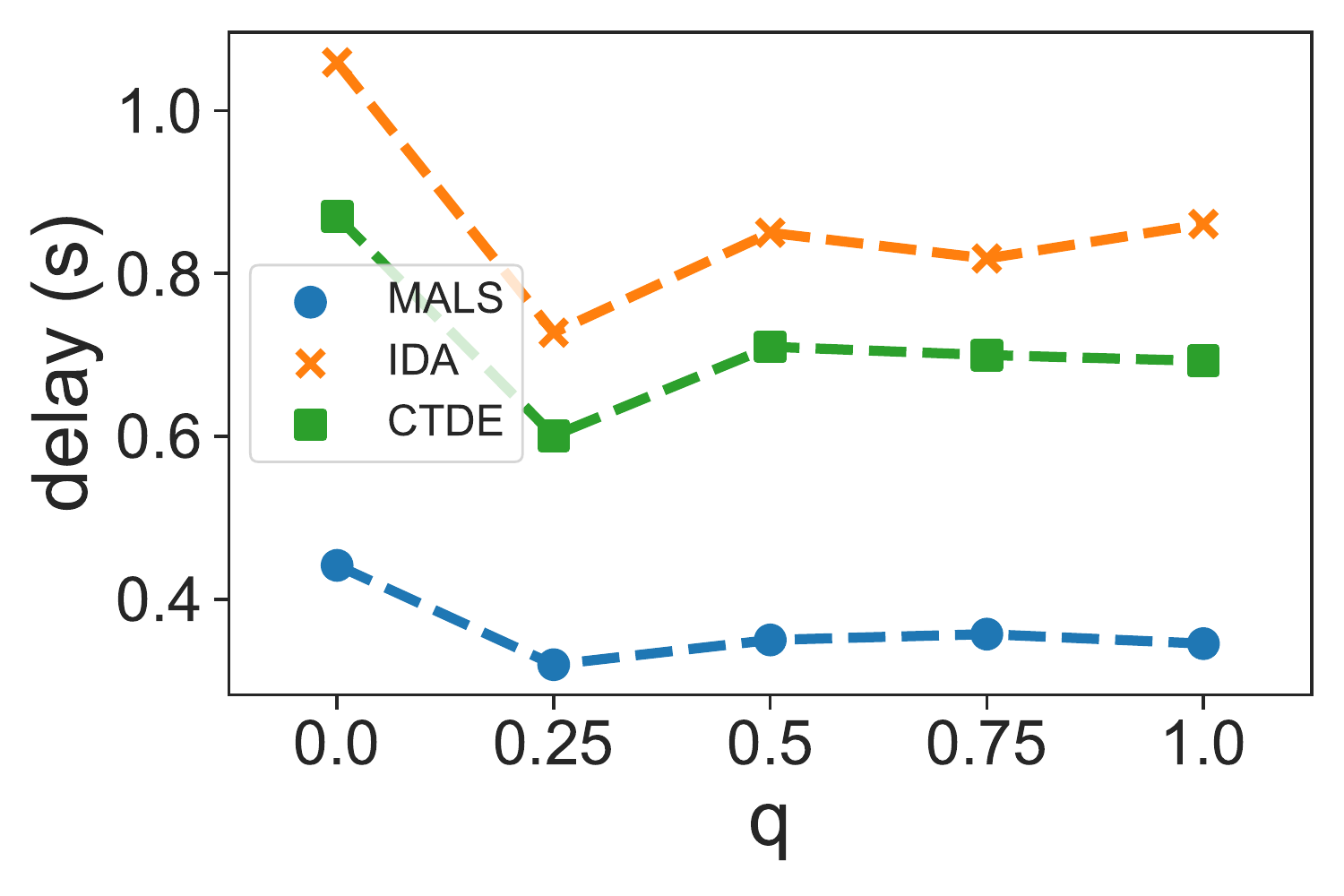}
\vspace{-10mm}
\end{minipage}%
}%
\subfigure[Battery usage with 7 UEs.]{
\begin{minipage}[t]{0.25\linewidth}
\centering
\includegraphics[width=1\linewidth]{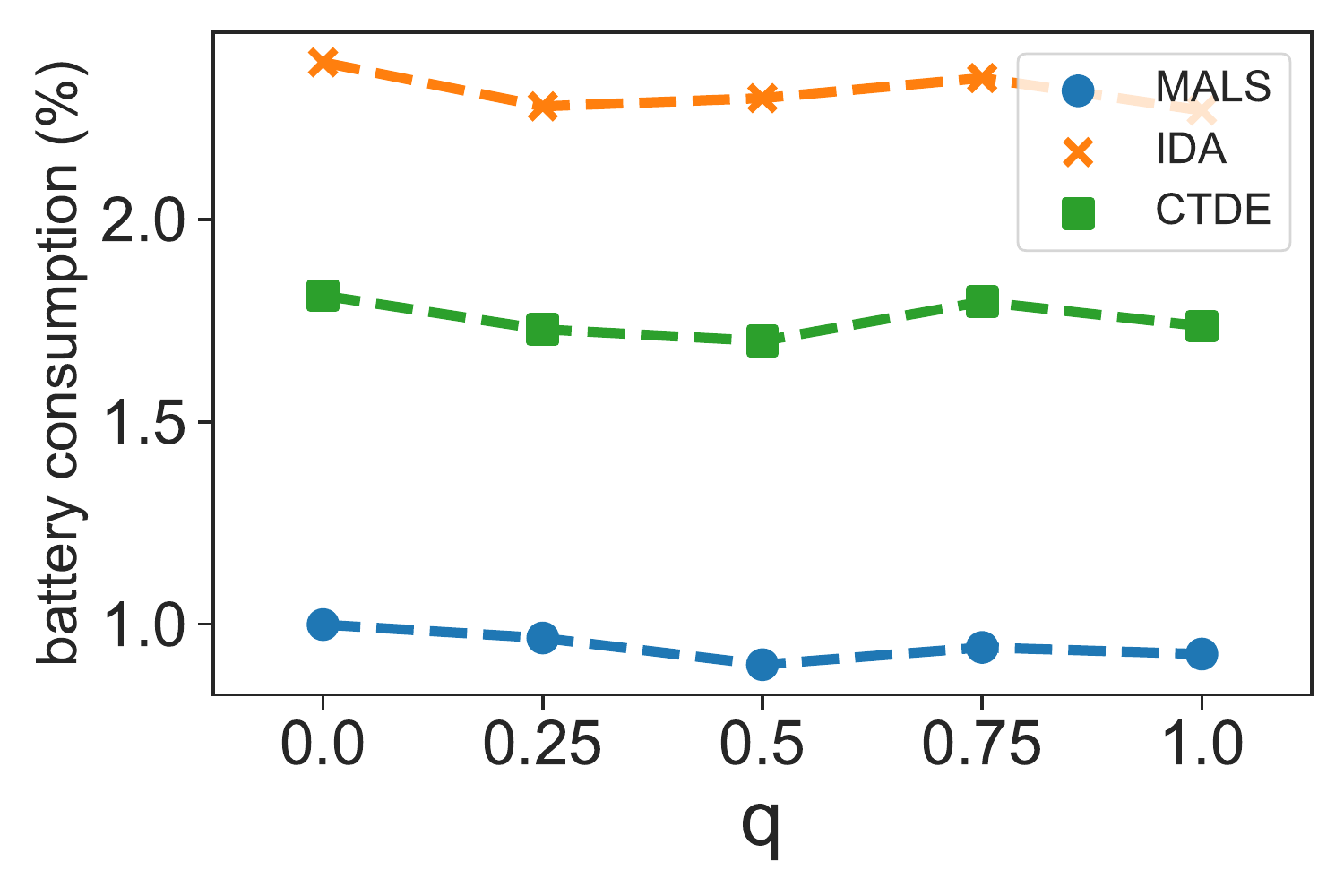}
\vspace{-10mm}
\end{minipage}%
}%

\subfigure[Downlink delay with 8 UEs.]{
\begin{minipage}[t]{0.25\linewidth}
\centering
\includegraphics[width=1\linewidth]{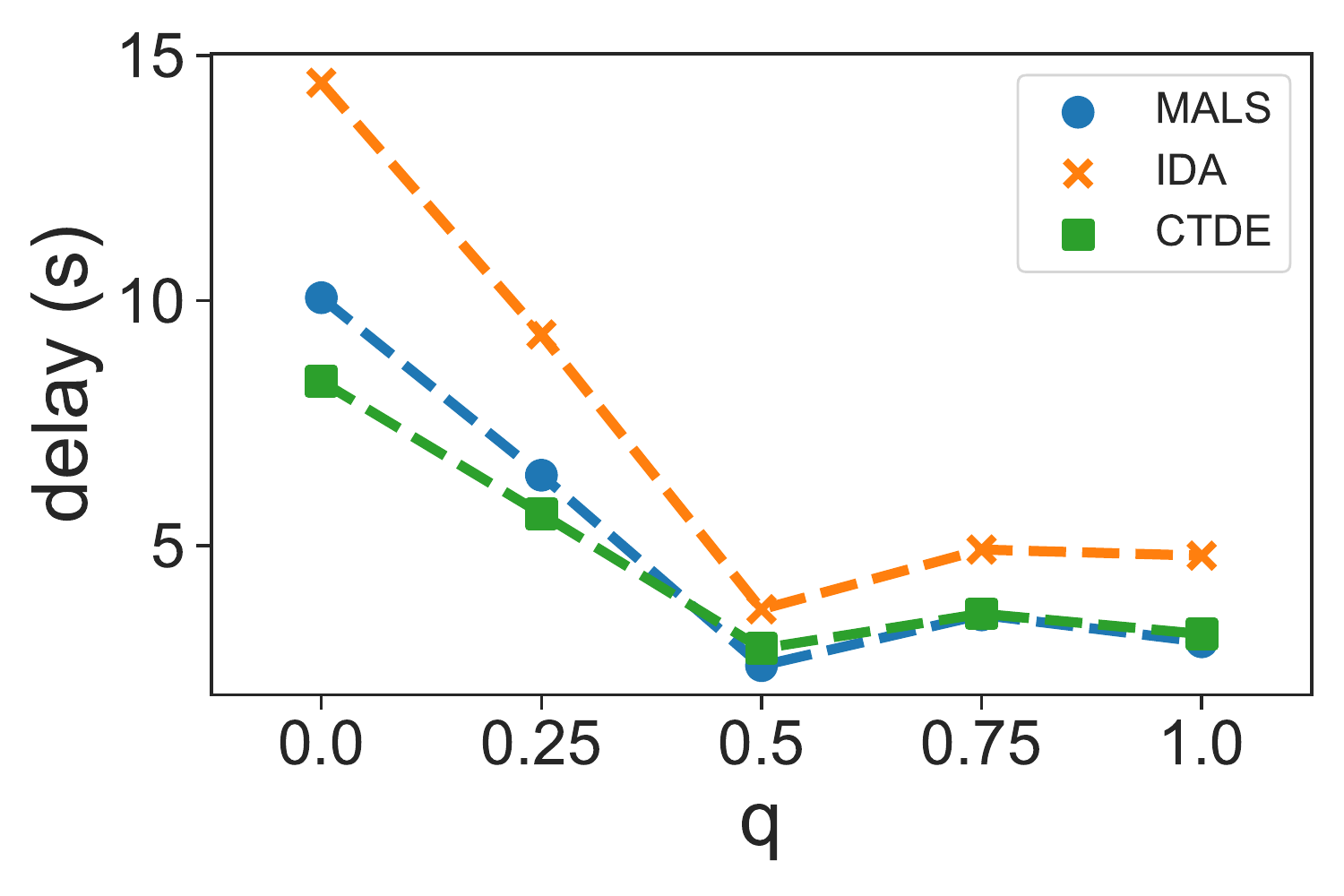}
\vspace{-10mm}
\end{minipage}
}%
\subfigure[Earning potential with  8 UEs.]{
\begin{minipage}[t]{0.25\linewidth}
\centering
\includegraphics[width=1\linewidth]{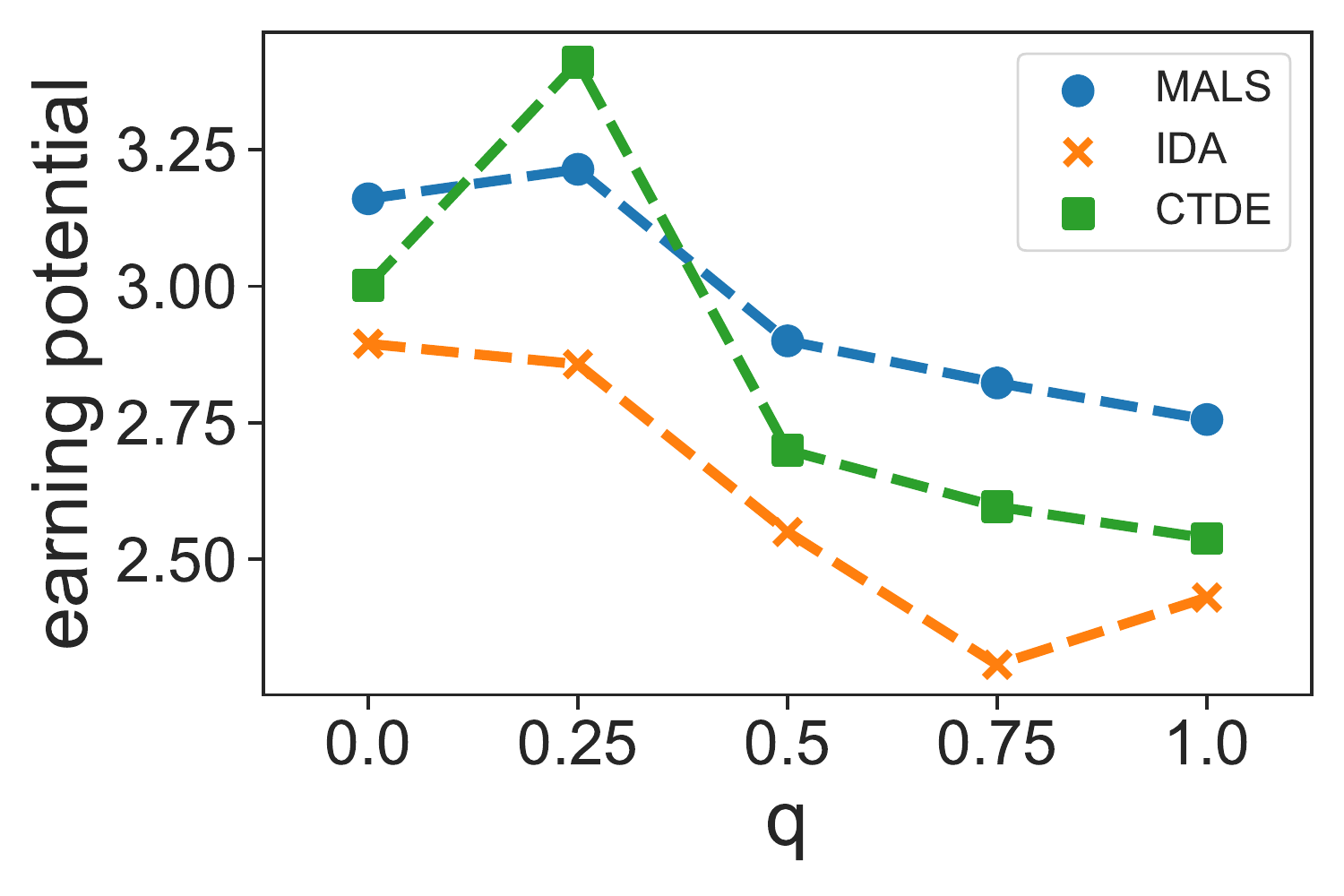}
\vspace{-10mm}
\end{minipage}%
}%
\subfigure[Up-link delay with 8 UEs.]{
\begin{minipage}[t]{0.25\linewidth}
\centering
\includegraphics[width=1\linewidth]{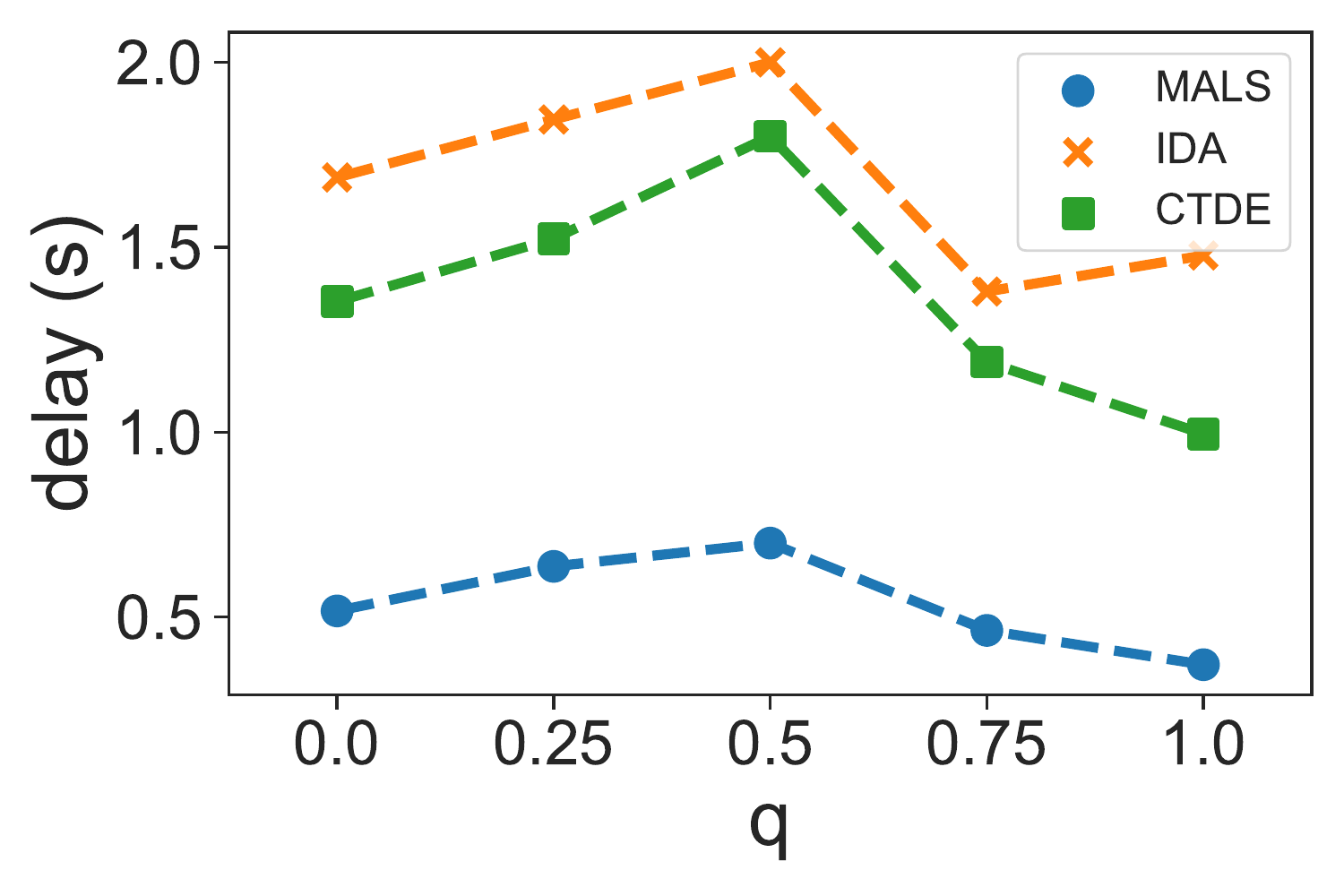}
\vspace{-10mm}
\end{minipage}%
}%
\subfigure[Battery usage with 8 UEs.]{
\begin{minipage}[t]{0.25\linewidth}
\centering
\includegraphics[width=1\linewidth]{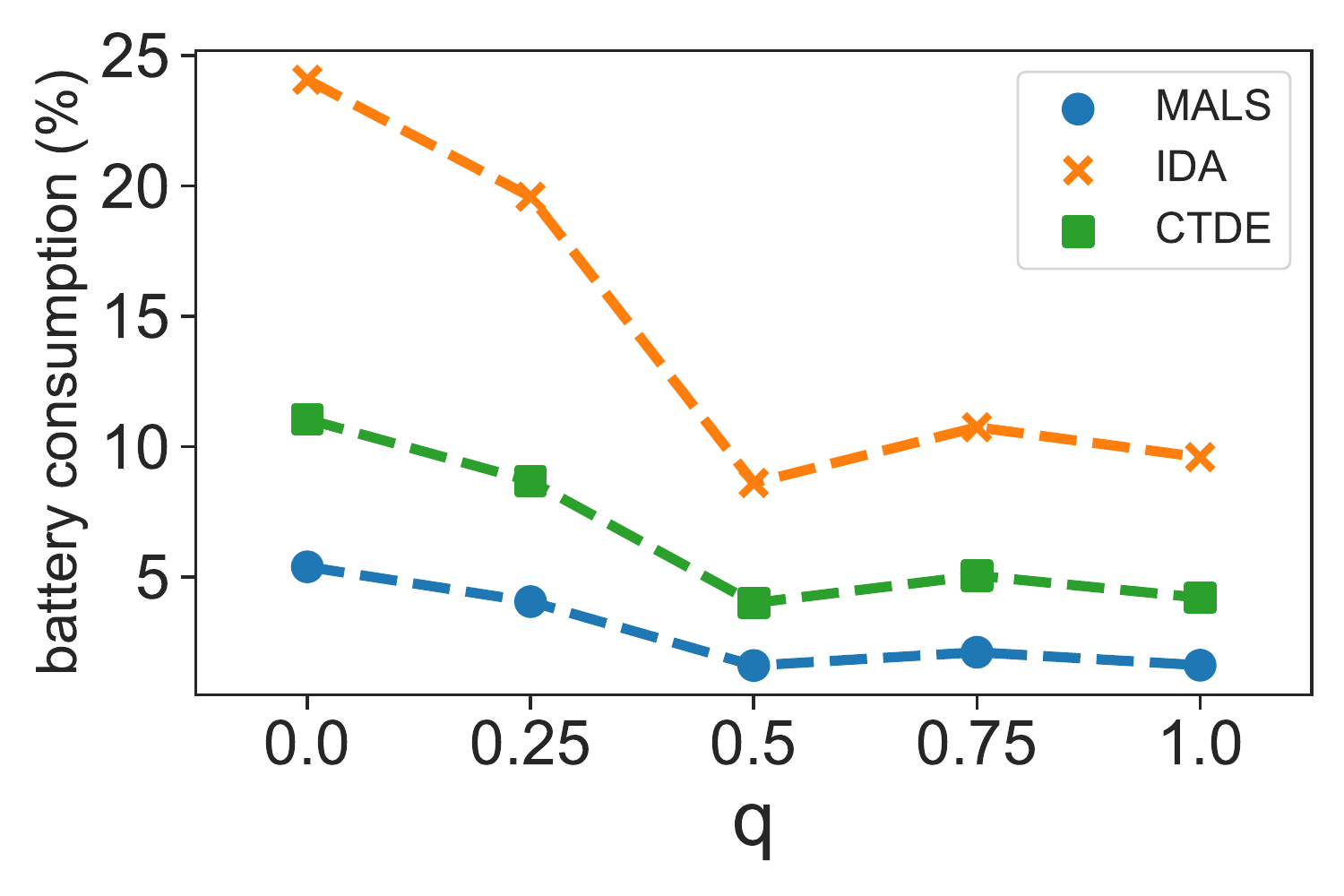}
\vspace{-10mm}
\end{minipage}
}%
\caption{Complete results - Metric values across different $q$ values.}
\label{fig:complete1}
\vspace{-0.5cm}
\end{figure*}

\begin{figure*}[t]

\centering
\subfigtopskip=2pt
\subfigbottomskip=2pt



\subfigure[Downlink delay with 6 UEs.]{
\begin{minipage}[t]{0.25\linewidth}
\centering
\includegraphics[width=1\linewidth]{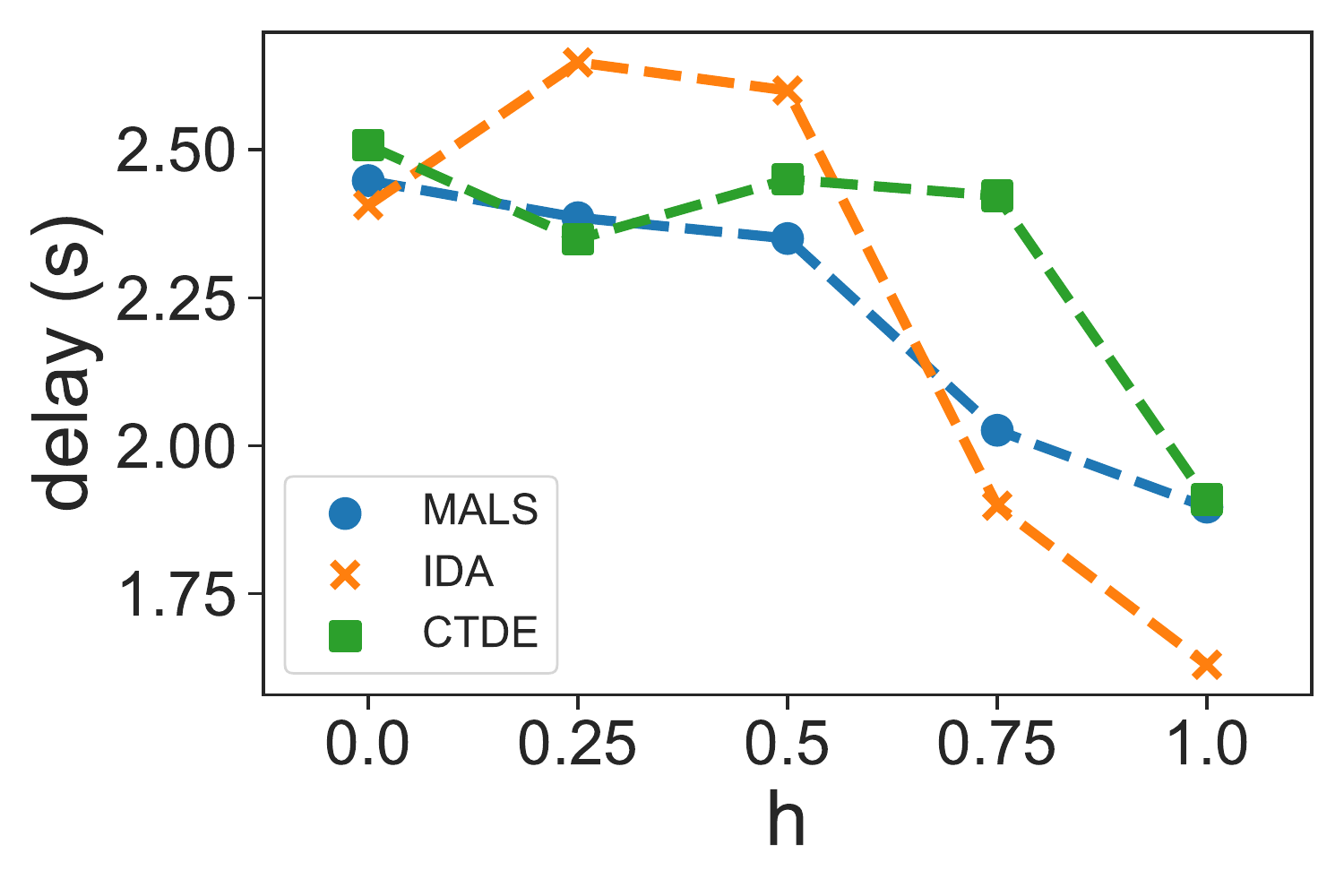}
\vspace{-10mm}
\end{minipage}
}%
\subfigure[Earning potential with  6 UEs.]{
\begin{minipage}[t]{0.25\linewidth}
\centering
\includegraphics[width=1\linewidth]{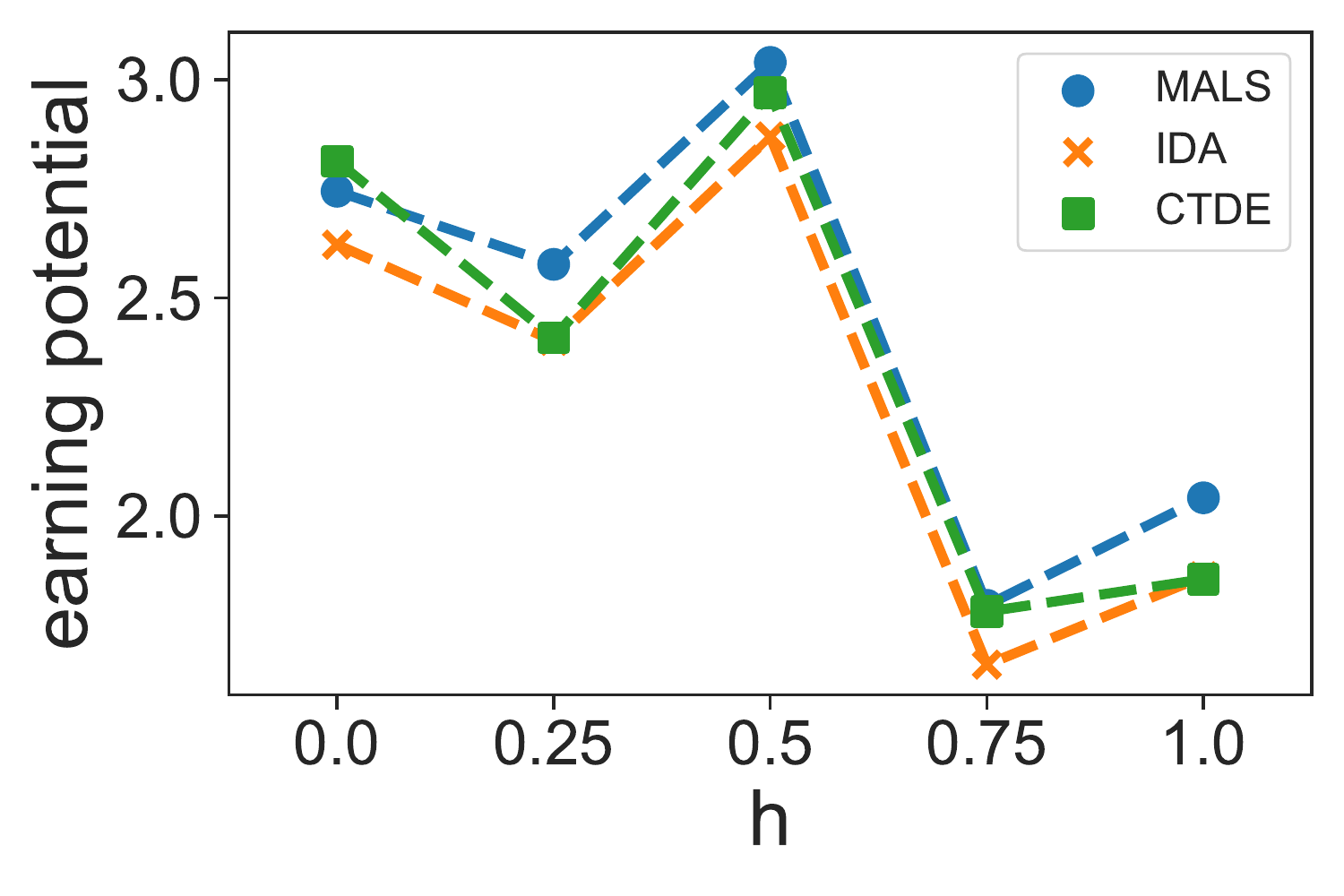}
\vspace{-10mm}
\end{minipage}
}%
\subfigure[Up-link delay with 6 UEs.]{
\begin{minipage}[t]{0.25\linewidth}
\centering
\includegraphics[width=1\linewidth]{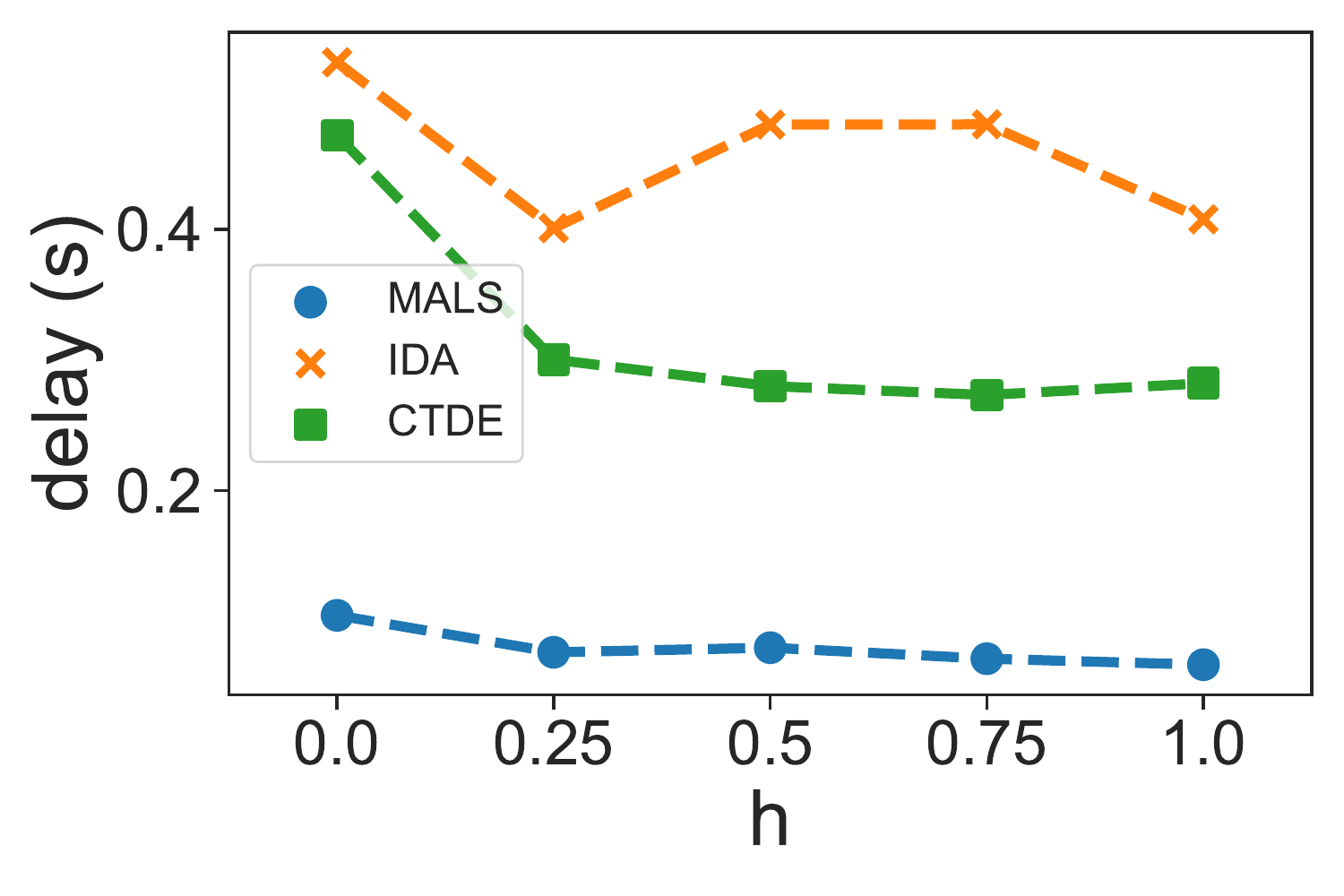}
\vspace{-10mm}
\end{minipage}
}%
\subfigure[Battery usage with 6 UEs.]{
\begin{minipage}[t]{0.25\linewidth}
\centering
\includegraphics[width=1\linewidth]{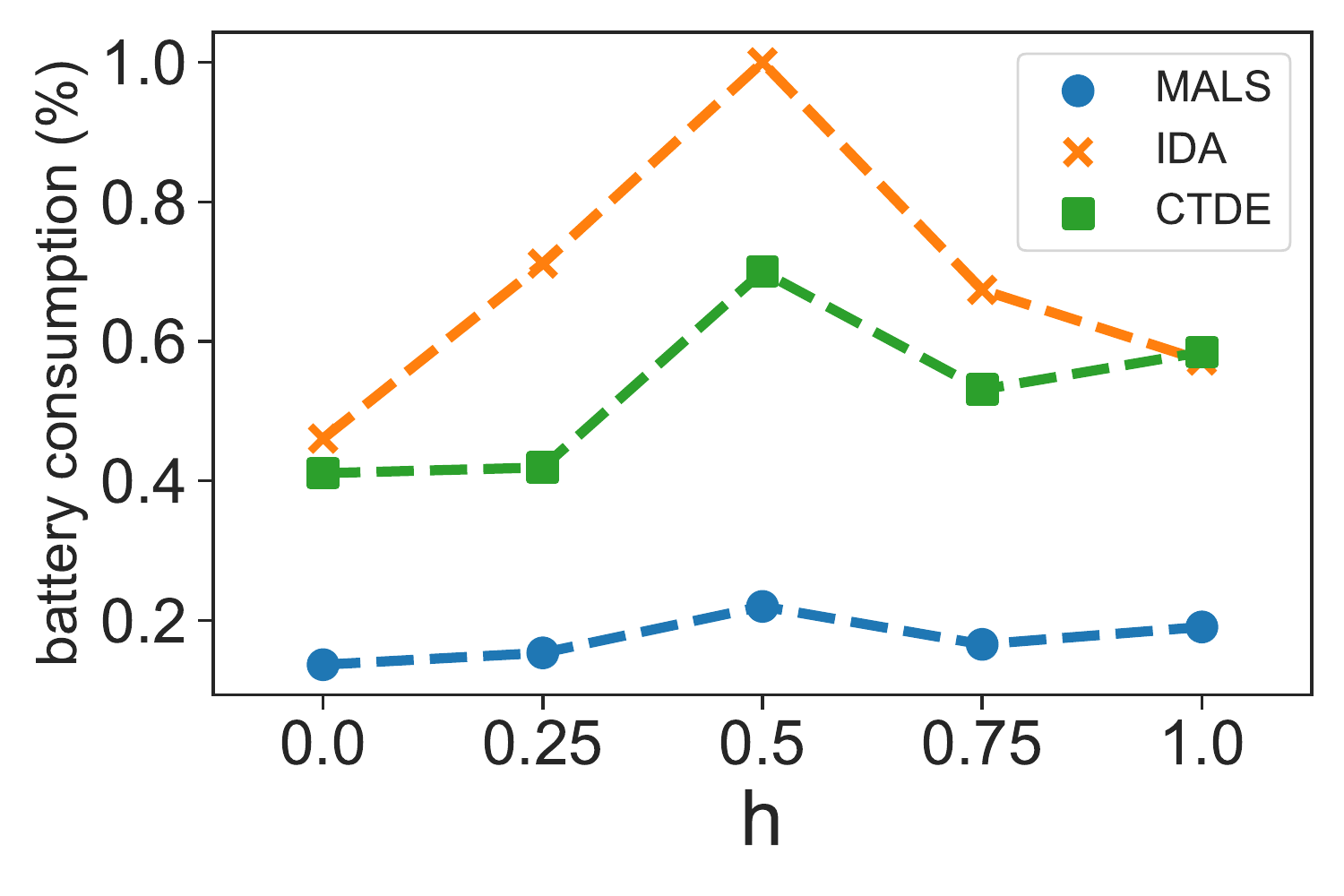}
\vspace{-10mm}
\end{minipage}%
}%

\subfigure[Downlink delay with 7 UEs.]{
\begin{minipage}[t]{0.25\linewidth}
\centering
\includegraphics[width=1\linewidth]{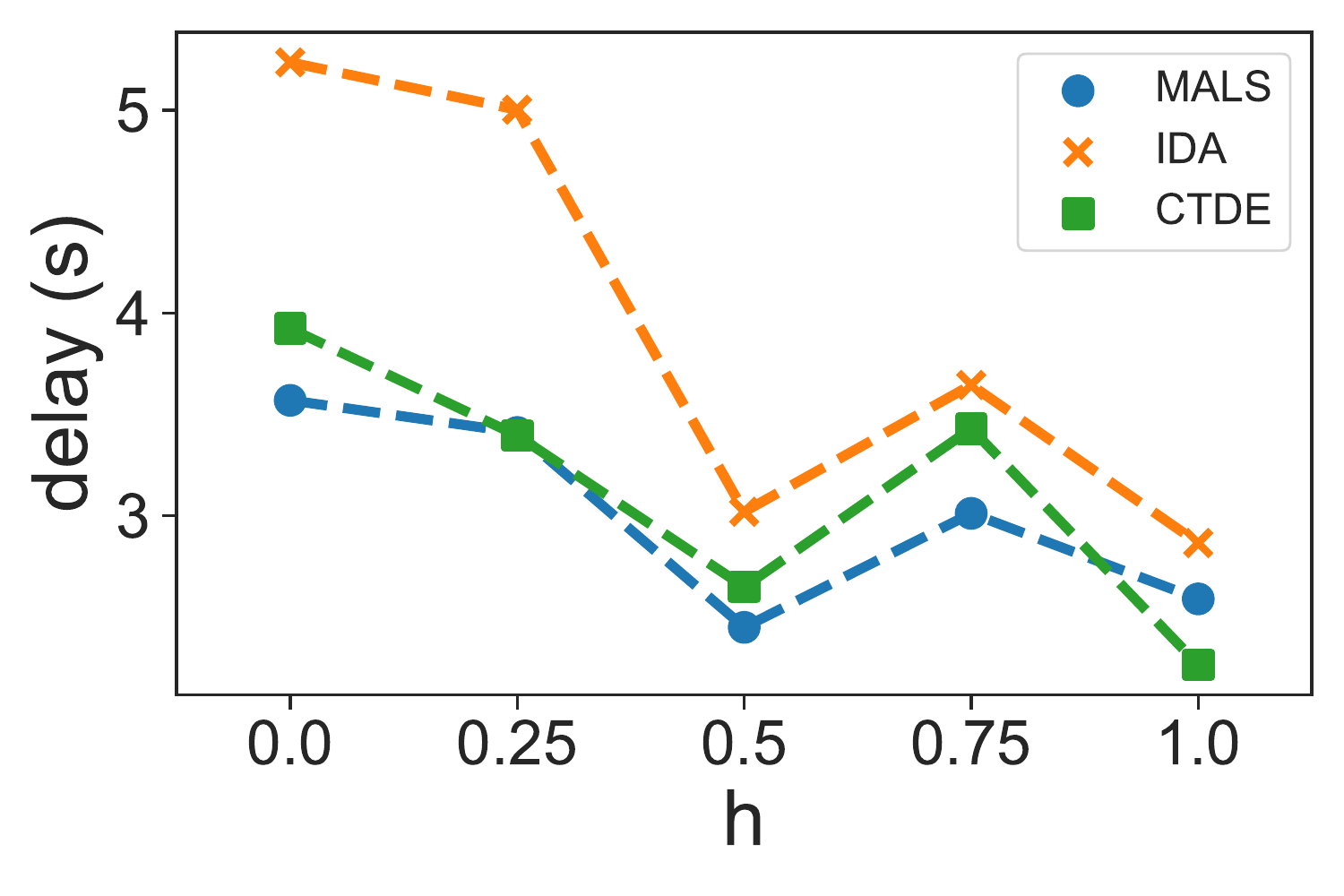}
\vspace{-10mm}
\end{minipage}
}%
\subfigure[Earning potential with 7 UEs.]{
\begin{minipage}[t]{0.25\linewidth}
\centering
\includegraphics[width=1\linewidth]{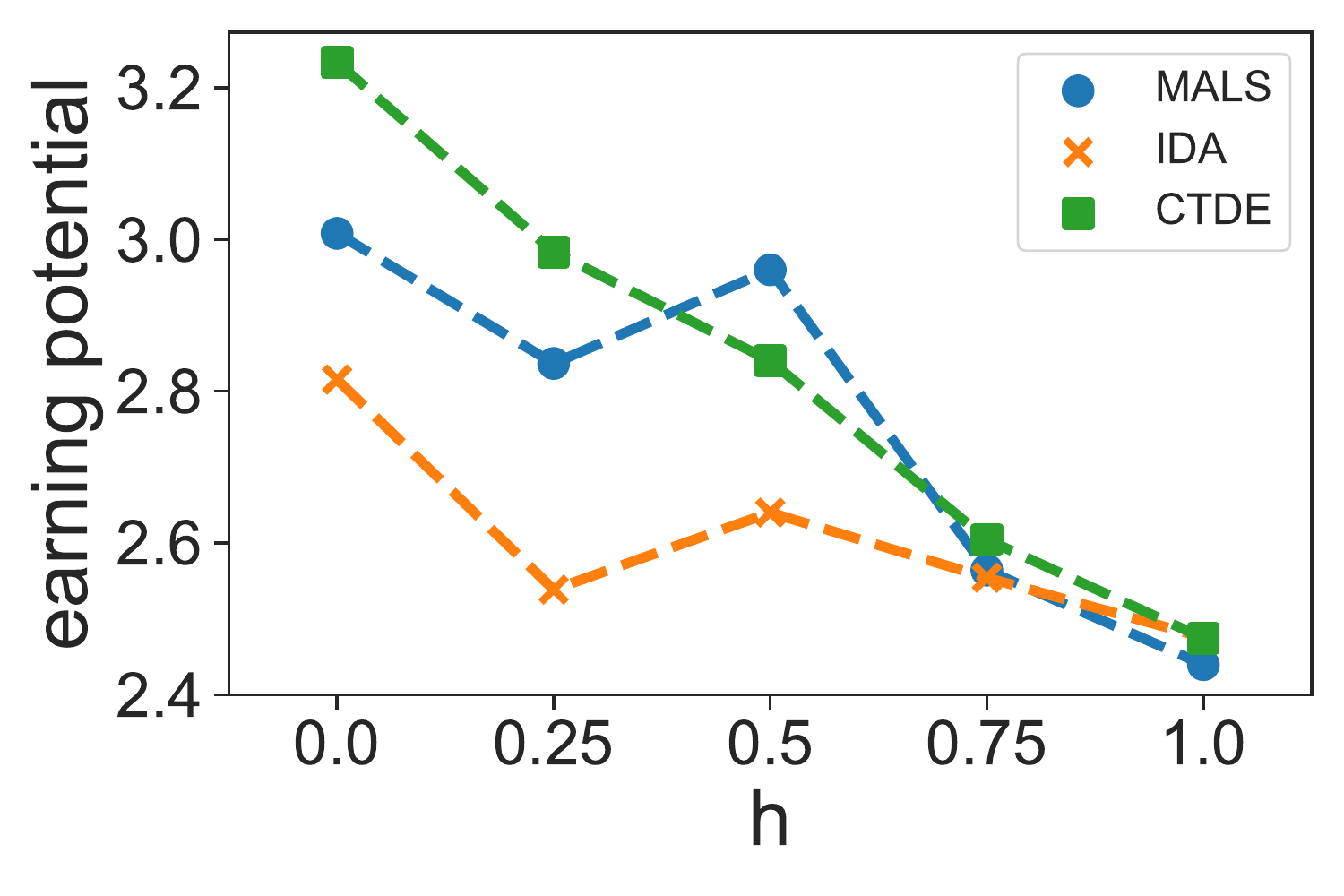}
\vspace{-10mm}
\end{minipage}
}%
\subfigure[Up-link delay with 7 UEs.]{
\begin{minipage}[t]{0.25\linewidth}
\centering
\includegraphics[width=1\linewidth]{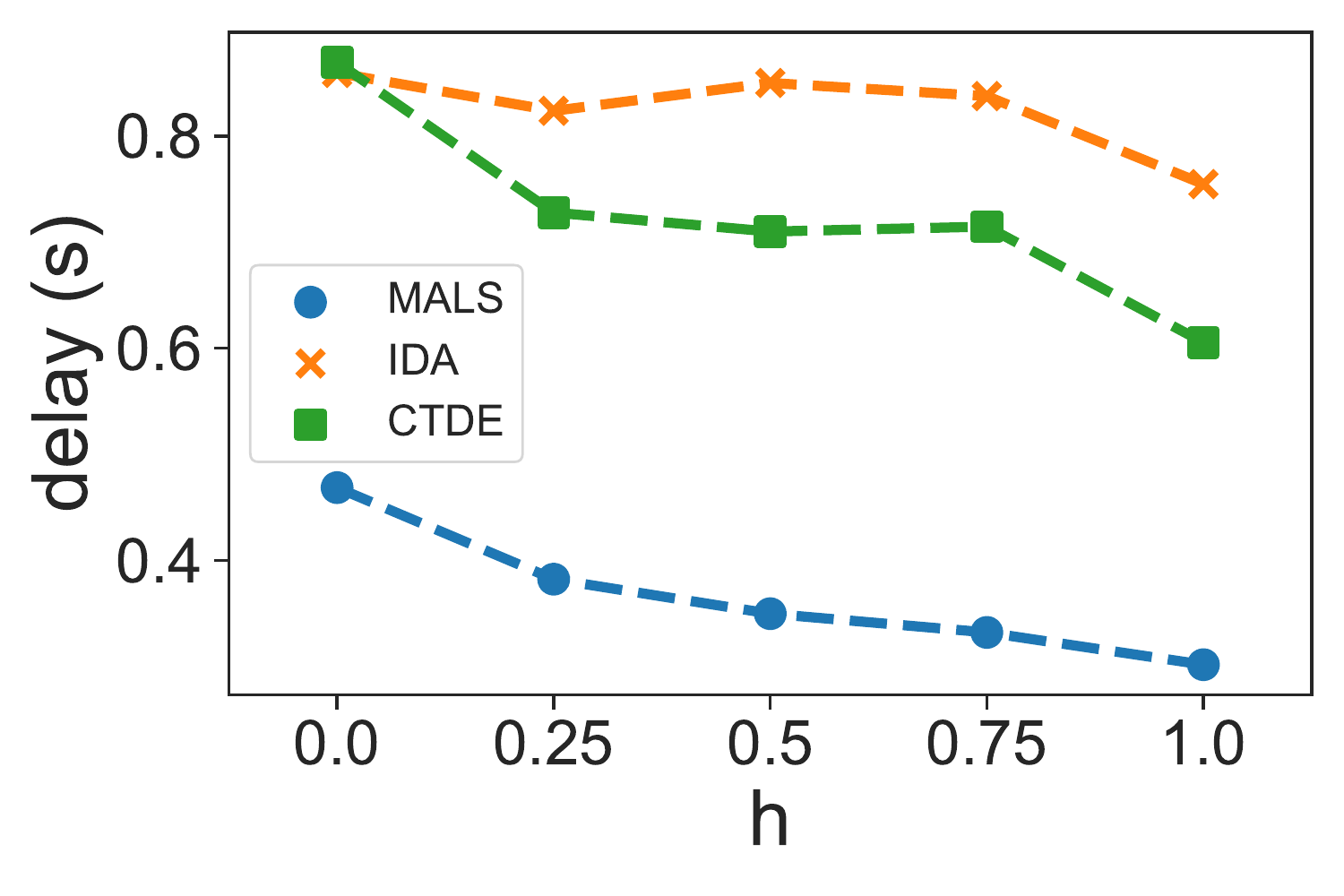}
\vspace{-10mm}
\end{minipage}%
}%
\subfigure[Battery usage with 7 UEs.]{
\begin{minipage}[t]{0.25\linewidth}
\centering
\includegraphics[width=1\linewidth]{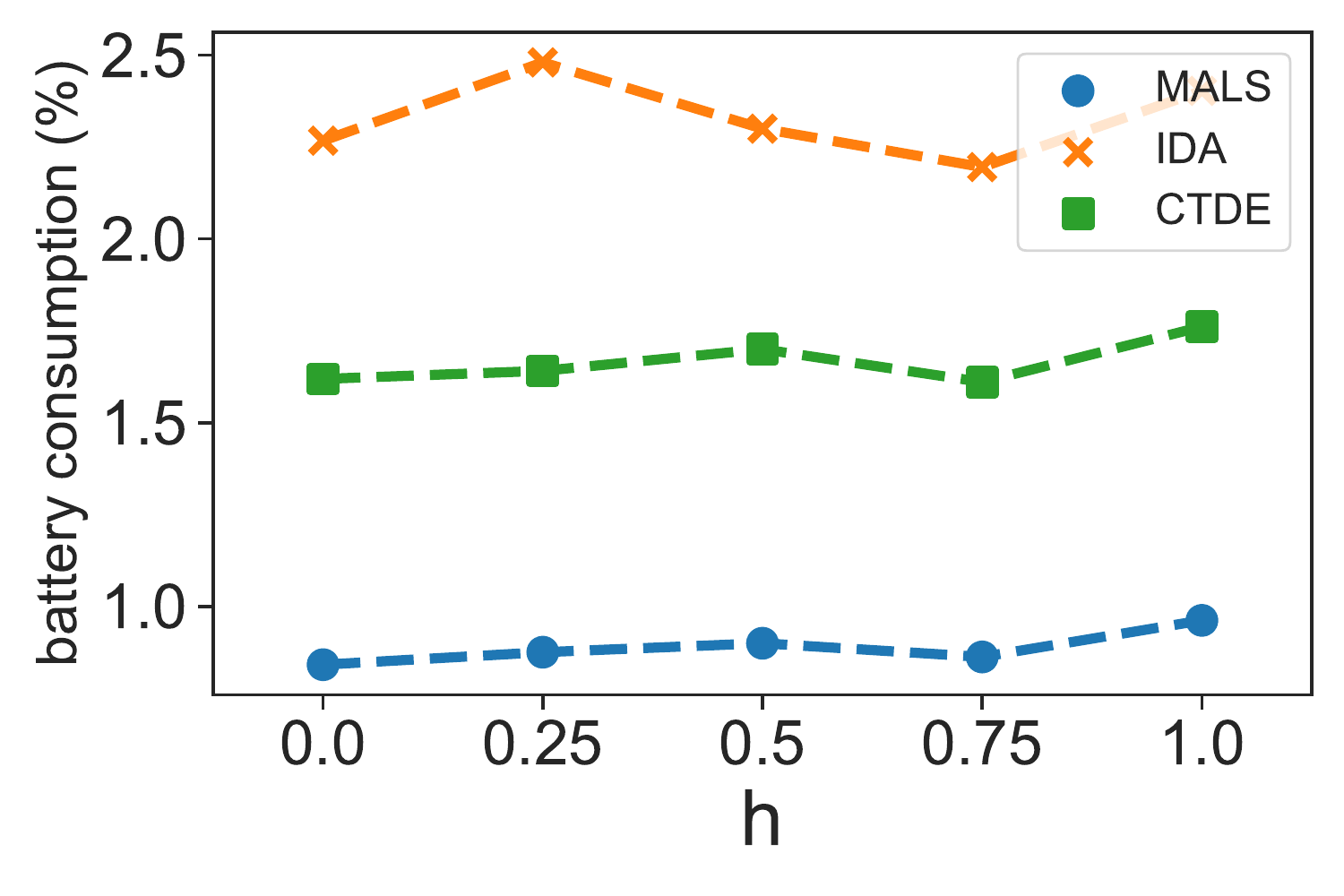}
\vspace{-10mm}
\end{minipage}%
}%

\subfigure[Downlink delay with 8 UEs.]{
\begin{minipage}[t]{0.25\linewidth}
\centering
\includegraphics[width=1\linewidth]{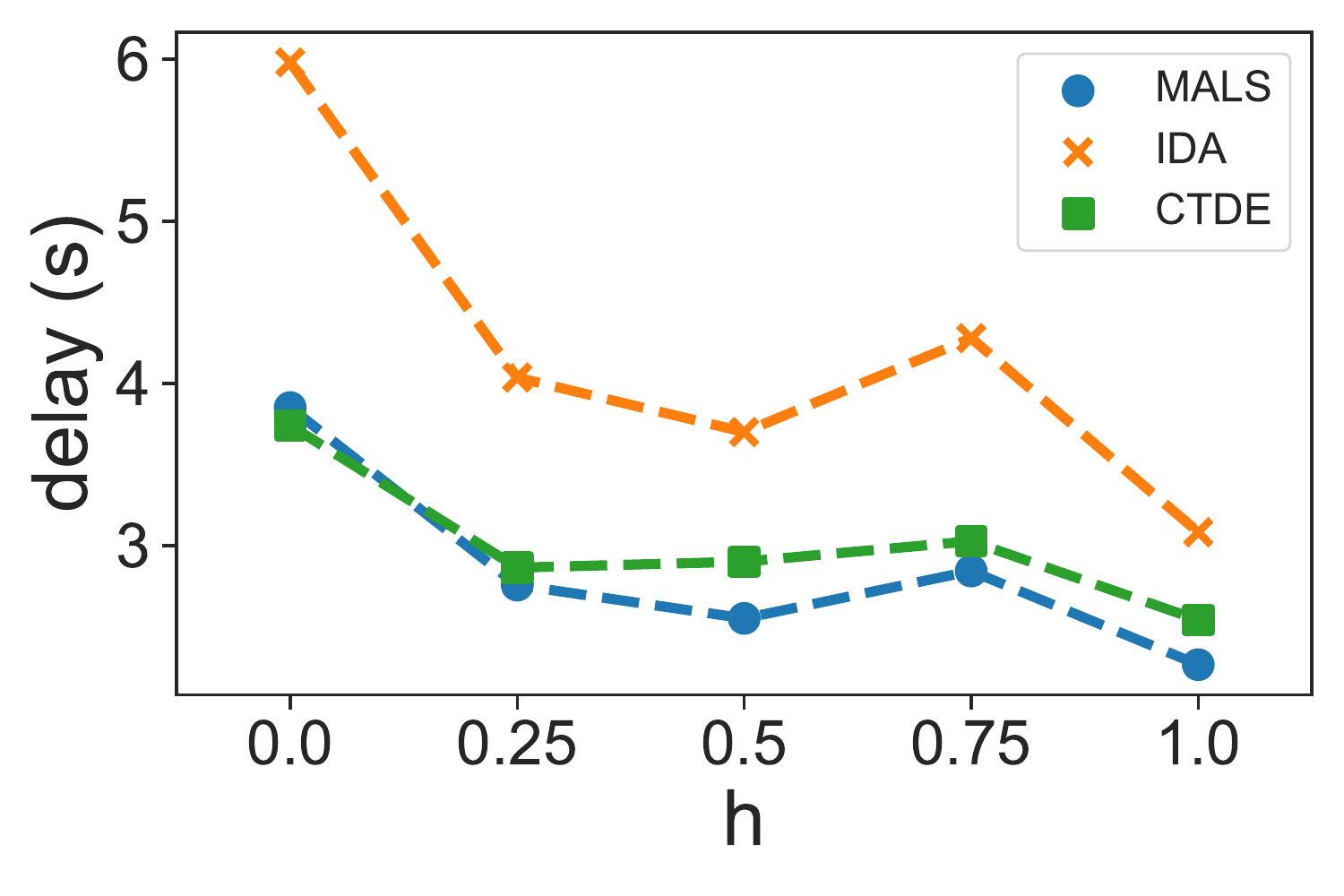}
\vspace{-10mm}
\end{minipage}
}%
\subfigure[Earning potential with 8 UEs.]{
\begin{minipage}[t]{0.25\linewidth}
\centering
\includegraphics[width=1\linewidth]{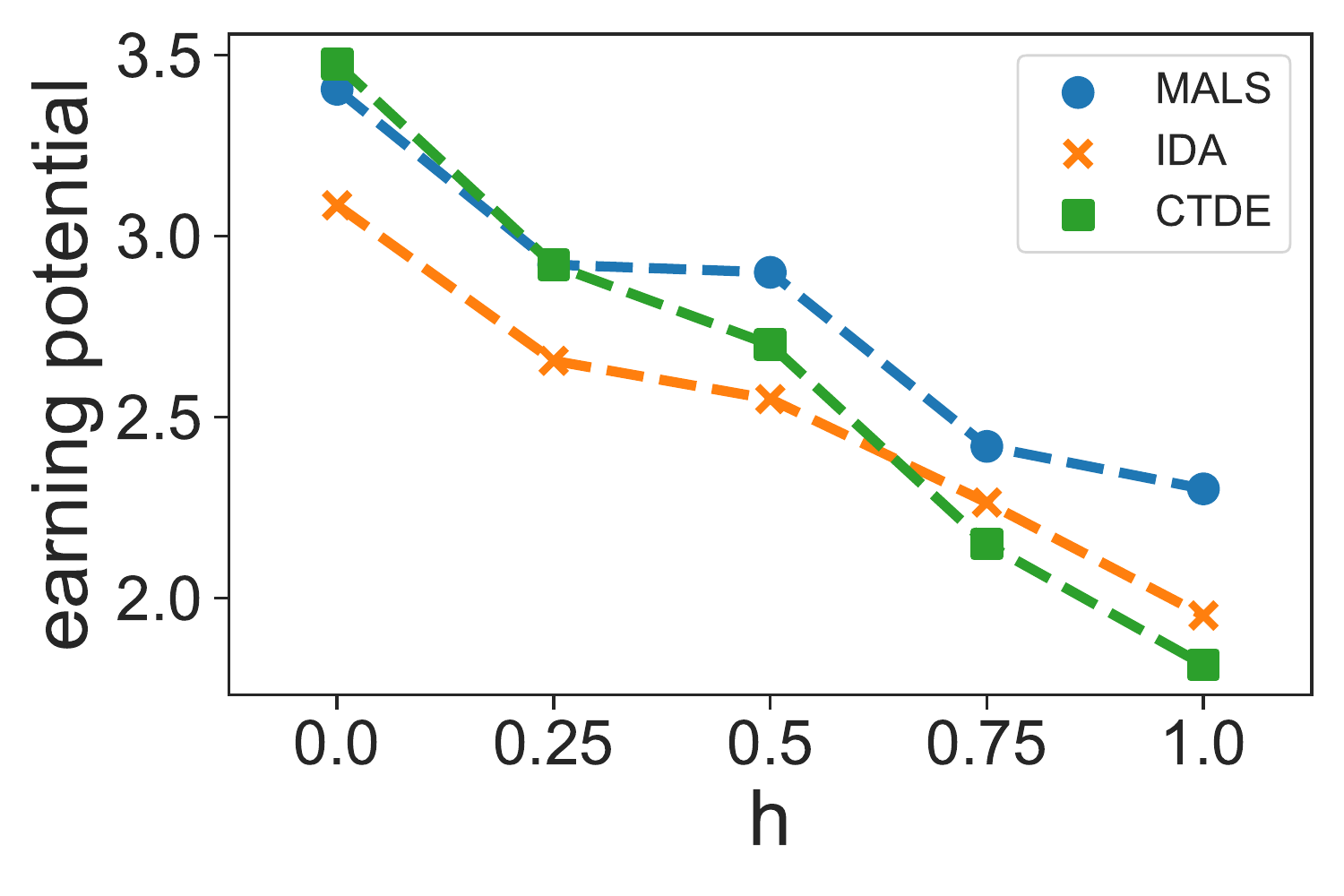}
\vspace{-10mm}
\end{minipage}%
}%
\subfigure[Up-link delay with 8 UEs.]{
\begin{minipage}[t]{0.25\linewidth}
\centering
\includegraphics[width=1\linewidth]{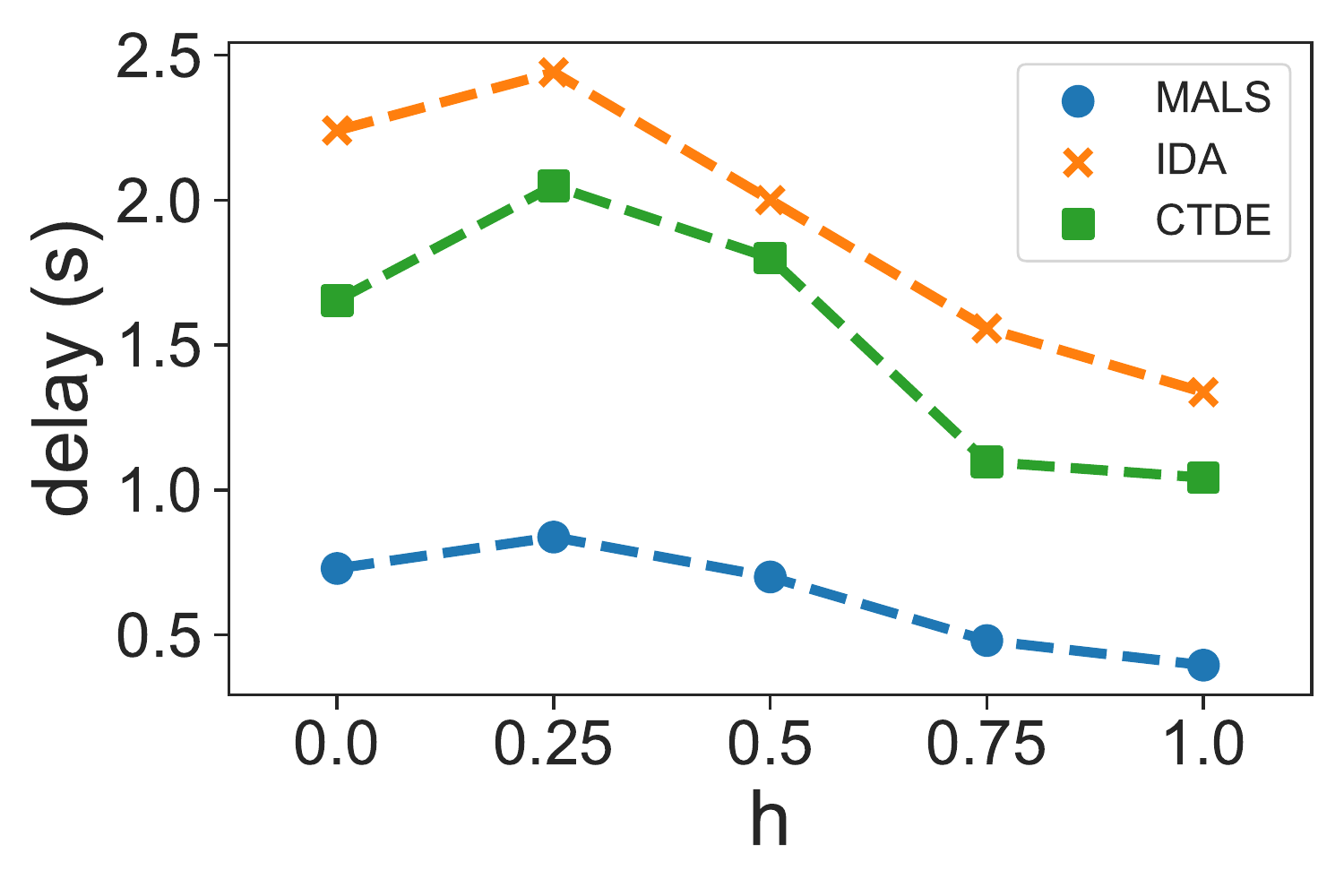}
\vspace{-10mm}
\end{minipage}%
}%
\subfigure[Battery usage with 8 UEs.]{
\begin{minipage}[t]{0.25\linewidth}
\centering
\includegraphics[width=1\linewidth]{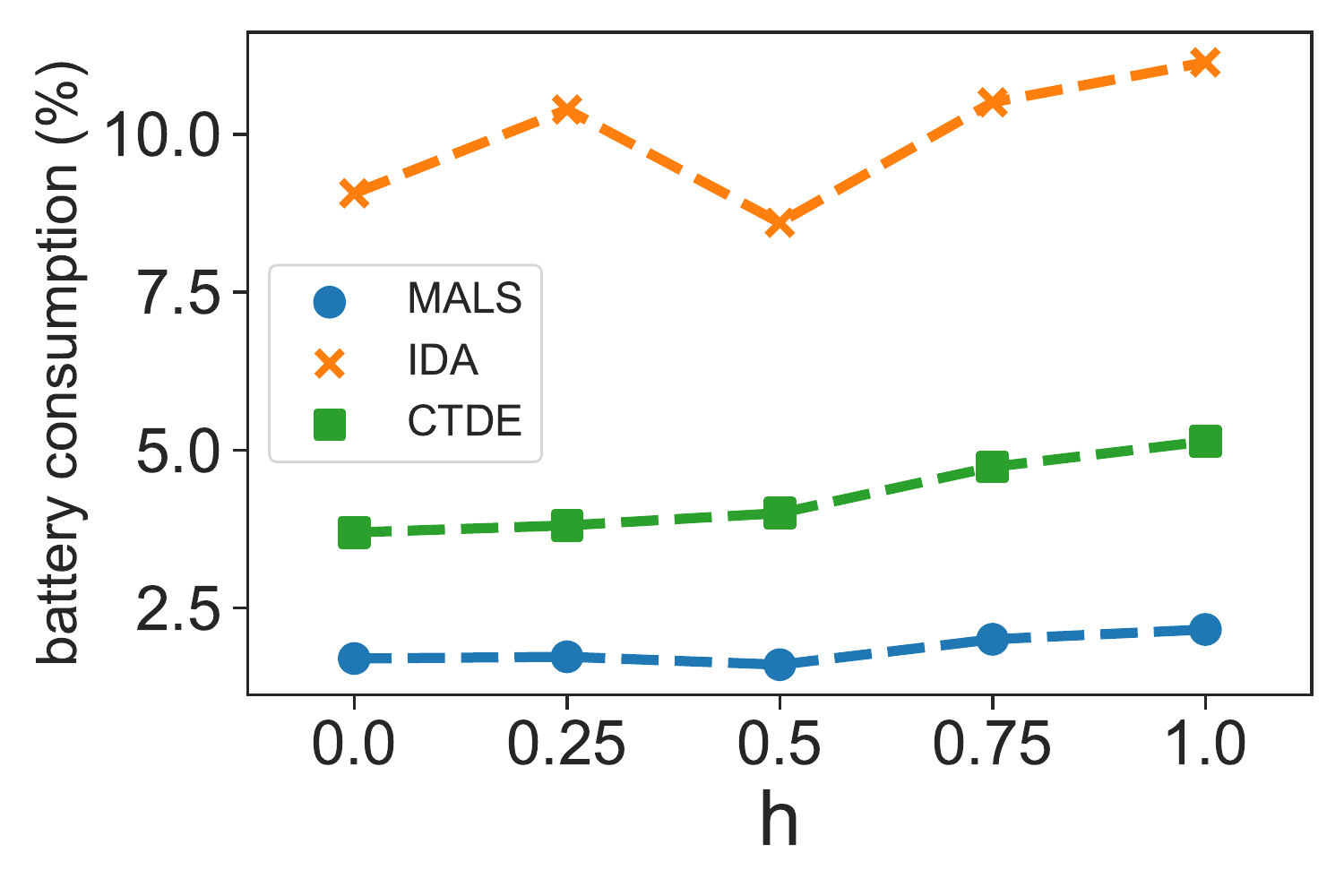}
\vspace{-10mm}
\end{minipage}
}%
\caption{Complete results - Metric values across different $h$ values.}
\label{fig:complete2}
\vspace{-0.5cm}
\end{figure*}

\begin{figure*}[t]

\centering
\subfigtopskip=2pt
\subfigbottomskip=2pt

\subfigure[Average downlink delay.]{
\begin{minipage}[t]{0.24\linewidth}
\centering
\includegraphics[width=1\linewidth]{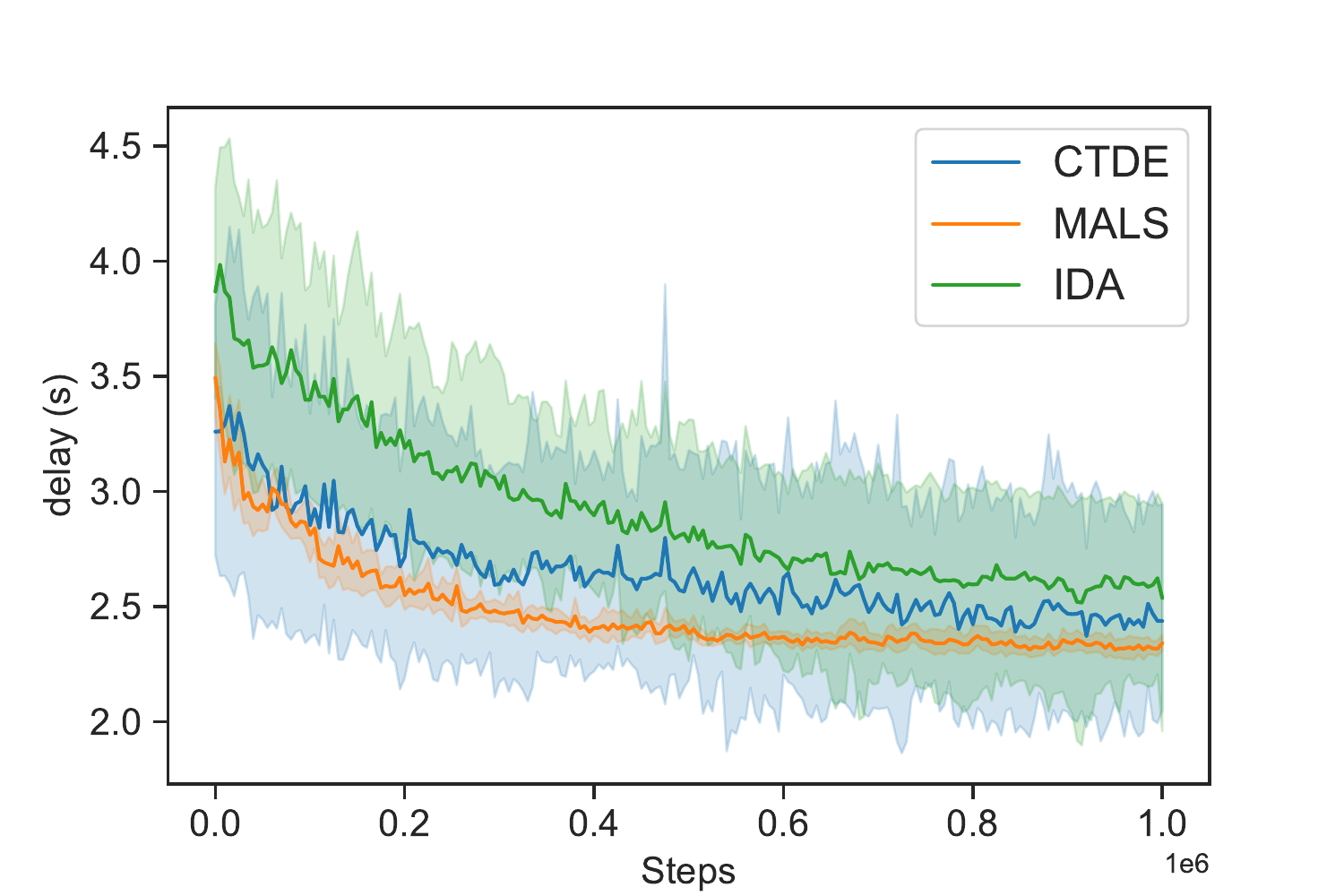}
\label{fig:delay_down_36}
\vspace{-10mm}
\end{minipage}%
}%
\subfigure[Average UE earning potential.]{
\begin{minipage}[t]{0.24\linewidth}
\centering
\includegraphics[width=1\linewidth]{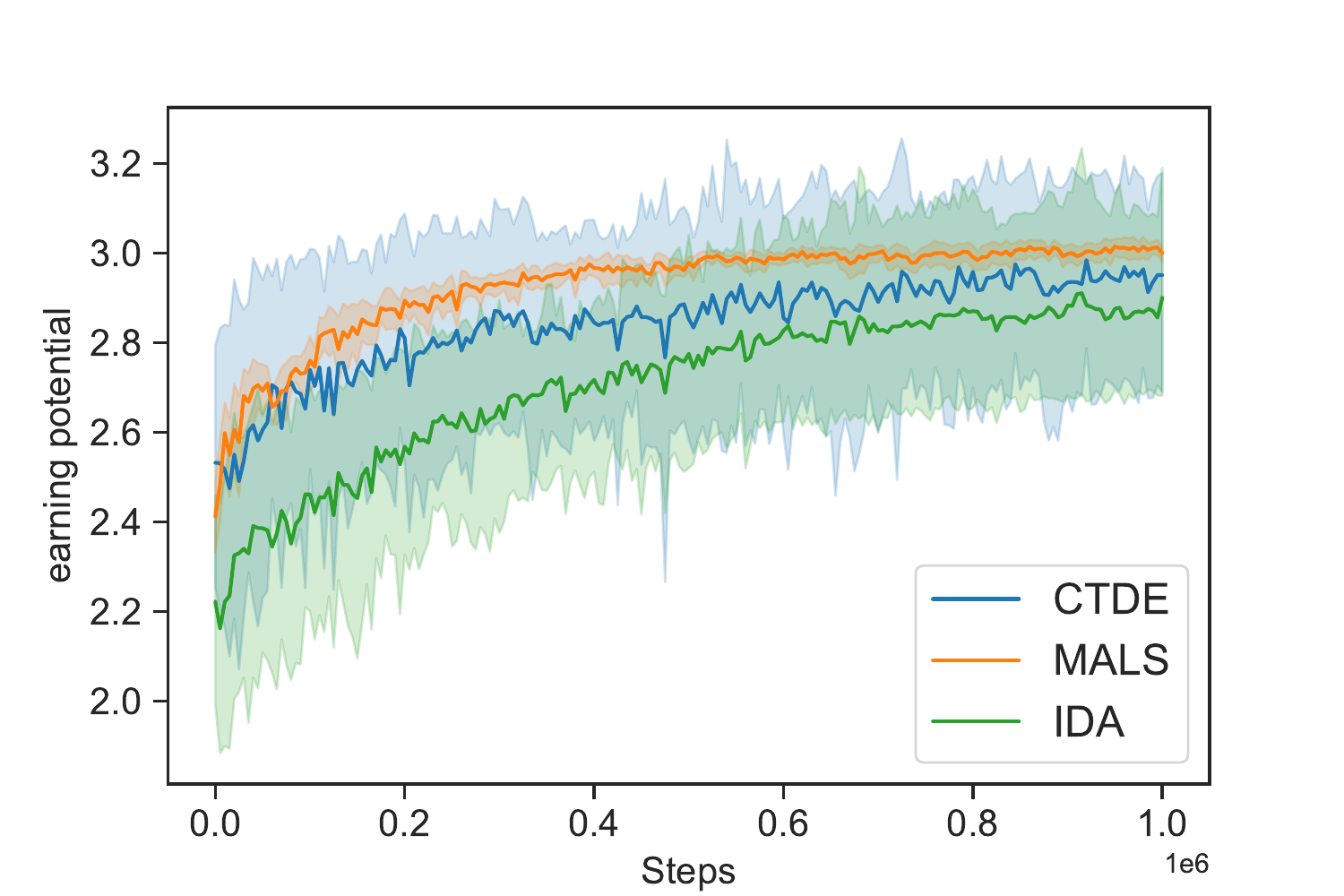}
\label{fig:earning_ability_36}
\vspace{-10mm}
\end{minipage}%
}%
\subfigure[Average uplink delay.]{
\begin{minipage}[t]{0.24\linewidth}
\centering
\includegraphics[width=1\linewidth]{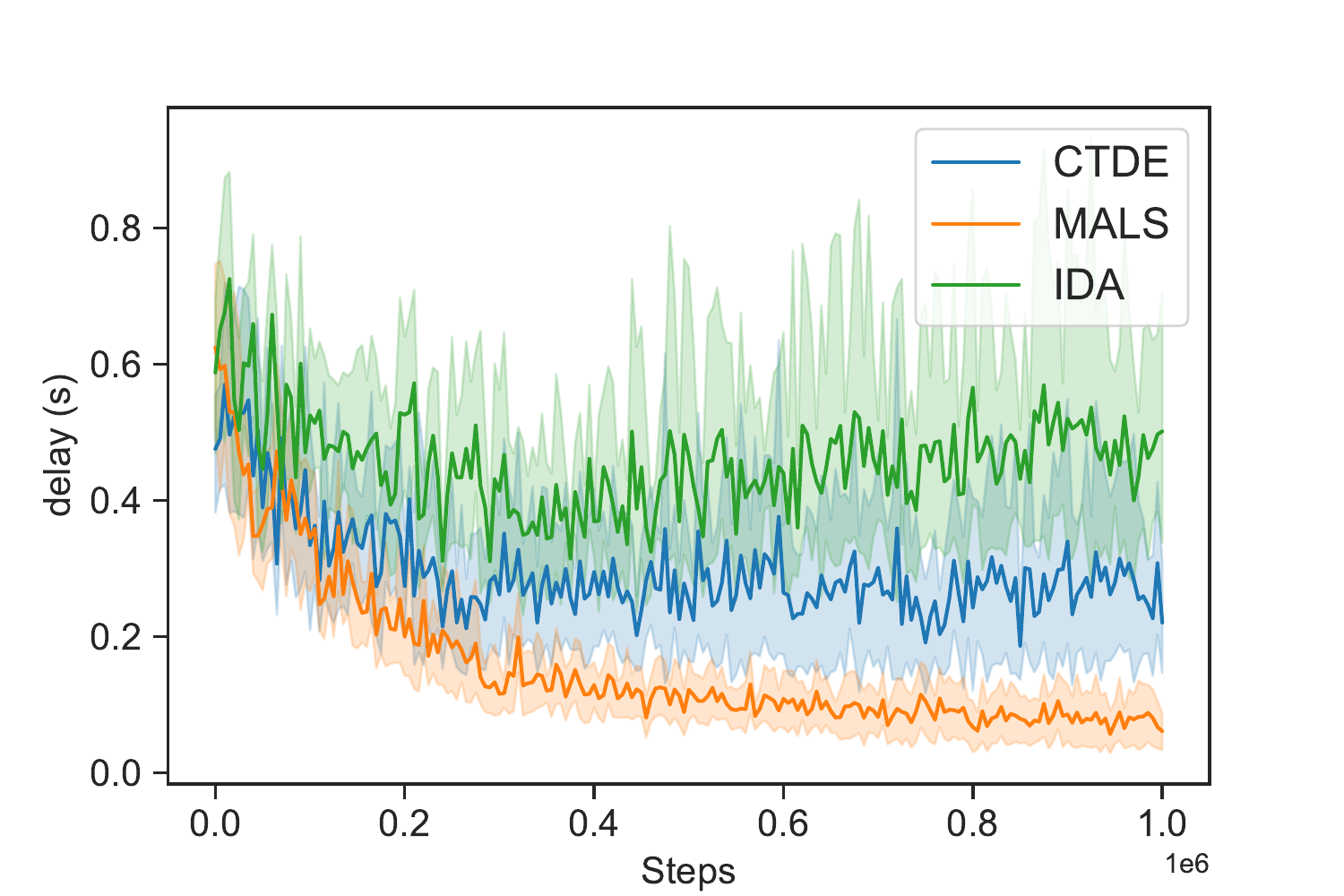}
\label{fig:log_delay_up_36}
\vspace{-10mm}
\end{minipage}
}%
\subfigure[Average battery consumption.]{
\begin{minipage}[t]{0.24\linewidth}
\centering
\includegraphics[width=1\linewidth]{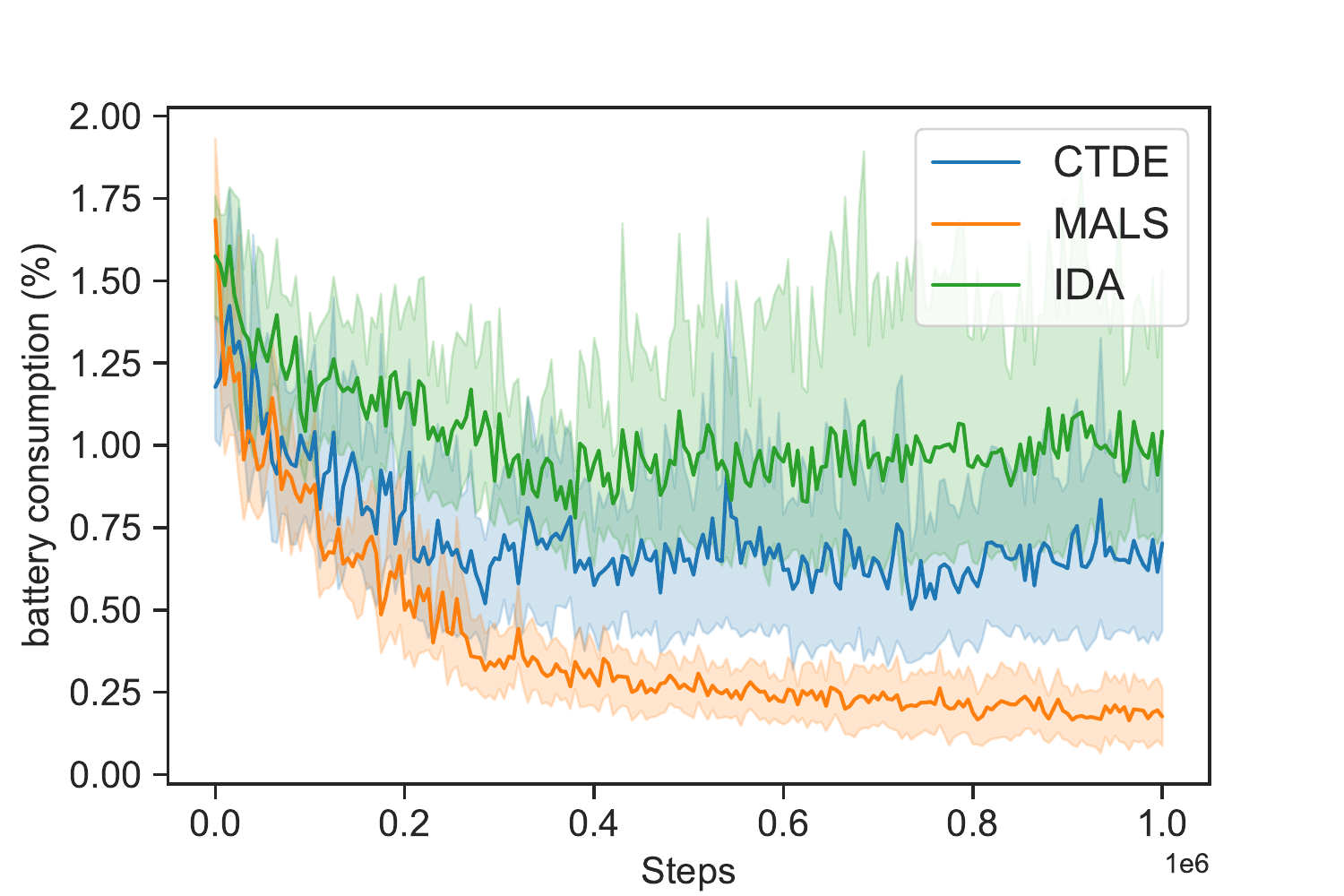}
\label{fig:log_battery_percent_consumed_36}
\vspace{-10mm}
\end{minipage}
}%
\caption{Key metrics performance obtained by MALS across training steps for the 4 MBS, 6 UE configuration.}
\label{fig:complete_weight_appendix_36}
\vspace{-0.5cm}
\end{figure*}

\begin{figure*}[t]

\centering
\subfigtopskip=2pt
\subfigbottomskip=2pt

\subfigure[Average downlink delay.]{
\begin{minipage}[t]{0.24\linewidth}
\centering
\includegraphics[width=1\linewidth]{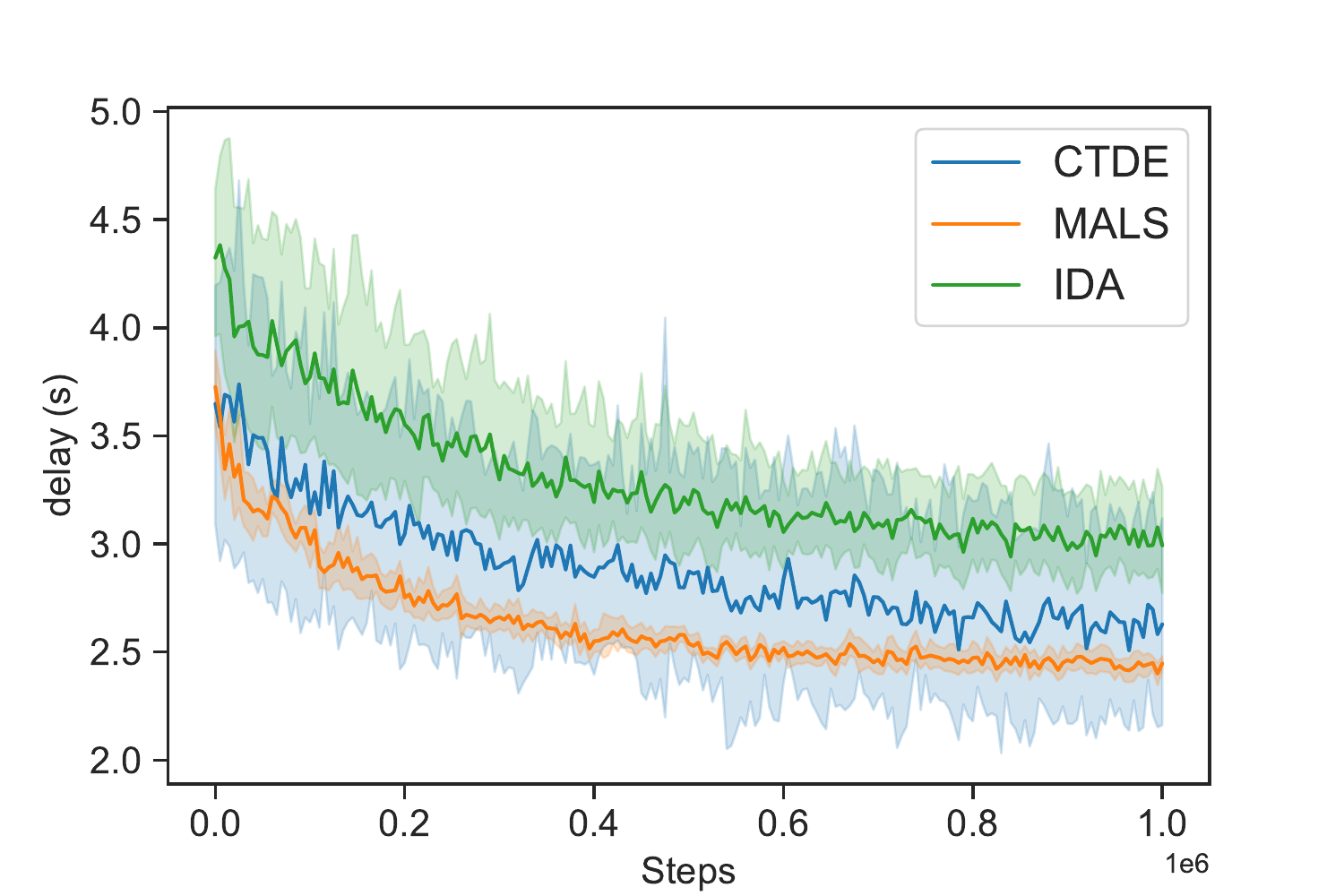}
\label{fig:delay_down_37}
\vspace{-10mm}
\end{minipage}%
}%
\subfigure[Average UE earning potential.]{
\begin{minipage}[t]{0.24\linewidth}
\centering
\includegraphics[width=1\linewidth]{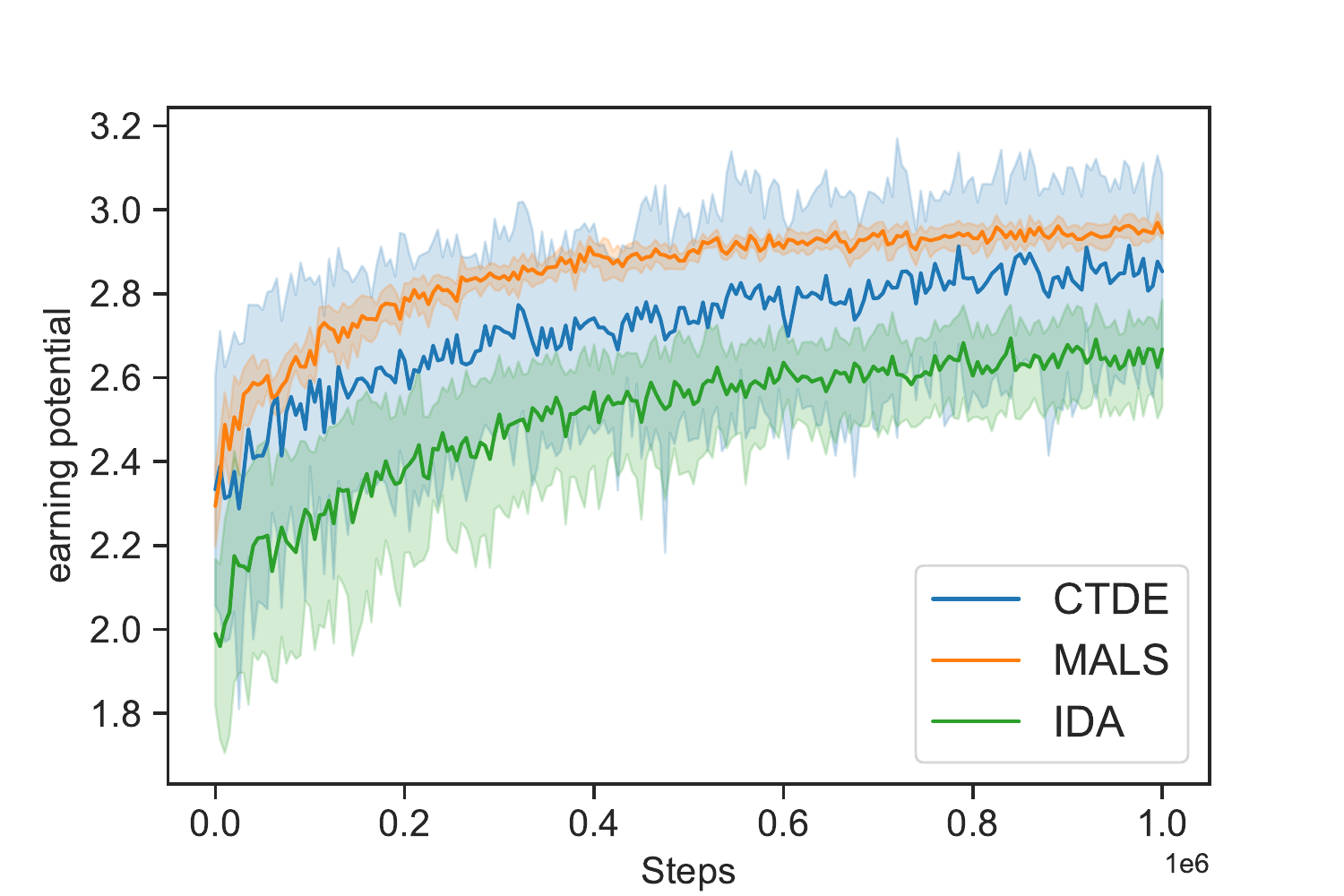}
\label{fig:earning_ability_37}
\vspace{-10mm}
\end{minipage}%
}%
\subfigure[Average uplink delay.]{
\begin{minipage}[t]{0.24\linewidth}
\centering
\includegraphics[width=1\linewidth]{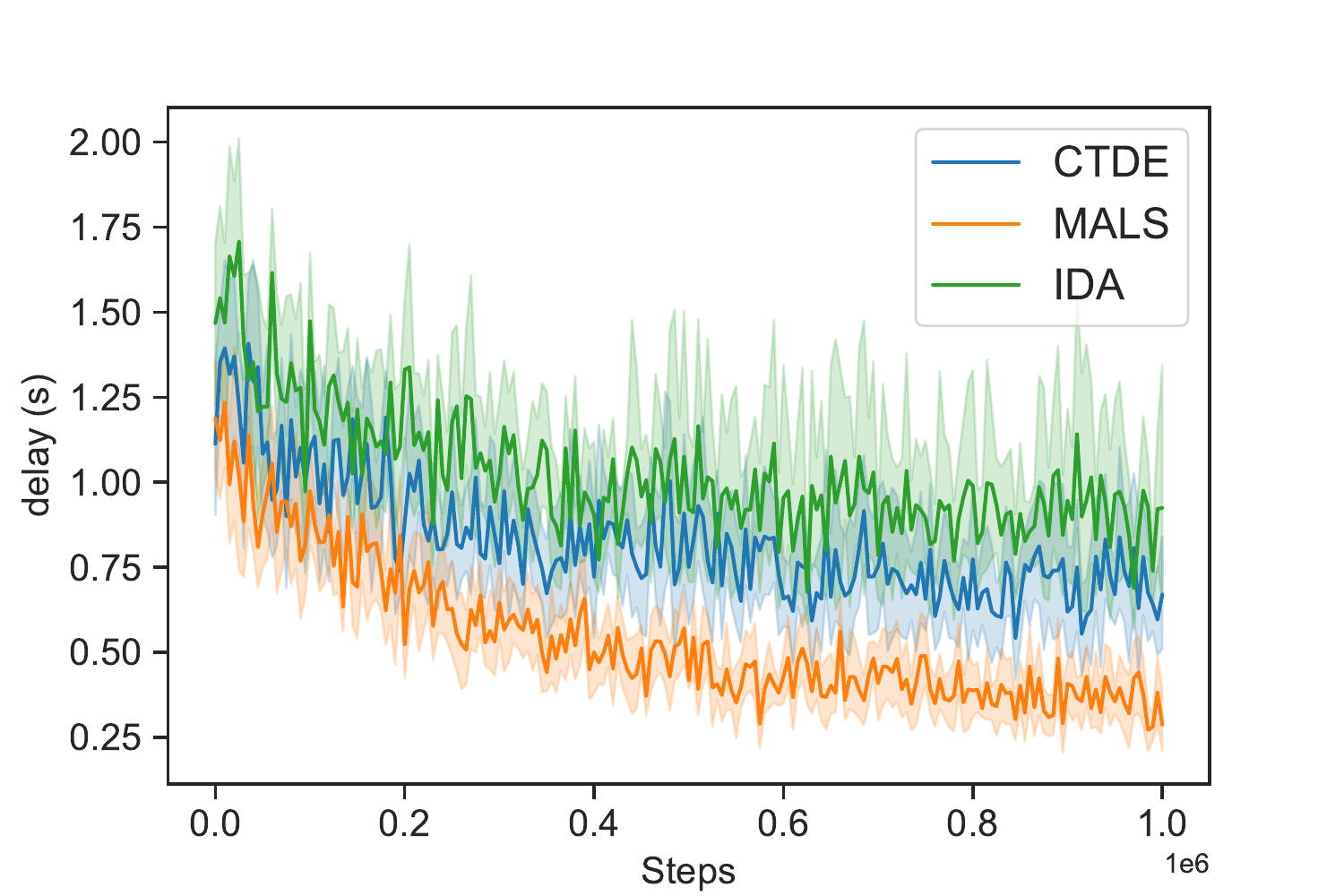}
\label{fig:log_delay_up_37}
\vspace{-10mm}
\end{minipage}
}%
\subfigure[Average battery consumption.]{
\begin{minipage}[t]{0.24\linewidth}
\centering
\includegraphics[width=1\linewidth]{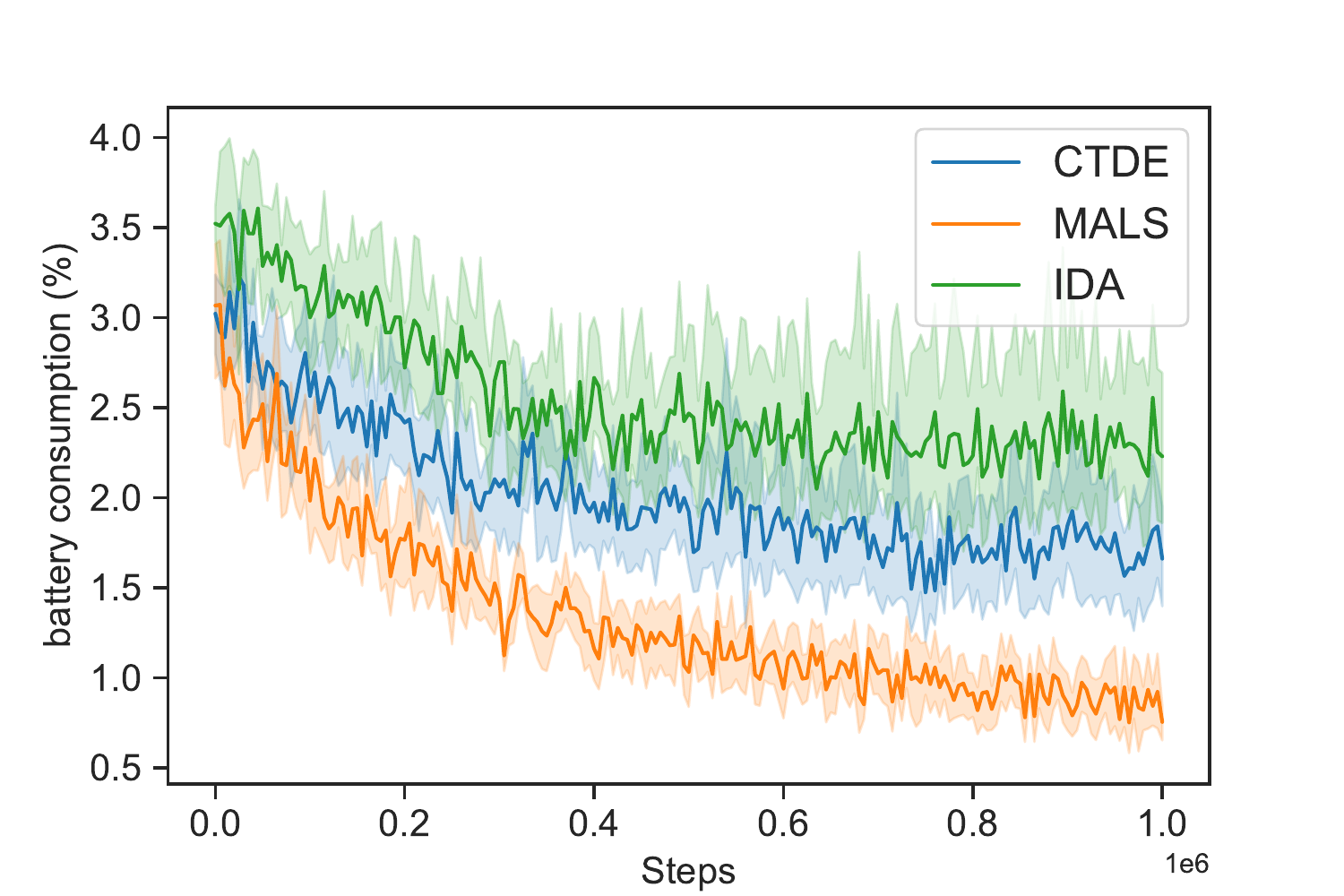}
\label{fig:log_battery_percent_consumed_37}
\vspace{-10mm}
\end{minipage}
}%
\caption{Key metrics performance obtained by MALS across training steps for the 4 MBS, 7 UE configuration.}
\label{fig:complete_weight_appendix}
\vspace{-0.5cm}
\end{figure*}

Fig.~\ref{fig:complete1} and \ref{fig:complete2} show how the performance metrics vary across different $q$ and $h$ values. The changes in $q$ and $h$ corresponds to changing weight emphasis between variables. Value of $h$ is $0.5$ when $q$ is varying, and vice versa. Figure~\ref{fig:complete_weight_appendix} and \ref{fig:complete_weight_appendix_36} shows the obtained metric values by the uplink and downlink agents in the 4 MBS, 7 UE and 4 MBS, 6 UE configuration, respectively.



\end{document}